\documentclass[aps,prd,a4paper,twocolumn,amsmath,showpacs,superscriptaddress,nofootinbib,preprintnumbers]{revtex4-1}
\pdfoutput=1
\usepackage{amssymb,amsmath,latexsym,mathrsfs}
\usepackage[sort&compress]{natbib}
\usepackage{graphicx,subfigure}
\usepackage{epsfig}
\usepackage{varioref,xr-hyper}
\usepackage{color}
\usepackage{multirow}
\usepackage{array}
\usepackage{hyperref}
\usepackage{wasysym}
\usepackage{color}
\usepackage{float}
\usepackage{xcolor}
\usepackage[utf8]{inputenc}
\usepackage[T1]{fontenc}
\hypersetup{colorlinks,linkcolor={blue},citecolor={blue},urlcolor={blue}}

\begin{document}

\title{On the dynamics of a dark sector coupling}

\author{Weiqiang Yang}
\email{d11102004@163.com}
\affiliation{Department of Physics, Liaoning Normal University, Dalian, 116029, P. R. China}

\author{Supriya Pan}
\email{supriya.maths@presiuniv.ac.in}
\affiliation{Department of Mathematics, Presidency University, 86/1 College Street, Kolkata
700073, India}
\affiliation{Institute of Systems Science, Durban University of Technology, PO Box 1334, Durban 4000, Republic of South Africa}

\author{Olga Mena}
\email{omena@ific.uv.es}
\affiliation{Instituto de F\'{i}sica Corpuscular (IFIC), University of Valencia-CSIC, Parc Cient\'{i}fic UV, c/ Cate\-dr\'{a}tico Jos\'{e} Beltr\'{a}n 2, E-46980 Paterna, Spain}

\author{Eleonora Di Valentino}
\email{e.divalentino@sheffield.ac.uk}
\affiliation{School of Mathematics and Statistics, University of Sheffield, Hounsfield Road, Sheffield S3 7RH, United Kingdom}

\begin{abstract}

Interacting dark energy models may play a crucial role in explaining several important observational issues in modern cosmology and also may provide a solution to current cosmological tensions.  Since the phenomenology of the dark sector could be extremely rich, one should not restrict the  interacting models to have a coupling parameter which is constant in cosmic time,  rather allow for  its dynamical behavior, as it is common practice in the literature when dealing with other dark energy properties, as  the dark energy equation of state. 
We present here a compendium of the current cosmological constraints on a large variety of interacting models, investigating scenarios where the coupling parameter of the interaction function and the dark energy equation of state 
can be either constant or dynamical. 
For the most general schemes, in which both the coupling parameter of the interaction function and the dark energy equation of state are dynamical, we find $95\%$~CL evidence for  a dark energy component at early times and slightly milder evidence for a dynamical dark coupling for the most complete observational data set exploited here, which includes CMB, BAO and Supernova Ia measurements.  Interestingly, there are some cases where a dark energy component different from the cosmological constant case at early times together with a coupling different from zero today, can alleviate both the $H_0$ and $S_8$ tension for the full dataset combination considered here. Due to the energy exchange among the dark sectors, the current values of the matter energy density and of the clustering parameter $\sigma_8$ are shifted from their $\Lambda$CDM-like values. This fact makes future surveys, especially those focused on weak lensing measurements, unique tools to test the nature and the couplings of the dark energy sector.

\end{abstract}

\maketitle
\tableofcontents

\section{Introduction}

 Interacting dark energy cosmologies~\cite{Amendola:1999er} are very appealing scenarios where to alleviate current cosmological problems, such as the cosmic coincidence (i.e. the so-called \emph{why now?} problem)~\cite{Wetterich:1994bg,Huey:2004qv,Cai:2004dk, Pavon:2005yx, Berger:2006db, delCampo:2006vv, delCampo:2008sr, delCampo:2008jx} and cosmological tensions, such as the one between CMB estimates~\cite{Planck:2018vyg,ACT:2020gnv,SPT-3G:2021eoc} and SH0ES (Supernovae and $H_0$ for the Equation of State of dark energy) measurements of the Hubble constant~\cite{Riess:2021jrx,Riess:2022mme}, with a significance of $\sim 5.3\sigma$ (see also Refs.~\cite{Verde:2019ivm,Knox:2019rjx,DiValentino:2020zio,Jedamzik:2020zmd,Riess:2019qba,DiValentino:2020vnx,DiValentino:2021izs,Perivolaropoulos:2021jda,Schoneberg:2021qvd,Shah:2021onj,DiValentino:2022fjm,Abdalla:2022yfr,Krishnan:2020vaf,Anchordoqui:2021gji,Colgain:2022nlb,Colgain:2022rxy,Naidoo:2022rda}).  Therefore, over the last several years, the intriguing possibility of an  interaction between the dark matter and dark energy fluids has been thoroughly investigated using different available cosmological  observations~\cite{Barrow:2006hia,Valiviita:2008iv,Gavela:2009cy,Gavela:2010tm,Li:2013bya,Salvatelli:2014zta,Wang:2014xca,Casas:2015qpa,vandeBruck:2016jgg,Yang:2016evp,Kumar:2016zpg, vandeBruck:2016hpz, Pan:2016ngu, Yang:2017yme,Sharov:2017iue, DiValentino:2017iww,Kumar:2017dnp, Mifsud:2017fsy, VanDeBruck:2017mua, Yang:2017ccc, Yang:2017zjs,Pan:2017ent, Yang:2018pej,Yang:2018ubt, Yang:2018uae,Yang:2018euj,Li:2018jiu, Yang:2018xlt,Yang:2018qec,Yang:2019uog,Li:2019ajo,DiValentino:2019ffd, Oikonomou:2019boy,DiValentino:2020kpf,Martinelli:2019dau,Kumar:2019wfs,Cheng:2019bkh,Lucca:2020zjb,Cid:2020kpp,Sa:2020fvn,BeltranJimenez:2020qdu,Wang:2021kxc,Kumar:2021eev,Nunes:2021zzi,DiValentino:2019jae,Yang:2019vni,Yang:2020uga,DiValentino:2020leo,Gomez-Valent:2020mqn,Yang:2019bpr,Pan:2020bur,Yao:2020hkw,Yao:2020pji,Mifsud:2019fut,Pan:2019gop,Paliathanasis:2019hbi,Yang:2019uzo,Yang:2020tax,Sa:2021eft,Kang:2021osc,daFonseca:2021imp,Yang:2021oxc,Lucca:2021dxo,Lucca:2021eqy,Gariazzo:2021qtg,Paliathanasis:2021egx,Mawas:2021tdx,Potting:2021bje,Harko:2022unn,Nunes:2022bhn,Yengejeh:2022tpa,Thipaksorn:2022yul,Landim:2022jgr,Yao:2022kub,Gomez-Valent:2022bku} (also see~\cite{Bolotin:2013jpa,Wang:2016lxa}, two review articles on interacting dark energy models). 
 
The basic underlying idea in these theories relies on the possible non-gravitational interaction between dark matter (DM) and dark energy (DE). Such an interaction can be characterized by a continuous flow of energy and/or momentum between these dark sectors. This energy flow modifies the expansion history of the universe both at the background and perturbation levels.  

The interaction function, also known as the coupling function, is the main feature of interacting dark energy theories: once the interaction function is prescribed, the dynamics of the universe can be determined either analytically or numerically.  Despite the fact that in the literature it is usually assumed a pure phenomenological approach for the dark sectors coupling, a class of interaction functions can be derived from the field theory perspective~\cite{vandeBruck:2015ida, Boehmer:2015kta,Boehmer:2015sha,Gleyzes:2015pma,DAmico:2016jbm,Pan:2020zza,Chatzidakis:2022mpf}. 
Thus, theoretically, the interaction functions can also be well-motivated.

The most exploited parameterization of the interaction function is $Q = 3 H \xi f(\rho_{\rm DM}, \rho_{\rm DE})$~\cite{Gavela:2009cy}, where $\xi$ is the coupling parameter that characterizes the strength of the interaction function, $H$ is the Hubble parameter of the FLRW universe and $f(\rho_{\rm DM}, \rho_{\rm DE})$ is any continuous function of 
$\rho_{\rm DM}$, $\rho_{\rm DE}$, i.e. of the energy densities of dark matter and dark energy. For simplicity, the coupling parameter, $\xi$, is commonly assumed to be independent of the cosmic time. This economical assumption has been however questioned in Refs.~\cite{Li:2011ga,Wang:2014xca,Yang:2019uzo,Yang:2020tax}, where it has been argued that there is no fundamental symmetry in nature forcing that coupling to be zero or constant. The assumption of a constant coupling parameter, together with a constant equation of state for the dark energy component provide only a minimal picture of a (possibly) very rich dark sector dynamics. A mandatory exercise is therefore to understand the quality of the fit of the different interacting scenarios to the observational constraints, assessing the preference for a constant \emph{versus} a dynamical coupling parameter in scenarios where the dark energy equation of state is either constant or dynamical. 
We follow a bottom up approach here, increasing step by step the complexity of the dark sectors, starting from the simplest scenario to arrive to the most general one. The main purpose is to scrutinize the quality of the fitting of the possible scenarios to the observational data, devoting special attention to their parameter degeneracies and their ability in solving the present cosmological tensions.

The study is presented as follows. Sec.~\ref{sec-efe} presents the basic equations of interacting scenarios. Sec.~\ref{sec-data} contains the description of the observational data and the fitting methodology applied to constrain the interacting models. In Sec.~\ref{sec-results} we discuss the results in the different scenarios, ordered from less to more complexity. Finally, we conclude the study in Sec.~\ref{sec-conclu}, discussing some remarks relevant for future work in interacting dark sector cosmologies.

\section{Interacting dark sector scenarios}
\label{sec-efe}

We consider a homogeneous and isotropic description of our universe which is well described by the Friedmann-Lema\^{i}tre-Robertson-Walker (FLRW) metric:

\begin{eqnarray}
d{\rm s}^2 =  - dt^2 + a^2 (t) \left[\frac{dr^2}{1-k r^2} + r^2 \left(d\theta ^2 + \sin^2 \theta d \phi^2 \right) \right]~,  \nonumber
\end{eqnarray}
expressed in terms of the comoving coordinates $(t, r, \theta, \phi)$ where $a (t)$ refers to the expansion scale factor of the universe and $k$ is the curvature scalar. The curvature scalar may take three distinct values, $k =0,~+1,~-1$, to represent three different geometries of the Universe, spatially flat, closed and open, respectively.  In the following, we shall work in a flat universe with an interaction between the pressure-less dark matter/cold dark matter (abbreviated as CDM) and the dark energy components, governed by the conservation equation:  

\begin{eqnarray}
\nabla^\mu (T^{\rm CDM}_{\mu\nu} + T^{\rm DE}_{\mu\nu})= 0~,\label{conservation}
\end{eqnarray}

\noindent 
which can be decoupled into two equations with the introduction of an interaction function $Q (t)$ as follows

\begin{eqnarray}
\dot{\rho}_{\rm CDM} + 3 H \rho_{\rm CDM} = - Q (t)~, \label{cons1}\\
\dot{\rho}_{\rm DE} + 3 H (1+w) \rho_{\rm DE}  =  Q (t)~;\label{cons2}
\end{eqnarray}

\noindent 
where $H \equiv \dot{a}/a$ is the Hubble rate of a FLRW flat universe and $w$ denotes the equation of state of the dark energy component. Let us note that for  $Q (t) > 0$, the energy transfer occurs from CDM to DE, while  $Q <0$  indicates the transfer of energy in the reverse direction, i.e.  from DE  to CDM. In this work we shall consider the following two well-known interaction functions: 
\begin{eqnarray}
&&{\rm Model\; 1:}\; Q_{A}  = 3 H  \xi  \rho_{\rm DE}~, \label{modelA}\\
&&{\rm Model\; 2:}\; Q_{B}  = 3 H \xi \left(\frac{\rho_{\rm CDM} \rho_{\rm DE}}{\rho_{\rm CDM} +\rho_{\rm DE}}\right)~;\label{modelB} 
\end{eqnarray}

\noindent 
where $\xi$ is a coupling parameter of the interaction functions and it could be either time-dependent or time-independent. Now, in agreement with the sign convention of the interaction function, $\xi >0$ (equivalently, $Q> 0$) means an energy transfer from CDM to DE and $\xi <0$ means that the energy flow is from DE to CDM.  Notice  that for some interaction models the energy density of dark matter and/or of dark energy could be negative~\cite{Pan:2020mst}. 
In this regard we do not impose any further constraints on the fluids and let the data discriminate between the most observationally favoured scenarios.  We believe this approach is appropriate because it avoids unwanted biases and also considers the rising interest in the community in the putative presence of a negative cosmological constant~\cite{Delubac:2014aqe,Poulin:2018zxs,Wang:2018fng,Visinelli:2019qqu,Calderon:2020hoc}. We therefore extract the constraints on the cosmological  parameters, assuming both the choices of the  interaction functions, namely, Model 1 of Eq.~(\ref{modelA}) and Model 2 of Eq.~(\ref{modelB}), where the coupling parameter of these two interaction models can be either constant or dynamical.  Notice that, as the dynamics of the interacting cosmologies will also be controlled by the dark energy equation of state parameter $w$,  different choices of the dark energy equation of state parameter will result into different constraints for the model parameters. Therefore, for completeness, it is essential to allow for freedom in the dark energy equation of state parameter. In what follows we shall specify the models that we will study in this work.

\subsection{Constant coupling}
\label{subsec-1}

The first scenario is the one where a vacuum energy component, characterized by the equation of state $w = -1$,   interacts with cold dark matter via the interaction functions $Q_A$ and $Q_B$, see Eqs.~(\ref{modelA}) and (\ref{modelB}) respectively, where $\xi$,  the coupling parameter, is constant in cosmic time (redshift).  We refer to these interactions as {\bf ``$\xi_0$IVS$Q_A$''}  and {\bf ``$\xi_0$IVS$Q_B$''}. 

The next interacting scenario is the one where the dark energy component, with a constant equation of state $w$ other than $w =-1$, interacts with cold dark matter through the interaction functions $Q_A$ or $Q_B$. We refer to these scenarios as ``$\xi_0w${\bf IDE}$Q_{A,B}$''. Due to the stability criteria of interacting cosmologies~\cite{Gavela:2009cy}, we have divided the models into two regions, accordingly to the values of the dark energy equation of state: quintessence ($w> -1$) and phantom ($w< -1$) models. We therefore divide the corresponding interacting scenarios as \textbf{``$\xi_0w_q$IDE$Q_{A,B}$''} and \textbf{``$\xi_0w_p$IDE$Q_{A,B}$''}, where the subscripts $q$ and $p$ of $w$ denote that the dark energy equation of state is either in the quintessence ($w_q> -1$) or in the phantom ($w_p <-1$) regimes.

 The most generalized interacting scenario in this category is the one where the dark energy equation of state is dynamical.  We shall assume the well-known CPL dark energy parametrization $w(a) = w_0 + w_a (1-a)$~\cite{Chevallier:2000qy,Linder:2002et}, which we reformulate as~\cite{Valiviita:2009nu}: 

\begin{eqnarray}\label{dyn-eos-1}
w(a)=w_0\; a+w_e (1-a),
\end{eqnarray}
in which it is straightforward to identify $w_e = w_0 + w_a$, providing the value of $w(a)$ in the early times, and $w_0$ giving the current value of $w(a)$.  In order to ensure the stability of the interaction models, we shall consider, in  the following, two different cases: \textit{(i)} $\xi_0> 0, w_0 > -1, w_e > -1$ (that is, where $w(a)$ remains in the quintessence regime);  and \textit{(ii)} $\xi_0 < 0, w_0 < -1, w_e < -1$ (i.e. where $w(a)$ remains in the phantom regime). Thus, we divide the corresponding interacting scenarios as \textbf{``$\xi_0w_0^{q} w_e^{q} $IDE$Q_{A,B}$''} and \textbf{``$\xi_0w_{0}^{p} w_e^{p}$IDE$Q_{A,B}$''} where the subscripts $q$ and $p$ of $w$ denote that the dark energy equation of state remains within the quintessence ($w_q> -1$) or the phantom ($w_p <-1$) regime.

\subsection{Dynamical coupling}
\label{subsec-2}

In this study we shall also consider interacting scenarios described by the interaction functions $Q_A$,  Eq.~(\ref{modelA}), and $Q_B$,  Eq.~(\ref{modelB}), where the coupling parameter $\xi$ is dynamical and takes the following simple form~\cite{Yang:2019uzo}: 

\begin{eqnarray}\label{dynamical-xi}
\xi (a) = \xi_0 + \xi_a \; (1-a)~, 
\end{eqnarray}
where $\xi_0$ and $\xi_a$ are both constant parameters.

In the very same spirit than in the constant coupling case,  we shall explore one scenario in which a vacuum energy  component with $w = -1$ interacts with cold dark matter (labelled as  {\bf ``$\xi_0\xi_a$IVS$Q_{A,B}$''}) and another one in which a dark energy fluid with a constant equation of state $w$ other than $w = -1$, interacts with the cold dark matter through the interaction functions $Q_A$ and $Q_B$. Following the stability criteria, we shall consider the following cases: \textit{(i)} $\xi_0 > 0, \xi_a > 0$ and $w>-1$; and \textit{(ii)} $\xi_0 < 0, \xi_a < 0$ and $w< -1$.  We divide the corresponding different parameter spaces as \textbf{``$\xi_0 \xi_a w_{q,p}$IDE$Q_A$''} and  \textbf{``$\xi_0 \xi_a w_{q,p}$IDE$Q_B$''}, where the subscripts $q$ and $p$ of $w$ denote that the dark energy equation of state is lying in the quintessence ($w_q> -1$) or in the phantom ($w_p <-1$) regimes. 
 
 Finally, we consider the most general interacting scenario, for which both the interaction rate $\xi (a)$ and the dark energy equation of state $w(a)$ are dynamical. We shall use the CPL  parameterization for $w(a)$ before described. 

The stability criteria also defines in this case certain allowed regions in the parameter space: \textit{(i)} $\xi_0> 0, \xi_a >0; w_0 > -1, w_e > -1$ (i.e. where $w_0$ and $w_e$ remain in the quintessence regimes); and \textit{(ii) } $\xi_0 < 0, \xi_a <0; w_0 < -1, w_e < -1$ (i.e. where $w_0$ and $w_e$ remain in the phantom regimes). The interacting scenarios will be defined as \textbf{``$w_0^{q,p} w_e^{q,p} \xi_0 \xi_a$IDE$Q_{A,B}$''} where the subscripts $q$ and $p$ of $w$ refer to either the quintessence ($w_q> -1$) or in the phantom ($w_p <-1$) regimes and $A, B$ to the interacting model $Q_A$ or $Q_B$, respectively.

\section{Methodology}
\label{sec-data}

In the following we shall briefly describe the observational datasets that are used to constrain all the interacting scenarios previously detailed.

\begin{itemize}

\item \textbf{Cosmic Microwave Background (CMB)}:  we shall consider the CMB measurements from Planck 2018 final release~\cite{Planck:2018vyg,Planck:2019nip}.
    
\item \textbf{Baryon Acoustic Oscillation (BAO)} distance measurements: measurements of BAO data from a number of large scale surveys are also considered in our analyses, namely, those from  6dFGS~\cite{Beutler:2011hx}, SDSS-MGS~\cite{Ross:2014qpa}, and BOSS DR12~\cite{Alam:2016hwk} (as used by the Planck 2018 collaboration~\cite{Planck:2018vyg}).

\item \textbf{Supernovae Type Ia (Pantheon)}: the Pantheon sample of the Supernovae Type Ia comprising 1048 data points in the redshift interval $z \in [0.01, 2.3]$~\cite{Scolnic:2017caz} have also been included in the analyses.

\end{itemize}
To derive cosmological constraints on the different interacting scenarios above detailed,  we make use of a modified version of the publicly available Monte-Carlo Markov Chain package~\texttt{CosmoMC}~\cite{Lewis:2002ah}. The \texttt{CosmoMC} package is furnished with a convergence diagnostic based on  the Gelman and Rubin statistic~\cite{Gelman:1992zz}.  In Table~\ref{tab.priors} we have shown the flat priors on various free parameters that we have considered during the statistical analyses.

\begin{table}[h]
\begin{center}
\renewcommand{\arraystretch}{1.5}
\begin{tabular}{c|cccc}
\hline
\textbf{Parameter}                    & \textbf{IVS}& \textbf{Quintessence}& \textbf{Phantom}\\
\hline\hline
$\Omega_{\rm b} h^2$         & $[0.005\,,\,0.1]$& $[0.005\,,\,0.1]$& $[0.005\,,\,0.1]$\\
$\Omega_{\rm c} h^2$       & $[0.001\,,\,0.99]$& $[0.001\,,\,0.99]$& $[0.001\,,\,0.99]$\\
$100\,\theta_{\rm {MC}}$             & $[0.5\,,\,10]$& $[0.5\,,\,10]$& $[0.5\,,\,10]$\\
$\tau$                       & $[0.01\,,\,0.8]$& $[0.01\,,\,0.8]$& $[0.01\,,\,0.8]$\\
$\log(10^{10}A_{\rm s})$         & $[1.61\,,\,3.91]$& $[1.61\,,\,3.91]$& $[1.61\,,\,3.91]$\\
$n_s$                        & $[0.8\,,\, 1.2]$& $[0.8\,,\, 1.2]$& $[0.8\,,\, 1.2]$\\
$w_0$                  &  -&  $[-1\,,\,1]$&  $[-3\,,\,-1]$\\
$w_e$                  &  -&  $[-1\,,\,1]$&  $[-3\,,\,-1]$\\
$\xi_0$       &  $[-1\,,\, 1]$&  $[-1\,,\, 0]$&  $[0\,,\, 1]$\\
$\xi_a$       &  $[-1\,,\, 1]$&  $[-1\,,\, 0]$&  $[0\,,\, 1]$\\
\hline\hline
\end{tabular}
\end{center}
\caption{List of the parameters varied and the external flat priors assumed.}
\label{tab.priors}
\end{table}

\section{Numerical analyses}
\label{sec-results}

\subsection{Constant coupling $\xi_0 $}

This section covers the observational constraints extracted out of all the interacting scenarios when the coupling parameter is constant. 

\begingroup                                                                                                                     
\begin{center}                                                                                                                  
\begin{table*}                                                                                                                   
\begin{tabular}{cccccccccccc}                                                                                                            
\hline\hline                                                                                                                    
Parameters & CMB & CMB+BAO & CMB+Pantheon & CMB+BAO+Pantheon    \\ \hline 

$\Omega_c h^2$ & $    <0.093\,<0.134$ & $    0.100_{-    0.016-    0.038}^{+    0.022+    0.035}$ & $    0.108_{-    0.011-    0.023}^{+    0.013+    0.022}$ & $    0.1121_{-    0.0070-    0.015}^{+    0.0080+    0.015}$ \\

$\Omega_b h^2$ & $    0.02230_{-    0.00015-    0.00029}^{+    0.00014+    0.00030}$ & $    0.02233_{-    0.00014-    0.00027}^{+    0.00014+    0.00028}$ & $    0.02230_{-    0.00014-    0.00029}^{+    0.00015+    0.00028}$ & $    0.02236_{-    0.00015-    0.00027}^{+    0.00014+    0.00028}$ \\

$100\theta_{MC}$ & $    1.0441_{-    0.0041-    0.0049}^{+    0.0026+    0.0055}$ & $    1.0419_{-    0.0013-    0.0021}^{+    0.0009+    0.0023}$ & $    1.04136_{-    0.00080-    0.0014}^{+    0.00066+    0.0015}$ & $    1.04132_{-    0.00076-    0.0014}^{+    0.00059+    0.0014}$ \\

$\tau$ & $    0.0542_{-    0.0079-    0.015}^{+    0.0075+    0.015}$ & $    0.0555_{-    0.0083-    0.016}^{+    0.0076+    0.016}$  & $    0.0542_{-    0.0079-    0.016}^{+    0.0083+    0.016}$ & $    0.0558_{-    0.0078-    0.015}^{+    0.0077+    0.016}$ \\

$n_s$ & $    0.9723_{-    0.0044-    0.0081}^{+    0.0043+    0.0083}$ & $    0.9734_{-    0.0040-    0.0078}^{+    0.0040+    0.0079}$ & $    0.9720_{-    0.0043-    0.0085}^{+    0.0042+    0.0084}$ & $    0.9738_{-    0.0043-    0.0075}^{+    0.0038+    0.0082}$ \\

${\rm{ln}}(10^{10} A_s)$ & $    3.054_{-    0.016-    0.030}^{+    0.015+    0.031}$ & $    3.056_{-    0.017-    0.032}^{+    0.016+    0.032}$  & $    3.055_{-    0.016-    0.031}^{+    0.016+    0.033}$  & $    3.057_{-    0.016-    0.032}^{+    0.016+    0.032}$ \\

$\xi_0$ & $    0.13_{-    0.08-    0.20}^{+    0.14+    0.17}$ &  $    0.059_{-    0.061-    0.10}^{+    0.053+    0.11}$ & $    0.039_{-    0.036-    0.069}^{+    0.035+    0.070}$ & $    0.032_{-    0.035-    0.069}^{+    0.034+    0.071}$  \\

$\Omega_{m0}$ & $    0.19_{-    0.14-    0.17}^{+    0.08+    0.19}$ &  $    0.261_{-    0.046-    0.099}^{+    0.056+    0.095}$ & $    0.283_{-    0.033-    0.064}^{+    0.033+    0.062}$  & $    0.285_{-    0.028-    0.060}^{+    0.032+    0.060}$  \\

$\sigma_8$ & $    1.6_{-    1.1-    1.3}^{+    0.3+    2.2}$ &  $    0.97_{-    0.20-    0.28}^{+    0.09+    0.36}$ & $    0.895_{-    0.098-    0.15}^{+    0.063+    0.17}$ & $    0.878_{-    0.092-    0.15}^{+    0.056+    0.17}$  \\

$H_0$ [km/s/Mpc] & $   70.8_{-    2.5-    5.9}^{+    4.3+    5.3}$ & $   68.8_{-    1.5-    2.6}^{+    1.3+    2.78}$  & $   68.0_{-    1.0-    1.9}^{+    1.0+    2.0}$  & $   68.17_{-    0.92-    1.6}^{+    0.81+    1.7}$  \\

$S_8$ & $    1.06_{-    0.30-    0.35}^{+    0.10+    0.54}$ & $    0.881_{-    0.071-    0.10}^{+    0.035+    0.13}$  & $    0.864_{-    0.038-    0.063}^{+    0.028+    0.067}$  & $    0.851_{-    0.038-    0.062}^{+    0.025+    0.068}$  \\

$r_{\rm{drag}}$ [Mpc] & $  147.08_{-    0.30-    0.59}^{+    0.31+    0.61}$ &  $  147.17_{-    0.28-    0.52}^{+    0.27+    0.55}$ & $  147.08_{-    0.30-    0.58}^{+    0.29+    0.59}$ & $  147.23_{-    0.27-    0.56}^{+    0.27+    0.53}$  \\
\hline\hline                                                                                                                    
\end{tabular}                                                                                                                   
\caption{Observational constraints on the interacting vacuum scenario \textbf{``$\xi_0$IVS$Q_A$''} obtained from  several observational datasets, namely, CMB, CMB+BAO, CMB+Pantheon and CMB+BAO+Pantheon are presented. }
\label{tab:IVS-ModelA-xi-cons}                                                                                                   
\end{table*}                                                                                                                     
\end{center}                                                                                                                    
\endgroup   
\begin{figure*}
    \centering
    \includegraphics[width=0.8\textwidth]{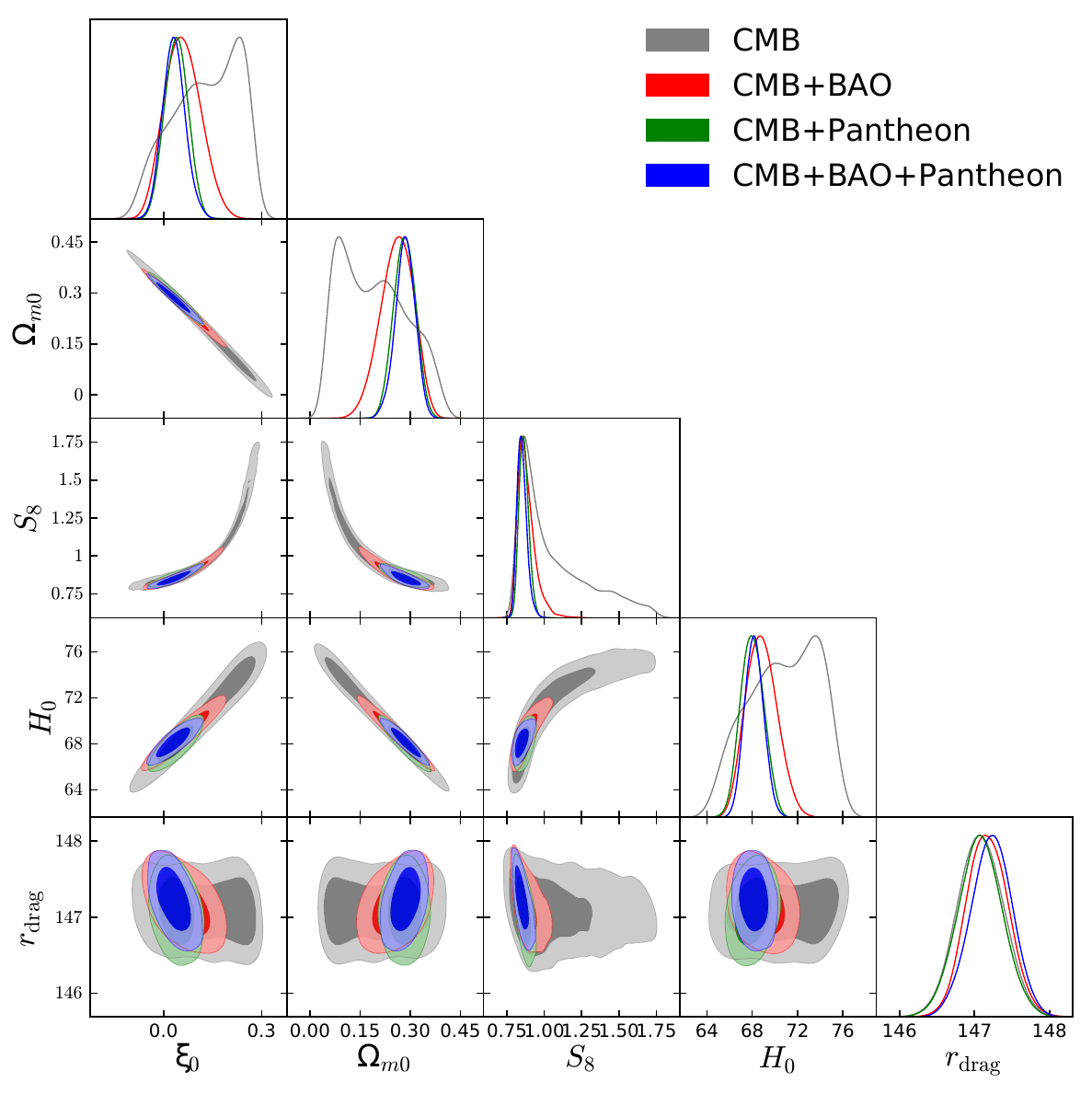}
    \caption{One-dimensional marginalized posterior distributions and two-dimensional joint contours for the most relevant parameters of the interacting vacuum scenario \textbf{``$\xi_0$IVS$Q_A$''} for several datasets, namely, CMB, CMB+BAO, CMB+Pantheon and CMB+BAO+Pantheon.  }
    \label{Fig:IVS-ModelA-xi-cons}
\end{figure*}
\begingroup                                                                                                                     
\begin{center}                                                                                                                  
\begin{table*}                                                                                                                   
\begin{tabular}{cccccccccccc}                                                                                                            
\hline\hline                                                                                                                    
Parameters & CMB & CMB+BAO & CMB+Pantheon & CMB+BAO+Pantheon    \\ \hline
$\Omega_c h^2$ & $    0.120_{-    0.033-    0.078}^{+    0.057+    0.070}$ & $    0.111_{-    0.010-    0.023}^{+    0.012+    0.020}$   & $    0.1125_{-    0.0086-    0.018}^{+    0.0093+    0.017}$  & $    0.1121_{-    0.0070-    0.015}^{+    0.0080+    0.015}$\\

$\Omega_b h^2$ & $    0.02230_{-    0.00015-    0.00031}^{+    0.00015+    0.00031}$ & $    0.02234_{-    0.00015-    0.00029}^{+    0.00015+    0.00030}$ & $    0.02230_{-    0.00015-    0.00030}^{+    0.00015+    0.00029}$  & $    0.02235_{-    0.00014-    0.00028}^{+    0.00014+    0.00028}$\\

$100\theta_{MC}$ & $    1.0408_{-    0.0032-    0.0039}^{+    0.0017+    0.0047}$ & $    1.04120_{-    0.00071-    0.0012}^{+    0.00060+    0.0013}$ & $    1.04107_{-    0.00057-    0.0011}^{+    0.00056+    0.0012}$  & $    1.04115_{-    0.00051-    0.00097}^{+    0.00045+    0.00099}$\\

$\tau$ & $    0.0544_{-    0.0083-    0.016}^{+    0.0077+    0.016}$ &  $    0.0554_{-    0.0082-    0.015}^{+    0.0075+    0.017}$ & $    0.0543_{-    0.0081-    0.015}^{+    0.0075+    0.016}$  & $    0.0553_{-    0.0082-    0.016}^{+    0.0076+    0.016}$\\

$n_s$ & $    0.9721_{-    0.0042-    0.0084}^{+    0.0042+    0.0084}$ & $    0.9734_{-    0.0042-    0.0080}^{+    0.0042+    0.0082}$  & $    0.9723_{-    0.0044-    0.0085}^{+    0.0045+    0.0091}$  & $    0.9734_{-    0.0042-    0.0079}^{+    0.0042+    0.0081}$\\

${\rm{ln}}(10^{10} A_s)$ & $    3.055_{-    0.017-    0.032}^{+    0.016+    0.032}$ & $    3.056_{-    0.017-    0.032}^{+    0.016+    0.033}$  & $    3.055_{-    0.015-    0.032}^{+    0.015+    0.032}$ & $    3.056_{-    0.016-    0.033}^{+    0.016+    0.03}$\\

$\xi_0$ & $    0.01_{-    0.41-    0.54}^{+    0.24+    0.63}$ & $    0.062_{-    0.093-    0.15}^{+    0.069+    0.17}$   & $    0.055_{-    0.070-    0.12}^{+    0.058+    0.13}$ & $    0.052_{-    0.061-    0.11}^{+    0.050+    0.11}$ \\

$\Omega_{m0}$ & $    0.35_{-    0.22-    0.27}^{+    0.12+    0.31}$ &  $    0.288_{-    0.034-    0.065}^{+    0.034+    0.065}$  & $    0.294_{-    0.028-    0.054}^{+    0.027+    0.056}$  & $    0.291_{-    0.022-    0.044}^{+    0.022+    0.044}$  \\

$\sigma_8$ & $    0.82_{-    0.21-    0.26}^{+    0.12+    0.30}$ &  $    0.843_{-    0.054-    0.086}^{+    0.043+    0.098}$  & $    0.841_{-    0.041-    0.072}^{+    0.034+    0.075}$ & $    0.838_{-    0.036-    0.066}^{+    0.031+    0.067}$  \\

$H_0$ [km/s/Mpc] & $   66.3_{-    6.1-   11}^{+    6.9+   11}$ & $   68.3_{-    1.3-    2.3}^{+    1.2+    2.5}$  & $   67.9_{-    1.1-    2.2}^{+    1.1+    2.2}$ & $   68.19_{-    0.87-    1.6}^{+    0.79+    1.6}$ \\

$S_8$ & $    0.823_{-    0.021-    0.12}^{+    0.060+    0.08}$ & $    0.823_{-    0.013-    0.026}^{+    0.013+    0.026}$  & $    0.830_{-    0.015-    0.030}^{+    0.016+    0.030}$  & $    0.824_{-    0.013-    0.024}^{+    0.013+    0.024}$ \\

$r_{\rm{drag}}$ [Mpc] & $  147.08_{-    0.30-    0.59}^{+    0.30+    0.61}$ & $  147.19_{-    0.29-    0.56}^{+    0.29+    0.57}$  & $  147.09_{-    0.30-    0.59}^{+    0.30+    0.60}$  & $  147.20_{-    0.28-    0.55}^{+    0.28+    0.54}$\\
\hline\hline                                                                                                                    
\end{tabular}                                                                                                                   
\caption{Observational constraints on the interacting vacuum scenario \textbf{``$\xi_0$IVS$Q_B$''} obtained from  several observational datasets, namely, CMB, CMB+BAO, CMB+Pantheon and CMB+BAO+Pantheon are presented.    }
\label{tab:IVS-ModelB-xi-cons}                                                                                                   
\end{table*}                                                                                                                     
\end{center}                                                                                                                    
\endgroup 
\begin{figure*}
    \centering
    \includegraphics[width=0.8\textwidth]{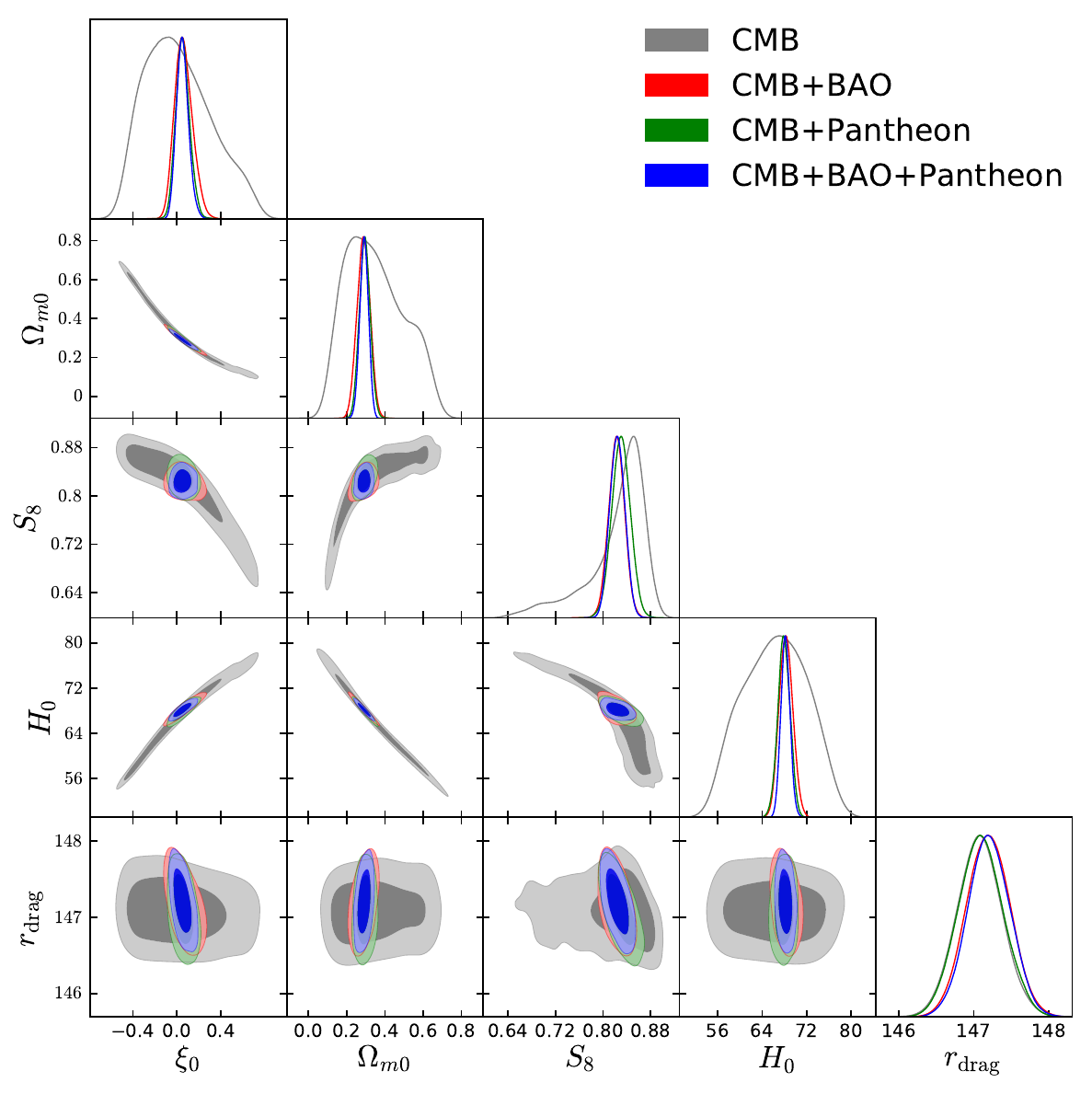}
    \caption{ One-dimensional marginalized posterior distributions and two-dimensional joint contours for the most relevant parameters of the interacting vacuum scenario \textbf{``$\xi_0$IVS$Q_B$''} for several datasets, namely, CMB, CMB+BAO, CMB+Pantheon and CMB+BAO+Pantheon.  }
    \label{Fig:IVS-ModelB-xi-cons}
\end{figure*}
\begingroup                                                                                                                     
\begin{center}                                                                                                                  
\begin{table*}                                                                                                                   
\begin{tabular}{cccccccccc}                                                                                                            
\hline\hline                                                                                                                    
Parameters & CMB & CMB+BAO & CMB+Pantheon & CMB+BAO+Pantheon \\ \hline
$\Omega_c h^2$ & $    0.134_{-    0.013-    0.015}^{+    0.007+    0.017}$ & $    0.1352_{-    0.0098-    0.014}^{+    0.0099+    0.014}$ & $    0.135_{-    0.011-    0.014}^{+    0.009+    0.015}$ & $    0.1350_{-    0.0098-    0.014}^{+    0.0098+    0.014}$ \\

$\Omega_b h^2$ & $    0.02239_{-    0.00015-    0.00030}^{+    0.00015+    0.00030}$ & $    0.02239_{-    0.00014-    0.00028}^{+    0.00014+    0.00028}$ & $    0.02236_{-    0.00015-    0.00029}^{+    0.00015+    0.00029}$ & $    0.02239_{-    0.00014-    0.00027}^{+    0.00014+    0.00028}$  \\

$100\theta_{MC}$ & $    1.04018_{-    0.00054-    0.0011}^{+    0.00065+    0.0011}$ & $    1.04013_{-    0.00055-    0.0010}^{+    0.00055+    0.0010}$  & $    1.04010_{-    0.00055-    0.00099}^{+    0.00055+    0.00098}$ & $    1.04015_{-    0.00055-    0.00099}^{+    0.00054+    0.00099}$  \\

$\tau$ & $    0.0540_{-    0.0074-    0.015}^{+    0.0074+    0.015}$ & $    0.0548_{-    0.0079-    0.015}^{+    0.0073+    0.016}$ & $    0.0543_{-    0.0075-    0.015}^{+    0.0075+    0.016}$ & $    0.0549_{-    0.0075-    0.015}^{+    0.0075+    0.016}$  \\

$n_s$ & $    0.9652_{-    0.0044-    0.0086}^{+    0.0043+    0.0086}$ &  $    0.9657_{-    0.0042-    0.0082}^{+    0.0042+    0.0081}$ & $    0.9647_{-    0.0044-    0.0084}^{+    0.0043+    0.0086}$  & $    0.9659_{-    0.0040-    0.0077}^{+    0.0040+    0.0078}$ \\

${\rm{ln}}(10^{10} A_s)$ & $    3.043_{-    0.015-    0.031}^{+    0.015+    0.029}$ & $    3.045_{-    0.016-    0.030}^{+    0.015+    0.032}$   & $    3.045_{-    0.015-    0.031}^{+    0.015+    0.032}$   & $    3.045_{-    0.015-    0.031}^{+    0.015+    0.032}$  \\

$w_p$ & $   -1.58_{-    0.34-    0.44}^{+    0.21+    0.49}$ &  $ -1.094_{-    0.040}^{+    0.070}\; > -1.193  $ & $   -1.087_{-    0.041-    0.076}^{+    0.049+    0.085}$ & $   -1.080_{-    0.038-    0.072}^{+    0.047+    0.079}$  \\

$\xi_0$ & $  > -0.051 > -0.090 $ &  $ -0.052_{-    0.023}^{+    0.046}\; > -0.100 $  & $   -0.051_{-    0.021}^{+    0.047}\; >-0.101$ & $   -0.052_{-    0.025}^{+    0.044}\; >-0.101$  \\

$\Omega_{m0}$ & $    0.226_{-    0.072-    0.09}^{+    0.031+    0.11}$ &  $    0.336_{-    0.025-    0.042}^{+    0.025+    0.044}$ & $    0.339_{-    0.024-    0.040}^{+    0.024+    0.041}$  & $    0.339_{-    0.023-    0.036}^{+    0.023+    0.037}$ \\

$\sigma_8$ & $    0.886_{-    0.089-    0.16}^{+    0.086+    0.15}$ & $    0.760_{-    0.047-    0.070}^{+    0.038+    0.074}$ & $    0.761_{-    0.046-    0.064}^{+    0.035+    0.068}$ & $    0.756_{-    0.046-    0.063}^{+    0.034+    0.067}$  \\

$H_0$ [km/s/Mpc] & $   85_{-    7-   17}^{+   13+   16}$ & $   68.7_{-    1.5-    2.5}^{+    1.1+    2.7}$ & $   68.3_{-    1.0-    1.9}^{+    1.0+    2.0}$ & $   68.32_{-    0.78-    1.5}^{+    0.77+    1.6}$ \\

$S_8$ & $    0.756_{-    0.034-    0.063}^{+    0.034+    0.064}$ & $    0.802_{-    0.020-    0.036}^{+    0.020+    0.036}$ & $    0.807_{-    0.021-    0.039}^{+    0.021+    0.040}$ & $    0.802_{-    0.019-    0.035}^{+    0.019+    0.036}$ \\

$r_{\rm{drag}}$ [Mpc] & $  147.08_{-    0.29-    0.58}^{+    0.29+    0.57}$ & $  147.11_{-    0.28-    0.54}^{+    0.27+    0.54}$ & $  147.05_{-    0.29-    0.56}^{+    0.29+    0.58}$  & $  147.14_{-    0.27-    0.52}^{+    0.27+    0.52]}$  \\
\hline\hline                                                                                                                    
\end{tabular}
\caption{Observational constraints on the interacting scenario \textbf{``$\xi_0 w_p$IDE$Q_A$''} obtained from  several observational datasets, namely, CMB, CMB+BAO, CMB+Pantheon and CMB+BAO+Pantheon are presented. }
\label{tab:IDEp-ModelA-xi-cons}                                                                
\end{table*}                                                                                                                     
\end{center}                                                                                                                    
\endgroup 
\begin{figure*}
\centering
\includegraphics[width=0.8\textwidth]{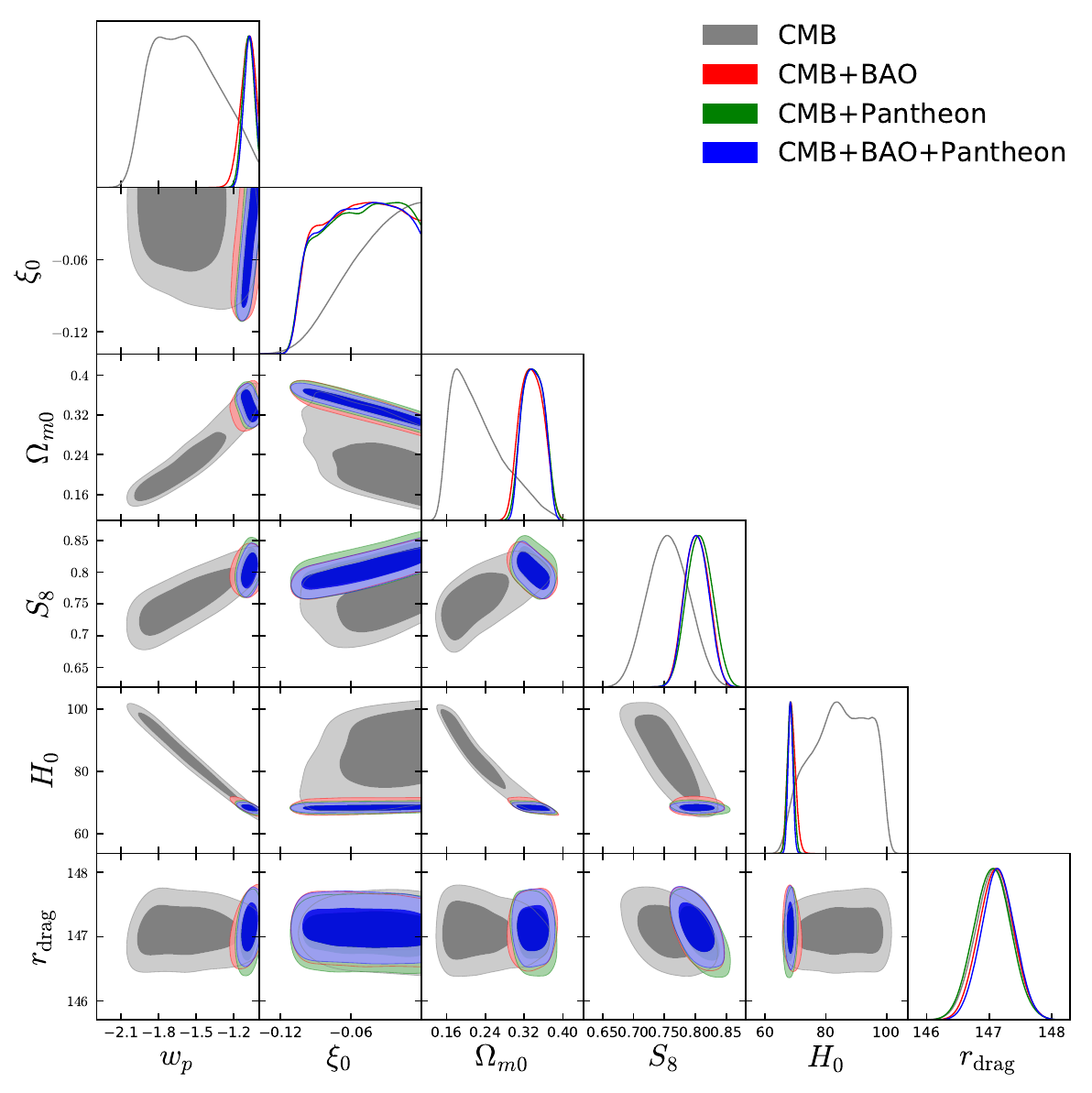}
    \caption{One dimensional marginalized posterior distributions and two-dimensional joint contours for the most relevant parameters of the interacting scenario \textbf{``$\xi_0 w_p$IDE$Q_A$''} for several datasets, namely, CMB, CMB+BAO, CMB+Pantheon and CMB+BAO+Pantheon.  }
    \label{Fig:IDEp-ModelA-xi-cons}
\end{figure*}                                                           
\begingroup                                                                                                                     
\begin{center}                                                                                                                  
\begin{table*}                                                                                                                   
\begin{tabular}{cccccccccccccccc}                                                                                               
\hline\hline                                                                                                                    
Parameters & CMB & CMB+BAO & CMB+Pantheon & CMB+BAO+Pantheon \\ \hline
$\Omega_c h^2$ & $   <0.083\;<0.115 $ & $    0.076_{-    0.018-    0.058}^{+    0.037+    0.046}$  & $    0.074_{-    0.019-    0.057}^{+    0.038+    0.046}$  & $    0.077_{-    0.016-    0.058}^{+    0.036+    0.044}$ \\

$\Omega_b h^2$ & $    0.02236_{-    0.00015-    0.00029}^{+    0.00015+    0.00029}$ & $    0.02239_{-    0.00014-    0.00027}^{+    0.00014+    0.00029}$  & $    0.02234_{-    0.00015-    0.00029}^{+    0.00015+    0.00029}$ & $    0.02238_{-    0.00014-    0.00027}^{+    0.00014+    0.00028}$  \\

$100\theta_{MC}$ & $    1.0448_{-    0.0037-    0.0043}^{+    0.0019+    0.0050}$ &  $    1.0438_{-    0.0025-    0.0031}^{+    0.0010+    0.0043}$  & $    1.0438_{-    0.0026-    0.0032}^{+    0.0011+    0.0043}$ & $    1.0437_{-    0.0024-    0.0031}^{+    0.0009+    0.0043}$ \\

$\tau$ & $    0.0544_{-    0.0074-    0.015}^{+    0.0071+    0.016}$ & $    0.0550_{-    0.0079-    0.015}^{+    0.0077+    0.017}$  & $    0.0538_{-    0.0075-    0.015}^{+    0.0074+    0.016}$ & $    0.0547_{-    0.0073-    0.015}^{+    0.0074+    0.016}$\\

$n_s$ & $    0.9650_{-    0.0042-    0.0084}^{+    0.0042+    0.0084}$ & $    0.9658_{-    0.0041-    0.0084}^{+    0.0042+    0.0082}$  & $    0.9643_{-    0.0043-    0.0082}^{+    0.0043+    0.0085}$ & $    0.9660_{-    0.0040-    0.0079}^{+    0.0040+    0.0078}$ \\

${\rm{ln}}(10^{10} A_s)$ & $    3.045_{-    0.015-    0.031}^{+    0.015+    0.031}$ & $    3.045_{-    0.016-    0.032}^{+    0.016+    0.034}$  & $    3.044_{-    0.015-    0.032}^{+    0.016+    0.033}$ & $    3.044_{-    0.015-    0.031}^{+    0.015+    0.033}$  \\

$w_q$ & $  <-0.897\;<-0.768$ & $  <-0.892\;<-0.801 $   & $ <-0.879\;<-0.793 $  & $ <-0.887\;<-0.792$ \\

$\xi_0$ & $  0.15_{- 0.07}^{+ 0.11}\;  < 0.27  $ & $ 0.119_{- 0.090}^{+    0.054}\; < 0.242 $  & $    0.125_{-    0.090-    0.12}^{+    0.056+    0.13}$ & $    0.115_{-    0.087-    0.12}^{+    0.047+    0.12}$ \\

$\Omega_{m0}$ & $    0.18_{-    0.12-    0.15}^{+    0.07+    0.16}$ & $    0.212_{-    0.047-    0.13}^{+    0.083+    0.11}$  & $    0.210_{-    0.047-    0.13}^{+    0.084+    0.11}$ & $    0.215_{-    0.038-    0.13}^{+    0.080+    0.10}$ \\

$\sigma_8$ & $    1.6_{-    1.0-    1.2}^{+    0.2+    2.1}$ & $    1.25_{-    0.50-    0.7}^{+    0.09+    1.1}$ & $    1.27_{-    0.51-    0.6}^{+    0.09+    1.1}$ & $    1.22_{-    0.46-    0.6}^{+    0.07+    1.1}$ \\

$H_0$ [km/s/Mpc] & $   69.3_{-    2.8-    6.5}^{+    3.9+    6.2}$ & $   68.4_{-    1.4-    2.5}^{+    1.3+    2.7}$ & $   68.1_{-    1.0-    2.0}^{+    1.0+    2.1}$ & $   68.2_{-    0.8-    1.5}^{+    0.8+    1.6}$ \\

$S_8$ & $    1.07_{-    0.27-    0.32}^{+    0.08+    0.52}$ & $    0.97_{-    0.16-    0.21}^{+    0.04+    0.35}$  & $    0.99_{-    0.17-    0.21}^{+    0.04+    0.35}$ & $    0.97_{-    0.15-    0.20}^{+    0.03+    0.33}$ \\

$r_{\rm{drag}}$ [Mpc] & $  147.04_{-    0.29-    0.58}^{+    0.30+    0.59}$ &  $  147.12_{-    0.27-    0.52}^{+    0.27+    0.55}$  & $  147.03_{-    0.29-    0.58}^{+    0.29+    0.58}$  & $  147.15_{-    0.27-    0.54}^{+    0.26+    0.52}$ \\
\hline\hline                                                                                                                    
\end{tabular}                                                                                                                   
\caption{Observational constraints on the interacting  scenario \textbf{``$\xi_0w_q$IDE$Q_A$''} obtained from  several observational datasets, namely, CMB, CMB+BAO, CMB+Pantheon and CMB+BAO+Pantheon are presented.   }
\label{tab:IDEq-ModelA-xi-cons}                                                                             
\end{table*}                                                                                            
\end{center}                                                     
\endgroup
\begin{figure*}
\centering
\includegraphics[width=0.8\textwidth]{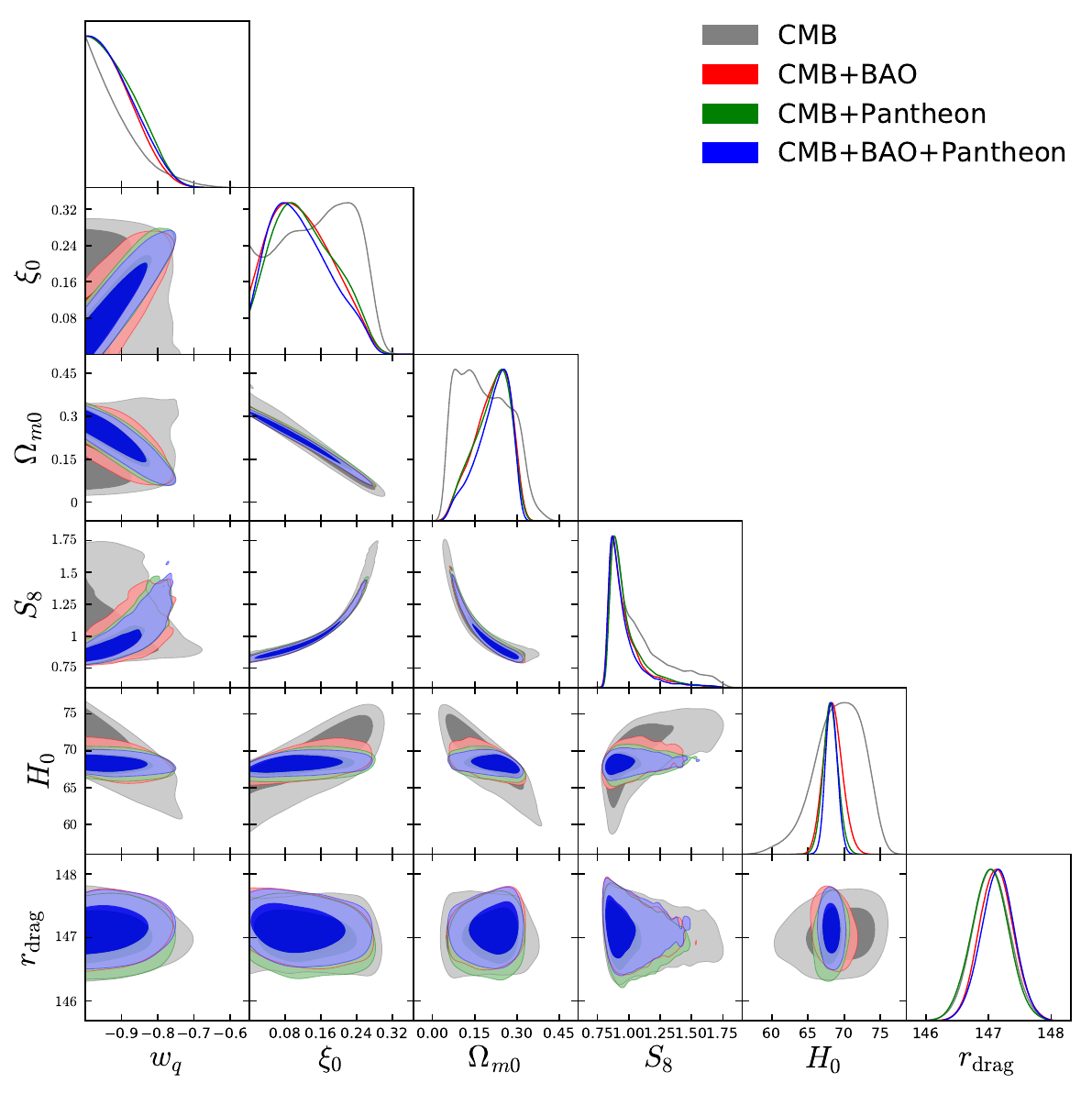}
    \caption{One dimensional marginalized posterior distributions and two-dimensional joint contours for the most relevant parameters of the interacting scenario \textbf{``$\xi_0 w_q$IDE$Q_A$''} for several datasets, namely, CMB, CMB+BAO, CMB+Pantheon and CMB+BAO+Pantheon.  }
    \label{Fig:IDEq-ModelA-xi-cons}
\end{figure*} 
\begingroup                                                                                             
\begin{center}                                                                                                                  
\begin{table*}                                                                                                                   
\begin{tabular}{cccccccc}                                                                                                            
\hline\hline                                                                                                                    
Parameters & CMB & CMB+BAO & CMB+Pantheon & CMB+BAO+Pantheon \\ \hline
$\Omega_c h^2$ & $    0.137_{-    0.018-    0.022}^{+    0.007+    0.032}$ & $    0.136_{-    0.016-    0.019}^{+    0.006+    0.025}$   & $    0.136_{-    0.016-    0.019}^{+    0.006+    0.027}$ & $    0.135_{-    0.014-    0.018}^{+    0.005+    0.024}$ \\

$\Omega_b h^2$ & $    0.02238_{-    0.00014-    0.00029}^{+    0.00014+    0.00028}$ & $    0.02236_{-    0.00014-    0.00029}^{+    0.00014+    0.00029}$   & $    0.02234_{-    0.00015-    0.00029}^{+    0.00015+    0.00029}$  & $    0.02237_{-    0.00014-    0.00027}^{+    0.00014+    0.00028}$ \\

$100\theta_{MC}$ & $    1.0400_{-    0.0005-    0.0017}^{+    0.0010+    0.0014}$ & $    1.04008_{-    0.00050-    0.0015}^{+    0.00084+    0.0012}$ & $    1.04003_{-    0.00050-    0.0015}^{+    0.00087+    0.0013}$ & $    1.04016_{-    0.00046-    0.0014}^{+    0.00080+    0.0012}$  \\

$\tau$ & $    0.0544_{-    0.0076-    0.015}^{+    0.0078+    0.016}$ &  $    0.0552_{-    0.0075-    0.015}^{+    0.0075+    0.015}$ & $    0.0548_{-    0.0084-    0.015}^{+    0.0074+    0.016}$ & $    0.0555_{-    0.0077-    0.015}^{+    0.0077+    0.016}$ \\

$n_s$ & $    0.9645_{-    0.0044-    0.0083}^{+    0.0044+    0.0086}$ & $    0.9644_{-    0.0041-    0.0081}^{+    0.0043+    0.0082}$  & $    0.9636_{-    0.0044-    0.0085}^{+    0.0043+    0.0083}$  & $    0.9649_{-    0.0039-    0.0077}^{+    0.0039+    0.0078}$ \\

${\rm{ln}}(10^{10} A_s)$ & $    3.045_{-    0.015-    0.032}^{+    0.015+    0.032}$ &  $    3.046_{-    0.015-    0.031}^{+    0.016+    0.032}$  & $    3.047_{-    0.017-    0.031}^{+    0.016+    0.033}$  & $    3.047_{-    0.016-    0.031}^{+    0.016+    0.032}$ \\

$w_p$ & $   -1.67_{-    0.33-    0.41}^{+    0.16+    0.52}$ & $ -1.111_{-    0.044}^{+    0.089}\; > -1.237 $  & $   -1.095_{-    0.035}^{+    0.070}\; >-1.20$  & $   -1.084_{-    0.032}^{+    0.064}\; >-1.18$  \\

$\xi_0$ & $ > -0.368 \; > -0.766 $ & $ > -0.309 \; > -0.619  $   & $ >-0.306\;>-0.657  $  & $  >-0.277\;>-0.591 $ \\

$\Omega_{m0}$ & $    0.216_{-    0.064-    0.08}^{+    0.026+    0.11}$ & $    0.334_{-    0.036-    0.052}^{+    0.020+    0.061}$ & $    0.342_{-    0.037-    0.050}^{+    0.018+    0.064}$  & $    0.337_{-    0.033-    0.044}^{+    0.015+    0.056}$ \\

$\sigma_8$ & $    0.92_{-    0.07-    0.16}^{+    0.10+    0.15}$ & $    0.774_{-    0.032-    0.092}^{+    0.052+    0.078}$  & $    0.769_{-    0.025-    0.090}^{+    0.053+    0.070}$ & $    0.771_{-    0.024-    0.084}^{+    0.050+    0.067}$  \\

$H_0$ [km/s/Mpc] & $   88_{-  6-   18}^{+   12+   14}$ & $   69.0_{-    1.5-    2.7}^{+    1.2+    2.8}$ & $   68.3_{-    1.0-    2.0}^{+    1.0+    2.0}$  & $   68.37_{-    0.85-    1.5}^{+    0.78+    1.6}$ \\

$S_8$ & $    0.764_{-    0.035-    0.053}^{+    0.024+    0.060}$ & $    0.814_{-    0.015-    0.036}^{+    0.019+    0.033}$ & $    0.819_{-    0.017-    0.038}^{+    0.020+    0.037}$  & $    0.816_{-    0.014-    0.036}^{+    0.019+    0.033}$ \\

$r_{\rm{drag}}$ [Mpc] & $  147.04_{-    0.32-    0.57}^{+    0.30+    0.59}$ &  $  147.07_{-    0.28-    0.54}^{+    0.28+    0.53}$  & $  147.00_{-    0.29-    0.57}^{+    0.29+    0.56}$ & $  147.10_{-    0.26-    0.52}^{+    0.26+    0.51}$   \\
\hline\hline                                                                                                                    
\end{tabular}                                                                                                                   
\caption{Observational constraints on the interacting scenario \textbf{``$\xi_0w_p$IDE$Q_B$''} obtained from  several observational datasets, namely, CMB, CMB+BAO, CMB+Pantheon and CMB+BAO+Pantheon are presented.  }\label{tab:IDEp-ModelB-xi-cons}                                                                                                   
\end{table*}                                                                                                                     
\end{center}                                                             
\endgroup                                                                                                   
\begingroup                                                                                                                     
\begin{center}                                                                                                                  
\begin{table*}                                                                                                                   
\begin{tabular}{cccccccc}                                                                                                            
\hline\hline                                                                                                                    
Parameters & CMB & CMB+BAO & CMB+Pantheon & CMB+BAO+Pantheon  \\ \hline
$\Omega_c h^2$ & $    0.1153_{-    0.0026-    0.0090}^{+    0.0053+    0.0072}$ & $    0.1129_{-    0.0028-    0.0090}^{+    0.0057+    0.0074}$  & $    0.1130_{-    0.0033-    0.0090}^{+    0.0056+    0.0079}$ & $    0.1125_{-    0.0031-    0.0084}^{+    0.0054+    0.0074}$  \\

$\Omega_b h^2$ & $    0.02236_{-    0.00015-    0.00029}^{+    0.00015+    0.00030}$ & $    0.02244_{-    0.00014-    0.00028}^{+    0.00015+    0.00027}$ & $    0.02240_{-    0.00014-    0.00029}^{+    0.00014+    0.00029}$   & $    0.02244_{-    0.00014-    0.00027}^{+    0.00014+    0.00027}$ \\

$100\theta_{MC}$ & $    1.04117_{-    0.00042-    0.00073}^{+    0.00036+    0.00080}$ & $    1.04139_{-    0.00040-    0.00070}^{+    0.00035+    0.00078}$ & $    1.04134_{-    0.00041-    0.00075}^{+    0.00037+    0.00078}$  & $    1.04140_{-    0.00037-    0.00070}^{+    0.00037+    0.00075}$  \\

$\tau$ & $    0.0538_{-    0.0074-    0.016}^{+    0.0074+    0.016}$ & $    0.0554_{-    0.0075-    0.014}^{+    0.0074+    0.016}$  & $    0.0543_{-    0.0076-    0.016}^{+    0.0078+    0.016}$ & $    0.0550_{-    0.0083-    0.015}^{+    0.0073+    0.016}$ \\

$n_s$ & $    0.9650_{-    0.0043-    0.0085}^{+    0.0043+    0.0087}$ & $    0.9680_{-    0.0040-    0.0075}^{+    0.0039+    0.0081}$  & $    0.9664_{-    0.0043-    0.0084}^{+    0.0043+    0.0084}$ & $    0.9678_{-    0.0038-    0.0076}^{+    0.0038+    0.0077}$  \\

${\rm{ln}}(10^{10} A_s)$ & $    3.044_{-    0.015-    0.032}^{+    0.016+    0.031}$ & $    3.044_{-    0.015-    0.030}^{+    0.015+    0.032}$ & $    3.043_{-    0.016-    0.032}^{+    0.016+    0.032}$ & $    3.044_{-    0.017-    0.031}^{+    0.016+    0.033}$ \\

$w_q$ & $ < -0.894 \; < - 0.759  $ &  $  < -0.955 \; < -0.908 $ & $ <-0.972\;<-0.939  $ & $  <-0.973\;<-0.944 $  \\

$\xi_0$ & $  < 0.097 \; < 0.203 $ & $  < 0.121 \; < 0.238  $ & $  <0.137\;<0.244  $ & $ <0.136\;<0.239   $ \\

$\Omega_{m0}$ & $    0.327_{-    0.033-    0.058}^{+    0.022+    0.066}$ &  $    0.301_{-    0.012-    0.030}^{+    0.017+    0.027}$ & $    0.299_{-    0.013-    0.031}^{+    0.016+    0.027}$ & $    0.296_{-    0.011-    0.026}^{+    0.015+    0.024}$ \\

$\sigma_8$ & $    0.809_{-    0.028-    0.066}^{+    0.031+    0.061}$ & $    0.826_{-    0.027-    0.044}^{+    0.018+    0.050}$ & $    0.834_{-    0.027-    0.042}^{+    0.017+    0.046}$ & $    0.832_{-    0.027-    0.040}^{+    0.017+    0.045}$  \\

$H_0$ [km/s/Mpc] & $   65.2_{-    1.5-    5.1}^{+    2.9+    4.1}$ & $   67.30_{-    0.76-    1.8}^{+    0.99+    1.7}$   & $   67.45_{-    0.75-    1.6}^{+    0.84+    1.5}$  & $   67.72_{-    0.61-    1.2}^{+    0.63+    1.2}$ \\

$S_8$ & $    0.842_{-    0.017-    0.033}^{+    0.018+    0.034}$ & $    0.826_{-    0.014-    0.027}^{+    0.014+    0.027}$  & $    0.833_{-    0.016-    0.030}^{+    0.016+    0.032}$ & $    0.826_{-    0.013-    0.026}^{+    0.013+    0.027}$  \\

$r_{\rm{drag}}$ [Mpc] & $  147.06_{-    0.30-    0.58}^{+    0.29+    0.58}$ & $  147.28_{-    0.26-    0.53}^{+    0.27+    0.51}$  & $  147.15_{-    0.29-    0.56}^{+    0.28+    0.55}$ & $  147.26_{-    0.25-    0.50}^{+    0.25+    0.50}$  \\
\hline\hline                                                                                                                    
\end{tabular}                                                                                                                   
\caption{Observational constraints on the interacting  scenario \textbf{``$\xi_0w_q$IDE$Q_B$''} obtained from  several observational datasets, namely, CMB, CMB+BAO, CMB+Pantheon and CMB+BAO+Pantheon are presented.  }
\label{tab:IDEq-ModelB-xi-cons}                                                                                                   
\end{table*}                                                                                                                     
\end{center}                                                                                                                    
\endgroup                                                                                                                       

\subsubsection{Constant dark energy equation of state}

The observational constraints on the interacting scenarios when the dark energy represents the vacuum energy (i.e. $w = -1$) are shown in Tabs.~\ref{tab:IVS-ModelA-xi-cons} and~\ref{tab:IVS-ModelB-xi-cons} and  Figs.~\ref{Fig:IVS-ModelA-xi-cons} and~\ref{Fig:IVS-ModelB-xi-cons}.

The constraints arising from CMB, CMB + BAO, CMB + Pantheon, and CMB + BAO + Pantheon  observations on the \textbf{$\xi_0$IVS$Q_A$} model are shown in  Tab.~\ref{tab:IVS-ModelA-xi-cons}. Even if for CMB alone there is a very mild evidence for a non-zero coupling ($\xi_0 = 0.132_{- 0.076}^{+    0.142}$ at 68\% CL), this evidence gets completely diluted when we include BAO measurements ($\xi_0 = 0.059_{-    0.061}^{+    0.053}$ at 68\% CL for CMB+BAO). A similar situation happens with $\Omega_{m0}$, $\sigma_8$ and $H_0$. The large shifts in the mean values of these parameters  obtained in the case of CMB data alone disappear when adding  BAO information. Namely, the lower value of $\Omega_{m0}$ (due to the energy flow from the dark matter sector to the dark energy one), the higher value of $\sigma_8$ (to compensate for the lower value of the matter energy density) and the higher value of the Hubble constant (to keep $\Omega_{m0}h^2$ unchanged) from CMB data alone analyses are restored to their $\Lambda$CDM values after the addition of BAO in the data analyses, see Fig.~\ref{Fig:IVS-ModelA-xi-cons}. For the CMB+Pantheon case, instead we observe a mild preference for a non-zero coupling ($\xi_0 = 0.039_{-    0.036}^{+    0.035}$ at 68\% CL) but  with the inclusion of BAO to CMB+Pantheon the coupling parameter is  back in agreement with zero ($\xi_0 = 0.032_{-0.035}^{+    0.034}$ at 68\% CL).

Table \ref{tab:IVS-ModelB-xi-cons} summarizes  the observational constraints on various parameters of the interacting scenario \textbf{$\xi_0$IVS$Q_B$} for various datasets, namely, CMB alone, CMB+BAO, CMB+Pantheon and CMB+BAO+Pantheon.  For all the datasets, the coupling parameter is statistically consistent with a vanishing value, i.e. with a non-interacting model  within $1\sigma$. 
Notice that the shift in the parameters is the opposite  of what we found for $Q_A$, see Fig.~\ref{Fig:IVS-ModelB-xi-cons}. In this case, the energy flow is such that the amount of dark matter today is higher, as the energy transfer is from the dark energy sector to the dark matter one. Consequently, the mean value of $\Omega_{m0}$ is higher and those of $\sigma_8$ and $H_0$ are  lower  compared to the interacting scenario with $Q_A$.  Nevertheless the addition of BAO measurements restores their usual (i.e. canonical, $\Lambda$CDM-like) values, as in the case of $Q_A$.

Table~\ref{tab:IDEp-ModelA-xi-cons}  and  Fig.~\ref{Fig:IDEp-ModelA-xi-cons} shows the observational constraints for the model \textbf{$\xi_0 w_p${\bf IDE}$Q_A$}, characterized by a parameter space given by $w_p <-1$, $\xi_0 <0$ and the coupling function $Q_A$, see Eq.~(\ref{modelA}). Note that, following section~\ref{subsec-1}, we use $w_p$ to represent the phantom dark energy equation of state and $w_q$ representing the quintessence dark energy equation of state. From the CMB data,  we do not find any evidence for a non-zero coupling.  However, what we observe for CMB data alone is a preference for phantom dark energy at more than $95\%$~CL ($w_{p} = - 1.58_{-    0.44}^{+    0.49}$ at 95\% CL) and a much larger value of the Hubble constant, as expected in these scenarios~\cite{DiValentino:2016hlg} and also noticed from the results depicted in  Fig.~\ref{Fig:IDEp-ModelA-xi-cons}. Once BAO observations are also considered, a very mild evidence for a non-vanishing coupling parameter is found ($\xi_0 = -0.052^{+0.046}_{-0.023}$ at 68\% CL for CMB+BAO). Also, the mean values of the parameters are shifted closer to their $\Lambda$CDM-like values and 
the preference for $w_p<-1$ is decreased($w_p = -1.094_{-    0.040}^{+    0.070}$ at 68\% CL for CMB+BAO). 
However, the value of $\Omega_{m0} $ ($\sigma_8$) is still mildly higher (lower) than in the canonical $\Lambda$CDM picture.  The inclusion of Pantheon data leads to very similar results to the CMB+BAO case except for $w_p$, which, in this case, remains in the phantom regime, $w_p = -1.087_{-0.076}^{+ 0.085}$ at 95\% CL for CMB+Pantheon. Finally, the constraints from CMB+BAO+Pantheon are very similar to those obtained with CMB+Pantheon,  i.e. show an indication for a phantom interacting scenario at $1\sigma$ level. However, we observe that this full dataset prefers a smaller value of $S_8$ compared to the Planck's estimation~\cite{Planck:2018vyg} (within the canonical $\Lambda$CDM picture), improving the consistency with weak lensing measurements~\cite{Heymans:2020gsg,KiDS:2020ghu,DES:2021vln,DES:2022ygi}. 

We present in Tab.~\ref{tab:IDEq-ModelA-xi-cons} and  Fig.~\ref{Fig:IDEq-ModelA-xi-cons} the constraints for the model $\xi_0 w_q${\bf IDE}$Q_A$, i.e. ($w_q >-1$,  $\xi_0 >0$ and the coupling function $Q_A$, see Eq.~(\ref{modelA})) for CMB and CMB+BAO, CMB+Pantheon and CMB+BAO+Pantheon measurements. For CMB alone we find a mild preference of a non-zero coupling at 68\% CL ($\xi_0 =  0.15_{- 0.07}^{+ 0.11}$ at 68\% CL). When BAO measurements are added to the CMB, the behavior of the coupling parameter remains the same, that means, we find a mild evidence of the coupling parameter at 68\% CL ($\xi_0 = 0.119_{- 0.090}^{+    0.054}$ at 68\% CL). Interestingly, for this case, the addition of BAO does not restore the values of the parameters (other than $w_q$) to their $\Lambda$CDM-like ones. Namely, the value of the matter density is much lower (due to the energy flow) and consequently, to leave unchanged the overall power spectra normalization, the clustering parameter $\sigma_8$ is required to be much higher than in the standard paradigm,  and its value is much relaxed, with very large error bars. The addition of Pantheon to CMB does not offer anything new compared to the CMB+BAO constraints except for some mild shifts in the mean values of the parameters and  a 95\%~CL significance for a coupling different from zero. Finally, the results for the combined analysis CMB+BAO+Pantheon are also similar to those of CMB+BAO.

Table~\ref{tab:IDEp-ModelB-xi-cons} presents the observational constraints of the model \textbf{$\xi_0 w_p${\bf IDE}$Q_B$}, characterized by a parameter space given by $w_p <-1$, $\xi_0 <0$ and the coupling function $Q_B$, see Eq.~(\ref{modelB}) for CMB, CMB+BAO, CMB+Pantheon and CMB+BAO+Pantheon. Focusing on the CMB constraints, we do not find any evidence for a non-zero value of the coupling parameter $\xi_0$ but the dark energy equation of state shows its phantom behaviour at more than 95\% CL ($w_p = -1.67_{-0.41}^{+0.52}$ at 95\% CL). Notice that a much larger value of the Hubble constant is also obtained, as expected in these phantom scenarios~\cite{DiValentino:2016hlg}.  When the BAO data are added to CMB, 
the mean values of some of the parameters are shifted to their $\Lambda$CDM-like values, and 
the preference for $w_p<-1$ is somehow decreased ($w_p =-1.111_{-    0.044}^{+    0.089}$ at 68\% CL for CMB+BAO). 
However, the value of $\Omega_{m0} $ ($\sigma_8$) is still mildly higher (lower) than in the $\Lambda$CDM model. For the remaining two combined datasets, namely, CMB+Pantheon and CMB+BAO+Pantheon, the results do not significantly differ compared to the constraints obtained with the CMB+BAO combination. 
This means that we do not find any evidence for a  non-zero coupling in the dark sector. The phantom nature of the dark energy equation of state  is evident at more than 68\% CL and for both the datasets the value of $\Omega_{m0} $ ($\sigma_8$) is still mildly higher (lower) than in the $\Lambda$CDM model. 

To conclude with the constant coupling and constant dark energy equation of state framework, we show in Tab.~\ref{tab:IDEq-ModelB-xi-cons} the constraints on the model \textbf{$\xi_0 w_q${\bf IDE}$Q_B$}, characterized by a parameter space given by $w_q >-1$, $\xi_0 > 0$ and the coupling function $Q_B$, see Eq.~(\ref{modelB}), for the observational datasets CMB, CMB+BAO, CMB+Pantheon and CMB+BAO+Pantheon. Starting with the CMB constraints, we note that the model does not lead to any non-zero value of the coupling parameter and it is very close to the non-interacting scenario. The Hubble constant takes a slightly lower value ($H_0 = 65.2_{-    1.5}^{+    2.9}$ km/s/Mpc at 68\% CL) compared to the $\Lambda$CDM value due to a slightly higher value of the matter density parameter and also to the fact that the dark energy equation of state is required to lie within the quintessence region. When the BAO observations are added to CMB, we find that the values of the parameters are very similar to their $\Lambda$CDM-like ones.  Our findings do not change for the remaining two cases with CMB+Pantheon and CMB+BAO+Pantheon datasets. 
\begingroup                                                                                                                     
\begin{center}                                                                                                                  
\begin{table*}                                                                                                                   
\begin{tabular}{cccccccccc}                                                                                                            
\hline\hline                                                                                                                    
Parameters & CMB & CMB+BAO & CMB+Pantheon & CMB+BAO+Pantheon \\ \hline
$\Omega_c h^2$ & $    0.137_{-    0.017-    0.019}^{+    0.008+    0.024}$ & $    0.137_{-    0.008-    0.016}^{+    0.011+    0.015}$  & $    0.138_{-    0.007-    0.016}^{+    0.012+    0.015}$ & $    0.136_{-    0.007-    0.015}^{+    0.011+    0.014}$ \\

$\Omega_b h^2$ & $    0.02241_{-    0.00015-    0.00029}^{+    0.00015+    0.00030}$ &  $    0.02237_{-    0.00014-    0.00028}^{+    0.00014+    0.00028}$  & $    0.02234_{-    0.00015-    0.00029}^{+    0.00015+    0.00029}$ & $    0.02238_{-    0.00014-    0.00028}^{+    0.00014+    0.00029}$  \\

$100\theta_{MC}$ & $    1.04007_{-    0.00061-    0.0013}^{+    0.00082+    0.0011}$ & $    1.04001_{-    0.00054-    0.00094}^{+    0.00054+    0.00098}$ & $    1.03993_{-    0.00063-    0.0010}^{+    0.00050+    0.0010}$  & $    1.04005_{-    0.00056-    0.00094}^{+    0.00051+    0.00098}$ \\

$\tau$ & $    0.0535_{-    0.0075-    0.015}^{+    0.0073+    0.016}$ &  $    0.0542_{-    0.0074-    0.015}^{+    0.0072+    0.015}$ & $    0.0538_{-    0.0075-    0.015}^{+    0.0076+    0.015}$  & $    0.0540_{-    0.0072-    0.014}^{+    0.0070+    0.016}$  \\

$n_s$ & $    0.9656_{-    0.0044-    0.0083}^{+    0.0044+    0.0087}$ & $    0.9652_{-    0.0043-    0.0082}^{+    0.0043+    0.0081}$ & $    0.9645_{-    0.0044-    0.0085}^{+    0.0044+    0.0085}$ & $    0.9653_{-    0.0042-    0.0080}^{+    0.0042+    0.0082}$  \\

${\rm{ln}}(10^{10} A_s)$ & $    3.042_{-    0.015-    0.032}^{+    0.015+    0.031}$ & $    3.044_{-    0.015-    0.031}^{+    0.015+    0.031}$ & $    3.044_{-    0.015-    0.031}^{+    0.016+    0.031}$ & $    3.044_{-    0.016-    0.032}^{+    0.015+    0.032}$  \\

$w_0^p$ & $ -1.84_{-    0.27}^{+    0.84} \; uncons.$ & $  > -1.205 \; >    -1.456 $ & $   >-1.142\;>-1.27$ & $  >-1.101\;>-1.19$  \\

$w_e^p$ & $ > -1.79\; > -2.64 $ & $ -1.11_{-    0.04}^{+    0.10}\; > -1.24 $ & $ >-1.135\;>-1.25 $ & $   -1.108_{-    0.047}^{+    0.082}\;>-1.22$ \\

$\xi_0$ & $ > -0.052 \; > -0.090  $ & $ > -0.095 \; > -0.103  $ & $   -0.060_{-    0.042-    0.048}^{+    0.025+    0.058}$  & $   -0.056_{-    0.038-    0.048}^{+    0.028+    0.055}$  \\

$\Omega_{m0}$ & $    0.211_{-    0.058-    0.071}^{+    0.024+    0.093}$ &  $    0.331_{-    0.025-    0.044}^{+    0.024+    0.044}$  & $    0.344_{-    0.026-    0.044}^{+    0.026+    0.045}$  & $    0.340_{-    0.019-    0.038}^{+    0.025+    0.035}$  \\

$\sigma_8$ & $    0.910_{-    0.080-    0.14}^{+    0.079+    0.14}$ &  $    0.764_{-    0.049-    0.070}^{+    0.034+    0.075}$ & $    0.754_{-    0.049-    0.063}^{+    0.029+    0.075}$  & $    0.754_{-    0.047-    0.061}^{+    0.030+    0.071}$  \\

$H_0$ [km/s/Mpc] & $   88_{-    5-   14}^{+   11+   13}$ &  $   69.7_{-    2.0-    3.1}^{+    1.2+    3.6}$ & $   68.5_{-    1.2-    2.2}^{+    1.1+    2.3}$ & $   68.52_{-    0.85-    1.4}^{+    0.74+    1.5}$ \\

$S_8$ & $    0.754_{-    0.032-    0.066}^{+    0.033+    0.061}$ & $    0.801_{-    0.020-    0.034}^{+    0.018+    0.037}$  & $    0.805_{-    0.022-    0.042}^{+    0.022+    0.041}$ & $    0.801_{-    0.020-    0.035}^{+    0.019+    0.037}$ \\

$r_{\rm{drag}}$ [Mpc] & $  147.09_{-    0.29-    0.57}^{+    0.29+    0.58}$ &  $  147.08_{-    0.28-    0.55}^{+    0.28+    0.55}$  & $  147.02_{-    0.29-    0.59}^{+    0.29+    0.60}$   & $  147.09_{-    0.28-    0.55}^{+    0.27+    0.55}$\\
\hline\hline                                                                                                                    
\end{tabular}                                                                                                                   
\caption{Observational constraints on the interacting  scenario \textbf{``$\xi_0w_0^{p}w_e^{p}$IDE$Q_A$''} obtained from  several observational datasets, namely, CMB, CMB+BAO, CMB+Pantheon and CMB+BAO+Pantheon are presented.  }
\label{tab:IDEp-ModelA-xi-cons-w-dyn}                                                                                           
\end{table*}                                                                                                                     
\end{center}                                                                                                                    
\endgroup
\begingroup                                                                                                                     
\begin{center}                                                                                                                  
\begin{table*}                                                                                                                   
\begin{tabular}{ccccccccccccccc}                                                                                                            
\hline\hline                                                                                                                    
Parameters & CMB & CMB+BAO & CMB+Pantheon & CMB+BAO+Pantheon  \\ \hline

$\Omega_c h^2$ & $   <0.072\;<0.110 $ & $    0.059_{-    0.030-    0.056}^{+    0.035+    0.050}$  & $  <0.054\;<0.095 $   & $    0.055_{-    0.034}^{+    0.033}\;<0.101$  \\

$\Omega_b h^2$ & $    0.02240_{-    0.00016-    0.00030}^{+    0.00016+    0.00031}$ & $    0.02243_{-    0.00015-    0.00030}^{+    0.00015+    0.00029}$ & $    0.02238_{-    0.00015-    0.00028}^{+    0.00015+    0.00028}$ & $    0.02242_{-    0.00014-    0.00028}^{+    0.00014+    0.00029}$ \\

$100\theta_{MC}$ & $    1.0453_{-    0.0032-    0.0043}^{+    0.0026+    0.0043}$ &  $    1.0449_{-    0.0029-    0.0038}^{+    0.0017+    0.0044}$ & $    1.0462_{-    0.0021-    0.0045}^{+    0.0032+    0.0042}$  & $    1.0453_{-    0.0031-    0.0038}^{+    0.0019+    0.0044}$  \\

$\tau$ & $    0.0525_{-    0.0076-    0.016}^{+    0.0075+    0.016}$ & $    0.0555_{-    0.0088-    0.016}^{+    0.0076+    0.017}$ & $    0.0540_{-    0.0080-    0.015}^{+    0.0076+    0.016}$ & $    0.0551_{-    0.0083-    0.014}^{+    0.0070+    0.017}$  \\

$n_s$ & $    0.9664_{-    0.0046-    0.0090}^{+    0.0046+    0.0092}$ & $    0.9678_{-    0.0042-    0.0086}^{+    0.0043+    0.0083}$  & $    0.9656_{-    0.0051-    0.0089}^{+    0.0047+    0.0093}$  & $    0.9671_{-    0.0041-    0.0082}^{+    0.0041+    0.0082}$  \\

${\rm{ln}}(10^{10} A_s)$ & $    3.040_{-    0.016-    0.033}^{+    0.016+    0.033}$ & $    3.045_{-    0.018-    0.032}^{+    0.016+    0.035}$  & $    3.044_{-    0.016-    0.031}^{+    0.016+    0.034}$ & $    3.045_{-    0.018-    0.031}^{+    0.015+    0.035}$  \\

$w_0^q$ & $ -0.59_{-    0.29}^{+    0.20} \; < -0.190  $ & $ -0.82_{-    0.09}^{+    0.10}\; < -0.687  $ & $   -0.80_{-    0.06-    0.18}^{+    0.11+    0.13}$  & $   -0.839_{-    0.065-    0.16}^{+    0.099+    0.12}$ \\

$w_e^q$ & $  < -0.897 \; < -0.769  $ & $ < -0.921 \; < -0.830  $  & $  <-0.931\;<-0.843 $ & $  <-0.910\;<-0.814 $ \\

$\xi_0$ & $ 0.25_{- 0.14}^{+  0.13} \;  < 0.458 $ & $  0.176_{- 0.098-    0.15}^{+    0.087+    0.15}$ & $    0.24_{-    0.07-    0.18}^{+    0.13+    0.16}$  & $    0.185_{-    0.093-    0.15}^{+    0.093+    0.15}$  \\

$\Omega_{m0}$ & $    0.20_{-    0.13-    0.16}^{+    0.07+    0.18}$ & $    0.181_{-    0.074-    0.13}^{+    0.073+    0.12}$ & $    0.142_{-    0.089-    0.11}^{+    0.043+    0.13}$  & $    0.168_{-    0.073-    0.12}^{+    0.071+    0.12}$  \\

$\sigma_8$ & $    1.7_{-    1.1-    1.2}^{+    0.3+    2.0}$ & $    1.55_{-    0.80-    1.0}^{+    0.20+    1.7}$ & $    2.1_{-    1.3-    1.5}^{+    0.5+    2.1}$   & $    1.66_{-    0.90-    1.1}^{+    0.24+    1.8}$  \\

$H_0$ [km/s/Mpc] & $   62.7_{-    4.8-    8.7}^{+    4.7+    8.6}$ & $   67.6_{-    1.5-    3.0}^{+    1.45+    3.0}$ & $   67.5_{-    1.0-    2.2}^{+    1.1+    2.1}$ & $   68.06_{-    0.81-    1.6}^{+    0.80+    1.6}$ \\

$S_8$ & $    1.20_{-    0.38-    0.43}^{+    0.14+    0.62}$ &  $    1.08_{-    0.26-    0.32}^{+    0.08+    0.49}$ & $    1.25_{-    0.40-    0.48}^{+    0.19+    0.60}$ & $    1.11_{-    0.29-    0.34}^{+    0.10+    0.52}$ \\

$r_{\rm{drag}}$ [Mpc] & $  147.11_{-    0.30-    0.59}^{+    0.30+    0.59}$ &  $  147.24_{-    0.28-    0.56}^{+    0.29+    0.59}$ & $  147.07_{-    0.31-    0.58}^{+    0.31+    0.62}$  & $  147.21_{-    0.27-    0.53}^{+    0.28+    0.52}$  \\
\hline\hline                                                                                                                    
\end{tabular}                                                                                                                   
\caption{Observational constraints on the interacting  scenario \textbf{``$\xi_0w_0^{q}w_e^{q}$IDE$Q_A$''} obtained from  several observational datasets, namely, CMB, CMB+BAO, CMB+Pantheon and CMB+BAO+Pantheon are presented.   }
\label{tab:IDEq-ModelA-xi-cons-w-dyn}                                                                                       
\end{table*}                                                                                                                     
\end{center}                                                                                                                    
\endgroup                                                              
\begingroup                                                                                                                     
\begin{center}                                                                                                                  
\begin{table*}                                                                                                                   
\begin{tabular}{ccccccccccccc}                                                                                                            
\hline\hline                                                                                                                    
Parameters & CMB & CMB+BAO & CMB+Pantheon & CMB+BAO+Pantheon \\ \hline
$\Omega_c h^2$ & $    0.144_{-    0.026-    0.029}^{+    0.012+    0.038}$ & $    0.139_{-    0.017-    0.021}^{+    0.008+    0.027}$ & $    0.142_{-    0.020-    0.024}^{+    0.010+    0.029}$ & $    0.137_{-    0.015-    0.019}^{+    0.007+    0.024}$  \\

$\Omega_b h^2$ & $    0.02238_{-    0.00017-    0.00031}^{+    0.00016+    0.00030}$ & $    0.02234_{-    0.00014-    0.00028}^{+    0.00014+    0.00029}$ & $    0.02232_{-    0.00015-    0.00030}^{+    0.00015+    0.00030}$ & $    0.02235_{-    0.00014-    0.00028}^{+    0.00014+    0.00028}$ \\

$100\theta_{MC}$ & $    1.0397_{-    0.0008-    0.0020}^{+    0.0013+    0.0017}$ & $    1.03992_{-    0.00055-    0.0015}^{+    0.00088+    0.0013}$  & $    1.0398_{-    0.0007-    0.0016}^{+    0.0010+    0.0014}$ & $    1.04000_{-    0.00051-    0.0014}^{+    0.00080+    0.0012}$  \\

$\tau$ & $    0.0538_{-    0.0077-    0.015}^{+    0.0078+    0.017}$ & $    0.0547_{-    0.0077-    0.015}^{+    0.0076+    0.016}$ & $    0.0550_{-    0.0084-    0.015}^{+    0.0073+    0.016}$  & $    0.0550_{-    0.0077-    0.014}^{+    0.0071+    0.015}$\\

$n_s$ & $    0.9648_{-    0.0042-    0.0086}^{+    0.0042+    0.0086}$ & $    0.9640_{-    0.0042-    0.0082}^{+    0.0042+    0.0084}$ & $    0.9633_{-    0.0041-    0.0080}^{+    0.0042+    0.0080}$  & $    0.9642_{-    0.0041-    0.0082}^{+    0.0041+    0.0082}$ \\

${\rm{ln}}(10^{10} A_s)$ & $    3.044_{-    0.016-    0.032}^{+    0.016+    0.033}$ & $    3.046_{-    0.016-    0.031}^{+    0.016+    0.033}$  & $    3.047_{-    0.017-    0.030}^{+    0.015+    0.032}$  & $    3.047_{-    0.016-    0.030}^{+    0.015+    0.030}$  \\

$w_0^p$ & $ -1.96_{-    0.53}^{+    0.73} \; uncons.$ & $ > -1.21 \; >   -1.50 $  & $  >-1.17\;>-1.43 $ & $  >-1.10\;>-1.21 $ \\

$w_e^p$ & $ -1.69_{-    0.18}^{+    0.68} \; > -2.61 $ & $ -1.13_{-    0.05}^{+    0.11} \; > -1.28 $  & $ >-1.15\;>-1.26 $ & $   -1.124_{-    0.059}^{+    0.094}\;>-1.25$  \\

$\xi_0$ & $  > -0.48 \; > -0.85  $ & $ > -0.36 \; > -0.67 $ & $   -0.33_{-    0.10}^{+    0.32}\;>-0.73$ & $   -0.27_{-    0.07}^{+    0.26}\;>-0.62$  \\

$\Omega_{m0}$ & $    0.218_{-    0.066-    0.09}^{+    0.025+    0.12}$ & $    0.333_{-    0.038-    0.054}^{+    0.024+    0.062}$  & $    0.354_{-    0.051-    0.064}^{+    0.025+    0.084}$ & $    0.340_{-    0.034-    0.047}^{+    0.017+    0.056}$  \\

$\sigma_8$ & $    0.91_{-    0.07-    0.18}^{+    0.11+    0.16}$ & $    0.774_{-    0.037-    0.094}^{+    0.058+    0.082}$  & $    0.755_{-    0.037-    0.10}^{+    0.066+    0.08}$  & $    0.767_{-    0.029-    0.084}^{+    0.052+    0.074}$ \\

$H_0$ [km/s/Mpc] & $   89_{-    5-   16}^{+   11+   13}$ & $   69.7_{- 2.0 -    3.2}^{+    1.3+    3.6}$  & $   68.3_{-    1.2-    2.5}^{+    1.2+    2.5}$ & $   68.60_{-    0.76-    1.5}^{+    0.76+    1.5}$ \\

$S_8$ & $    0.761_{-    0.032-    0.052}^{+    0.024+    0.056}$ & $    0.813_{-    0.015-    0.040}^{+    0.022+    0.036}$  & $    0.816_{-    0.017-    0.038}^{+    0.021+    0.037}$ & $    0.815_{-    0.015-    0.038}^{+    0.019+    0.033}$ \\

$r_{\rm{drag}}$ [Mpc] & $  147.06_{-    0.28-    0.58}^{+    0.29+    0.57}$ & $  147.03_{-    0.29-    0.56}^{+    0.29+    0.56}$  & $  146.98_{-    0.28-    0.55}^{+    0.28+    0.56}$  & $  147.03_{-    0.28-    0.56}^{+    0.28+    0.54}$  \\
\hline\hline                                                                                                                    
\end{tabular} 
\caption{Observational constraints on the interacting  scenario \textbf{``$\xi_0w_0^{p}w_e^{p}$IDE$Q_B$''} obtained from  several observational datasets, namely, CMB, CMB+BAO, CMB+Pantheon and CMB+BAO+Pantheon are presented.  }
\label{tab:IDEp-ModelB-xi-cons-w-dyn}                                                                                       
\end{table*}                                                                                                                     
\end{center}                                                                                                                    
\endgroup                                                                 
\begingroup                                                                                                                     
\begin{center}                                                                                                                  
\begin{table*}                                                                                                                   
\begin{tabular}{cccccccccccc}                                                                                                            
\hline\hline                                                                                                                    
Parameters & CMB & CMB+BAO & CMB+Pantheon & CMB+BAO+Pantheon \\ \hline

$\Omega_c h^2$ & $    0.1164_{-    0.0021-    0.0075}^{+    0.0042+    0.0060}$ & $    0.1132_{-    0.0028-    0.0080}^{+    0.0049+    0.0068}$ & $    0.1128_{-    0.0035-    0.0086}^{+    0.0054+    0.0079}$  & $    0.1121_{-    0.0033-    0.0083}^{+    0.0053+    0.0075}$ \\

$\Omega_b h^2$ & $    0.02235_{-    0.00014-    0.00028}^{+    0.00016+    0.00026}$ & $    0.02247_{-    0.00014-    0.00027}^{+    0.00014+    0.00029}$ & $    0.02240_{-    0.00014-    0.00027}^{+    0.00014+    0.00030}$   & $    0.02245_{-    0.00014-    0.00026}^{+    0.00014+    0.00027}$ \\

$100\theta_{MC}$ & $    1.04110_{-    0.00036-    0.00070}^{+    0.00036+    0.00071}$ & $    1.04140_{-    0.00036-    0.00067}^{+    0.00036+    0.00070}$ & $    1.04135_{-    0.00037-    0.00072}^{+    0.00037+    0.00078}$ & $    1.04142_{-    0.00039-    0.00071}^{+    0.00036+    0.00076}$ \\

$\tau$ & $    0.0543_{-    0.0073-    0.016}^{+    0.0075+    0.016}$ & $    0.0559_{-    0.0086-    0.015}^{+    0.0074+    0.016}$  & $    0.0544_{-    0.0081-    0.015}^{+    0.0075+    0.016}$  & $    0.0552_{-    0.0073-    0.015}^{+    0.0072+    0.016}$\\

$n_s$ & $    0.9650_{-    0.0041-    0.0086}^{+    0.0046+    0.0083}$ & $    0.9688_{-    0.0040-    0.0077}^{+    0.0039+    0.0079}$ & $    0.9666_{-    0.0041-    0.0084}^{+    0.0041+    0.0085}$ & $    0.9686_{-    0.0038-    0.0076}^{+    0.0039+    0.0075}$  \\

${\rm{ln}}(10^{10} A_s)$ & $    3.045_{-    0.015-    0.031}^{+    0.015+    0.032}$ & $    3.044_{-    0.018-    0.030}^{+    0.015+    0.032}$ & $    3.043_{-    0.015-    0.032}^{+    0.016+    0.032}$  & $    3.044_{-    0.017-    0.031}^{+    0.015+    0.033}$  \\

$w_0^q$ & $   < -0.722 \; < -0.359  $ & $ < -0.888\; < -0.750 $ & $ <-0.963\;<-0.917  $   & $ <-0.957\; <-0.907 $  \\

$w_e^q$ & $ < -0.868 \; < -0.714 $ & $ < -0.952 \; < -0.892 $ & $  <-0.941\;<-0.861 $   & $   <-0.957\;<-0.903$  \\

$\xi_0$ & $  < 0.080 \; < 0.17  $ & $  < 0.112 \; < 0.21  $ & $ <0.145\;<0.23$   & $    0.108_{-    0.103}^{+    0.034}\;<0.23$ \\

$\Omega_{m0}$ & $    0.356_{-    0.048-    0.071}^{+    0.026+    0.089}$ & $    0.309_{-    0.016-    0.032}^{+    0.015+    0.032}$ & $    0.303_{-    0.014-    0.029}^{+    0.016+    0.027}$ & $    0.297_{-    0.012-    0.026}^{+    0.014+    0.024}$  \\

$\sigma_8$ & $    0.780_{-    0.028-    0.081}^{+    0.042+    0.068}$ & $    0.812_{-    0.026-    0.045}^{+    0.022+    0.050}$  & $    0.829_{-    0.025-    0.039}^{+    0.017+    0.042}$ & $    0.828_{-    0.025-    0.039}^{+    0.017+    0.043}$  \\

$H_0$ [km/s/Mpc] & $   62.8_{-    2.1-    6.5}^{+    3.9+    5.3}$ & $   66.4_{-    0.9-    2.5}^{+    1.4+    2.2}$  & $   67.02_{-    0.75-    1.7}^{+    0.96+    1.5}$ & $   67.4_{-    0.63-    1.3}^{+    0.64+    1.3}$   \\

$S_8$ & $    0.847_{-    0.018-    0.034}^{+    0.018+    0.034}$ & $    0.824_{-    0.014-    0.026}^{+    0.013+    0.027}$ & $    0.832_{-    0.015-    0.030}^{+    0.015+    0.032}$  & $    0.824_{-    0.013-    0.026}^{+    0.013+    0.026}$  \\

$r_{\rm{drag}}$ [Mpc] & $  147.05_{-    0.32-    0.57}^{+    0.29+    0.58}$ & $  147.34_{-    0.26-    0.51}^{+    0.26+    0.50}$ & $  147.18_{-    0.28-    0.55}^{+    0.29+    0.55}$ & $  147.32_{-    0.25-    0.49}^{+    0.25+    0.50}$  \\

\hline\hline                                                                                                                    
\end{tabular}                                                                                                                   
\caption{Observational constraints on the interacting  scenario \textbf{``$\xi_0w_0^{q}w_e^{q}$IDE$Q_B$''} obtained from  several observational datasets, namely, CMB, CMB+BAO, CMB+Pantheon and CMB+BAO+Pantheon are presented.   }
\label{tab:IDEq-ModelB-xi-cons-w-dyn}                                                                                                   
\end{table*}                                                                                                                     
\end{center}                                                                                                                    
\endgroup                                                      
\subsubsection{Dynamical dark energy equation of state}

We extend in this section the previous scenarios by allowing the dark energy equation of state to be dynamical, adopting the most common choice in the literature, see Eq.~(\ref{dyn-eos-1}). The stability of these interacting scenarios strongly relies on the parameter space. We shall follow here the analysis of Ref.~\citep{Gavela:2009cy}, dividing the parameter space into two distinct regions, namely, $\xi_0 < 0, w_0 < -1, w_e < -1$ and $\xi_0> 0, w_0 > -1, w_e > -1$, where the cosmological evolution will be stable. In the following, we shall describe the constraints for all the interacting scenarios.

We start by presenting in Tab.~\ref{tab:IDEp-ModelA-xi-cons-w-dyn} the constraints on the interacting scenario $\xi_0w_0^{p}w_e^{p}${\bf IDE}$Q_A$,  characterized by the parameter space $\xi_0 < 0, w_0^{p} < -1, w_e^{p} < -1$ and the interaction function $Q_A$, Eq.~(\ref{modelA}), by making use of the CMB, CMB+BAO, CMB+Pantheon and CMB+BAO+Pantheon datasets. 
For CMB alone, we do not find any indication for a non-zero coupling in the dark sector and we also do not get any evidence for phantom dark energy at the present time: indeed, $w_0^p$ remains unconstrained at 95\% CL. On the other hand, the evidence for a dark energy component  at early times (different from pure vacuum energy) is also not present. The strong degeneracy between the dark energy equation of state parameter and the Hubble constant leads to a much larger value of $H_0$, and consequently, to a much lower value of $\Omega_{m0}$, to leave unchanged $\Omega_m h^2$. When the BAO observations are added to CMB measurements, we still do not find any evidence of a non-zero coupling parameter but the mean values of some of the  parameters are shifted closer to their $\Lambda$CDM-like values, leaving a  signal for a dark energy component at early times (different from the minimal cosmological constant) at 68\% CL ($w_e^p = -1.11^{+0.10}_{-0.04}$ at 68\% CL  for CMB+BAO) which is of phantom nature,  and solving the Hubble constant tension $H_0$ within $2\sigma$.
Interestingly, the addition of Pantheon to CMB data shows an evidence for a non-zero coupling in the dark sector at more than 95\% CL ($\xi_0 =  -0.060_{-0.048}^{+  0.058}$ at 95\% CL for CMB+Pantheon) and the mean values of some of the parameters are shifted closer to their $\Lambda$CDM-like values.  The inclusion of both BAO and Pantheon observations to CMB does not change the overall picture except for an evidence of early dark energy (different from a cosmological constant at early times) at 68\% CL compared to the CMB+Pantheon constraints. Similarly to the CMB+Pantheon case, the coupling parameter is found to be non-zero at more than 95\% CL ($\xi_0 = -0.056_{-  0.048}^{+  0.055}$ at 95\% CL for CMB+BAO+Pantheon). It is important to note that for this full dataset combination the Hubble constant tension $H_0$ is alleviated at $2.5\sigma$, and we also obtain a lower value for $S_8$, improving the consistency with weak lensing data~\cite{Heymans:2020gsg,KiDS:2020ghu,DES:2021vln,DES:2022ygi}.

Table~\ref{tab:IDEq-ModelA-xi-cons-w-dyn} corresponds to the constraints on the interacting scenario $\xi_0 w_0^{q} w_e^{q}${\bf IDE}$Q_A$ characterized by the parameter space $\xi_0> 0, w_0^q > -1, w_e^q > -1$  and the interaction function $Q_A$, given in Eq. (\ref{modelA}), for CMB, CMB+BAO, CMB+Pantheon and CMB+BAO+Pantheon datasets. We find that CMB alone shows a preference of a non-zero coupling at 68\% CL ($\xi_0 = 0.25_{- 0.14}^{+  0.13}$ at 68\% CL) while this preference is diluted within 95\% CL. When BAO data are added to CMB alone, the preference of a non-zero coupling is pronounced as we can see from its 95\% CL constraints: $\xi_ 0 = 0.176_{- 0.15}^{+    0.15}$ at 95\% CL for CMB+BAO. The preference for a non-zero coupling at more than 95\% CL is also found for CMB+Pantheon and CMB+BAO+Pantheon. Notice the extreme low values of $\Omega_{c}h^2$ (and, consequently, the very high values of $\sigma_8$), which are a consequence of two effects. On one hand, the values of $\Omega_{m0}$ are very small, due to the energy flow from the dark matter sector to the dark energy one. On the other hand, since the equation of state parameter is required to lie within the quintessence region, the values of $H_0$ are lower. The product of these two parameters (i.e. $\Omega_{m0}$ and $H_0$) is consequently much lower.

In Tables~\ref{tab:IDEp-ModelB-xi-cons-w-dyn} and~\ref{tab:IDEq-ModelB-xi-cons-w-dyn} we have summarized the constraints for the interaction scenarios driven by the interaction function $Q_B$, Eq.~(\ref{modelB}), for different parameter spaces. More concretely, Tab.~\ref{tab:IDEp-ModelB-xi-cons-w-dyn} corresponds to the interacting scenario $\xi_0w_0^{p}w_e^{p}${\bf IDE}$Q_B$ characterized by the parameter space $\xi_0 < 0, w_0^p < -1, w_e^p < -1$. Focusing on the CMB constraints, notice that we do not find any evidence for a non-zero coupling in the dark sector. Further, the present value of the dark energy equation of state lies in the phantom regime at 68\% CL ($w_0^p = -1.96_{-    0.53}^{+    0.73}$ at 68\% CL for CMB alone) while it remains unconstrained at 95\% CL. Due to the strong degeneracy between the dark energy equation of state and the Hubble constant, a very high value of $H_0$ is also found.  A mild  evidence for an early dark energy component (which is not vacuum energy at early times) for CMB data alone appears ($ w_e^p = -1.69_{-    0.18}^{+    0.68}$ at 68\% CL). When BAO data are added to CMB observations, the mean values of some of the parameters are shifted to their $\Lambda$CDM values, except for the mild preference for a dark energy component different from the standard cosmological constant case  at early times ($w_e\neq-1$), which still persists ($w_e ^p =  -1.13_{-    0.05}^{+    0.11}$ at 68\% CL, CMB+BAO), and the value of $\Omega_{m0} $ ($\sigma_8$), which are higher (lower) than in the $\Lambda$CDM picture. Interestingly, the inclusion of Pantheon data to CMB measurements leads a non-null value of the coupling parameter ($\xi_0  = -0.33_{-    0.10}^{+    0.32}$ at 68\% CL, CMB+Pantheon). Again we see that for CMB+Pantheon dataset, the value of $\Omega_{m0} $ ($\sigma_8$) is still mildly higher (lower) than in the $\Lambda$CDM picture.
Finally, for CMB+BAO+Pantheon, we have the same indication for a non-zero coupling at $1\sigma$ together with an early dark energy parameter ($w_{e}^p =  -1.124_{-    0.059}^{+    0.094}$ at 68\% CL), which shows a mild preference for early dark energy instead than for vacuum energy, compared to the constraints from CMB+Pantheon.

In Table~\ref{tab:IDEq-ModelB-xi-cons-w-dyn} we summarize the constraints on the scenario  $\xi_0w_0^{q}w_e^{q}${\bf IDE}$Q_B$ characterized by the parameter space $\xi_0> 0, w_0^q > -1, w_e^q > -1$  arising from CMB, CMB+BAO, CMB+Pantheon and CMB+BAO+Pantheon datasets. For CMB alone, we do not find any  indication for a non-zero coupling in the dark sector. The matter density parameter takes a very high value due to the  energy flow and also by the fact that $w(a)>-1$. As a consequence, the Hubble constant assumes a relatively small value, to leave $\Omega_c h^2$ unchanged.  When the BAO observations are added to CMB, the mean values of the parameters are shifted toward their $\Lambda$CDM values and again no evidence for a non vanishing coupling parameter is found,  and similar conclusions hold for CMB+Pantheon.  For the remaining CMB+BAO+Pantheon,  we obtain instead very similar results to those obtained with CMB+BAO  and CMB+Pantheon, with the exception  of $1\sigma$ preference for a coupling different from zero.
\begingroup                                                                                                                     
\begin{center}                                                                                                                  
\begin{table*}                                                                                                                   
\begin{tabular}{ccccccccc}                                                                                                            
\hline\hline                                                                                                                    
Parameters & CMB & CMB+BAO & CMB+Pantheon & CMB+BAO+Pantheon \\ \hline
$\Omega_c h^2$ & $  <0.092\;<0.137 $ &  $    0.119_{-    0.014-    0.040}^{+    0.026+    0.034}$ & $    0.110_{-    0.017-    0.037}^{+    0.021+    0.036}$ & $    0.117_{-    0.014-    0.031}^{+    0.018+    0.031}$\\

$\Omega_b h^2$ & $    0.02231_{-    0.00016-    0.00028}^{+    0.00014+    0.00030}$ & $    0.02232_{-    0.00014-    0.00030}^{+    0.00014+    0.00027}$ & $    0.02228_{-    0.00015-    0.00030}^{+    0.00015+    0.00028}$ & $    0.02232_{-    0.00014-    0.00029}^{+    0.00015+    0.00028}$\\

$100\theta_{MC}$ & $    1.0441_{-    0.0036-    0.0052}^{+    0.0030+    0.0054}$ & $    1.0408_{-    0.0015-    0.0020}^{+    0.0008+    0.00234}$ & $    1.0412_{-    0.0013-    0.0022}^{+    0.0010+    0.0022}$ & $    1.0409_{-    0.0011-    0.0019}^{+    0.0009+    0.0019}$ \\

$\tau$ & $    0.0547_{-    0.0075-    0.016}^{+    0.0077+    0.015}$ & $    0.0549_{-    0.0088-    0.015}^{+    0.0077+    0.016}$  & $    0.0545_{-    0.0086-    0.015}^{+    0.0079+    0.015}$ & $    0.0552_{-    0.0074-    0.016}^{+    0.0077+    0.016}$ \\

$n_s$ & $    0.9724_{-    0.0043-    0.0082}^{+    0.0044+    0.0085}$ & $    0.9732_{-    0.0039-    0.0080}^{+    0.0041+    0.0078}$ & $    0.9718_{-    0.0042-    0.0081}^{+    0.0042+    0.0086}$   & $    0.9728_{-    0.0042-    0.0081}^{+    0.0041+    0.0082}$\\

${\rm{ln}}(10^{10} A_s)$ & $    3.056_{-    0.015-    0.031}^{+    0.015+    0.032}$ & $    3.055_{-    0.018-    0.031}^{+    0.016+    0.034}$ & $    3.056_{-    0.016-    0.032}^{+    0.016+    0.031}$ & $    3.056_{-    0.015-    0.032}^{+    0.015+    0.030}$  \\

$\xi_0$ & $    0.09_{-    0.15-    0.27}^{+    0.15+    0.28}$ & $   -0.05_{-    0.14-    0.21}^{+    0.10+    0.23}$  & $    0.01_{-    0.13-    0.25}^{+    0.14+    0.24}$ & $   -0.04_{-    0.10-    0.20}^{+    0.11+    0.20}$ \\

$\xi_a$ & $ 0.19_{-    0.47}^{+    0.38} \; > -0.421 $ & $    0.21_{-    0.27-    0.51}^{+    0.27+    0.52}$ & $    0.10_{-    0.38-    0.63}^{+    0.30+    0.68}$  & $    0.19_{-    0.29-    0.50}^{+    0.23+    0.53}$  \\

$\Omega_{m0}$ & $    0.19_{-    0.13-    0.16}^{+    0.07+    0.20}$ & $    0.306_{-    0.039-    0.10}^{+    0.063+    0.09}$ & $    0.286_{-    0.038-    0.087}^{+    0.048+    0.082}$ & $    0.300_{-    0.035-    0.075}^{+    0.043+    0.073}$   \\

$\sigma_8$ & $    1.6_{-    1.1-    1.3}^{+    0.3+    2.1}$ & $    0.83_{-    0.16-    0.21}^{+    0.06+    0.28}$  & $    0.89_{-    0.16-    0.23}^{+    0.08+    0.28}$  & $    0.84_{-    0.12-    0.18}^{+    0.06+    0.21}$  \\

$H_0$ [km/s/Mpc] & $   71.8_{-    3.2-    7.2}^{+    4.3+    6.2}$ & $   68.2_{-    1.2-    2.2}^{+    1.2+    2.4}$ & $   68.2_{-    1.1-    2.3}^{+    1.1+    2.2}$ & $   68.30_{-    0.90-    1.6}^{+    0.82+    1.7}$  \\

$S_8$ & $    1.04_{-    0.29-    0.35}^{+    0.10+    0.53}$ & $    0.828_{-    0.065-    0.09}^{+    0.032+    0.12}$  & $    0.860_{-    0.076-    0.12}^{+    0.045+    0.13}$  & $    0.831_{-    0.052-    0.085}^{+    0.034+    0.094}$  \\

$r_{\rm{drag}}$ [Mpc] & $  147.08_{-    0.30-    0.58}^{+    0.30+    0.59}$ & $  147.15_{-    0.28-    0.58}^{+    0.29+    0.55}$ & $  147.04_{-    0.29-    0.60}^{+    0.30+    0.61}$  & $  147.14_{-    0.29-    0.56}^{+    0.28+    0.55}$  \\

\hline\hline                                                
\end{tabular}
\caption{Observational constraints on the interacting  scenario \textbf{``$\xi_0 \xi_a$IVS$Q_A$''} obtained from  several observational datasets, namely, CMB, CMB+BAO, CMB+Pantheon and CMB+BAO+Pantheon are presented.  }                                       \label{tab:IVS-ModelA-xi-dyn}                                                                                                  
\end{table*}                                                                                                                     
\end{center}                                                                                                                    
\endgroup                                                                                                                       
\begingroup                                                                                                                     
\begin{center}                                                                                                                  
\begin{table*}                                                                                                                   
\begin{tabular}{ccccccccc}                                                                                                            
\hline\hline                                                                                                                    
Parameters & CMB & CMB+BAO & CMB+Pantheon & CMB+BAO+Pantheon   \\ \hline

$\Omega_c h^2$ & $    0.118_{-    0.036-    0.054}^{+    0.028+    0.056}$ & $    0.123_{-    0.017-    0.035}^{+    0.018+    0.034}$ & $    0.117_{-    0.011-    0.024}^{+    0.013+    0.022}$  & $    0.122_{-    0.011-    0.021}^{+    0.011+    0.021}$ \\

$\Omega_b h^2$ & $    0.02230_{-    0.00014-    0.00028}^{+    0.00014+    0.00027}$ & $    0.02233_{-    0.00015-    0.00030}^{+    0.00015+    0.00029}$ & $    0.02229_{-    0.00015-    0.00029}^{+    0.00015+    0.00031}$  & $    0.02232_{-    0.00014-    0.00029}^{+    0.00014+    0.00029}$ \\

$100\theta_{MC}$ & $    1.0408_{-    0.0017-    0.0029}^{+    0.0018+    0.0030}$ & $    1.0405_{-    0.0010-    0.0018}^{+    0.0009+    0.0020}$ & $    1.04080_{-    0.00071-    0.0013}^{+    0.00070+    0.0014}$ & $    1.04058_{-    0.00068-    0.0012}^{+    0.00066+    0.0013}$  \\

$\tau$ & $    0.0548_{-    0.0082-    0.015}^{+    0.0074+    0.016}$ & $    0.0550_{-    0.0075-    0.015}^{+    0.0073+    0.016}$ & $    0.0548_{-    0.0080-    0.016}^{+    0.0079+    0.017}$ & $    0.0551_{-    0.0075-    0.016}^{+    0.0076+    0.017}$ \\

$n_s$ & $    0.9723_{-    0.0042-    0.0082}^{+    0.0043+    0.0084}$ & $    0.9732_{-    0.0042-    0.0080}^{+    0.0041+    0.0082}$  & $    0.9721_{-    0.0044-    0.0084}^{+    0.0044+    0.0086}$  & $    0.9730_{-    0.0041-    0.0079}^{+    0.0040+    0.0082}$ \\

${\rm{ln}}(10^{10} A_s)$ & $    3.056_{-    0.016-    0.032}^{+    0.016+    0.034}$ & $    3.055_{-    0.015-    0.031}^{+    0.015+    0.033}$ & $    3.056_{-    0.017-    0.032}^{+    0.016+    0.034}$ & $    3.056_{-    0.016-    0.032}^{+    0.015+    0.033}$  \\

$\xi_0$ & $   -0.04_{-    0.31-    0.45}^{+    0.19+    0.50}$ & $   -0.14_{-    0.23-    0.39}^{+    0.19+    0.41}$  & $   -0.07_{-    0.21-    0.34}^{+    0.16+    0.35}$   & $   -0.13_{-    0.18-    0.31}^{+    0.16+    0.32}$ \\

$\xi_a$ & $    0.20_{-    0.34-    0.59}^{+    0.26+    0.68}$ & $    0.37_{-    0.33-    0.56}^{+    0.41+    0.63}$  & $    0.30_{-    0.38-    0.62}^{+    0.41+    0.70}$  & $    0.38_{-    0.34-    0.54}^{+    0.37+    0.62}$  \\

$\Omega_{m0}$ & $    0.32_{-    0.13-    0.17}^{+    0.07+    0.22}$ & $    0.319_{-    0.047-    0.094}^{+    0.046+    0.093}$ & $    0.301_{-    0.032-    0.062}^{+    0.031+    0.063}$ & $    0.311_{-    0.030-    0.055}^{+    0.028+    0.058}$ \\

$\sigma_8$ & $    0.83_{-    0.11-    0.20}^{+    0.12+    0.20}$ & $    0.813_{-    0.063-    0.12}^{+    0.054+    0.12}$ & $    0.837_{-    0.043-    0.074}^{+    0.038+    0.083}$ & $    0.820_{-    0.035-    0.070}^{+    0.035+    0.072}$  \\

$H_0$ [km/s/Mpc] & $   67.7_{-    3.3-    7.9}^{+    4.9+    6.9}$ & $   67.9_{-    1.3-    2.7}^{+    1.3+    2.6}$ & $   68.3_{-    1.1-    2.3}^{+    1.2+    2.3}$ & $   68.17_{-    0.81-    1.6}^{+    0.80+    1.6}$ \\

$S_8$ & $    0.828_{-    0.022-    0.068}^{+    0.038+    0.054}$ &  $    0.831_{-    0.015-    0.032}^{+    0.017+    0.029}$ & $    0.835_{-    0.017-    0.034}^{+    0.017+    0.033}$ & $    0.833_{-    0.015-    0.030}^{+    0.015+    0.029}$ \\

$r_{\rm{drag}}$ [Mpc] & $  147.09_{-    0.29-    0.59}^{+    0.28+    0.60}$ & $  147.17_{-    0.28-    0.55}^{+    0.27+    0.56}$  & $  147.07_{-    0.30-    0.59}^{+    0.30+    0.59}$  & $  147.17_{-    0.28-    0.54}^{+    0.28+    0.56}$ \\

\hline\hline                                                                                                                    
\end{tabular}                                                                                                                   
\caption{Observational constraints on the interacting  scenario \textbf{``$\xi_0 \xi_a$IVS$Q_B$''} obtained from  several observational datasets, namely, CMB, CMB+BAO, CMB+Pantheon and CMB+BAO+Pantheon are presented.  } 
\label{tab:IVS-ModelB-xi-dyn}                                                                                               
\end{table*}                                                                                                                     
\end{center}                                                                                                                    
\endgroup                                                                                                                                                                                                                                              

\subsection{Dynamical coupling $\xi (a)$}

In the following sections we shall present the interacting cosmologies for the two interaction functions $Q_A$ and $Q_B$ of Eqs.~(\ref{modelA}) and (\ref{modelB}) assuming that a time-dependent coupling parameter $\xi(a)$, see Eq.~(\ref{dynamical-xi}).

\begingroup                                                                                                                     
\begin{center}                                                                                                                  
\begin{table*}                                                                                                                   
\begin{tabular}{cccccccc}                                                                                                            
\hline\hline                                                                                                                    
Parameters & CMB & CMB+BAO & CMB+Pantheon & CMB+BAO+Pantheon \\ \hline
$\Omega_c h^2$ & $    0.149_{-    0.008-    0.021}^{+    0.012+    0.018}$ & $    0.141_{-    0.006-    0.015}^{+    0.010+    0.013}$  & $    0.1444_{-    0.0042-    0.015}^{+    0.0092+    0.012}$ & $    0.139_{-    0.007-    0.015}^{+    0.010+    0.014}$  \\

$\Omega_b h^2$ & $    0.02246_{-    0.00016-    0.00030}^{+    0.00016+    0.00031}$ & $    0.02246_{-    0.00015-    0.00029}^{+    0.00015+    0.00030}$  & $    0.02244_{-    0.00015-    0.00030}^{+    0.00016+    0.00030}$ & $    0.02246_{-    0.00015-    0.00028}^{+    0.00014+    0.00030}$ \\

$100\theta_{MC}$ & $    1.03949_{-    0.00069-    0.0010}^{+    0.00042+    0.0012}$ & $    1.03989_{-    0.00054-    0.00087}^{+    0.00046+    0.00096}$  & $    1.03966_{-    0.00054-    0.0009}^{+    0.00043+    0.0010}$ & $    1.04001_{-    0.00048-    0.00086}^{+    0.00048+    0.00089}$ \\

$\tau$ & $    0.0522_{-    0.0074-    0.016}^{+    0.0079+    0.015}$ & $    0.0544_{-    0.0080-    0.015}^{+    0.0073+    0.016}$  & $    0.0527_{-    0.0081-    0.015}^{+    0.0076+    0.016}$ & $    0.0552_{-    0.0083-    0.015}^{+    0.0076+    0.016}$ \\

$n_s$ & $    0.9672_{-    0.0041-    0.0086}^{+    0.0044+    0.0087}$ & $    0.9676_{-    0.0045-    0.0086}^{+    0.0043+    0.0085}$  & $    0.9666_{-    0.0044-    0.0087}^{+    0.0048+    0.0083}$ & $    0.9680_{-    0.0044-    0.0081}^{+    0.0044+    0.0087}$ \\

${\rm{ln}}(10^{10} A_s)$ & $    3.038_{-    0.015-    0.031}^{+    0.017+    0.029}$ & $    3.042_{-    0.016-    0.031}^{+    0.015+    0.033}$  & $    3.039_{-    0.017-    0.032}^{+    0.016+    0.032}$ & $    3.044_{-    0.016-    0.031}^{+    0.016+    0.033}$ \\

{\color{red} $w_p$} & $ -1.83_{-    0.38}^{+    0.54} \; > -2.63 $ & $  -1.21_{-    0.08-    0.22}^{+    0.14+    0.20}$  & $   -1.203_{-    0.082-    0.16}^{+    0.092+    0.16}$ & $   -1.131_{-    0.052-    0.11}^{+    0.065+    0.11}$ \\

$\xi_0$ & $ > -0.030 \; > -0.062 $ & $ > -0.048 \; > -0.082 $  & $  >-0.035\;>-0.070 $ & $  >-0.055\;>-0.087 $ \\

$\xi_a$ & $  > -0.106 \; > -0.162  $ & $ > -0.051 \; > -0.095  $ & $ -0.071_{-    0.040}^{+    0.049}\; > -0.136 $  & $ >-0.066\;>-0.073  $ \\

$\Omega_{m0}$ & $    0.266_{-    0.091-    0.12}^{+    0.051+    0.14}$ & $    0.340_{-    0.021-    0.040}^{+    0.021+    0.039}$  & $    0.372_{-    0.020-    0.049}^{+    0.028+    0.045}$ & $    0.348_{-    0.018-    0.036}^{+    0.021+    0.034}$  \\

$\sigma_8$ & $    0.853_{-    0.088-    0.14}^{+    0.075+    0.15}$ & $    0.763_{-    0.035-    0.058}^{+    0.030+    0.060}$  & $    0.748_{-    0.027-    0.043}^{+    0.019+    0.048}$ & $    0.751_{-    0.035-    0.054}^{+    0.028+    0.057}$ \\

$H_0$ [km/s/Mpc] & $   82_{-   11-   19}^{+   12+   19}$ & $   69.4_{-    1.9-    3.2}^{+    1.5+    3.4}$  & $   67.1_{-    1.1-    2.2}^{+    1.1+    2.2}$ & $   68.19_{-    0.78-    1.5}^{+    0.78+    1.6}$ \\

$S_8$ & $    0.789_{-    0.042-    0.067}^{+    0.038+    0.070}$ & $    0.811_{-    0.016-    0.033}^{+    0.018+    0.032}$  & $    0.832_{-    0.020-    0.044}^{+    0.024+    0.043}$   & $    0.808_{-    0.018-    0.036}^{+    0.018+    0.034}$  \\

$r_{\rm{drag}}$ [Mpc] & $  147.16_{-    0.29-    0.59}^{+    0.29+    0.57}$ & $  147.25_{-    0.29-    0.54}^{+    0.29+    0.56}$   & $  147.14_{-    0.29-    0.58}^{+    0.30+    0.59}$  & $  147.28_{-    0.28-    0.54}^{+    0.28+    0.58}$  \\ 
\hline\hline                                                                                                                    
\end{tabular}                                       
\caption{Observational constraints on the interacting  scenario \textbf{``$\xi_0 \xi_a w_p$IDE$Q_A$''} obtained from  several observational datasets, namely, CMB, CMB+BAO, CMB+Pantheon and CMB+BAO+Pantheon are presented.    }                                \label{tab:IDEp-ModelA-xi-dyn}                            
\end{table*}                                                                                                                     
\end{center}                                                                                                                    
\endgroup                                                                                                                       
\begingroup                                                                                                                     
\begin{center}                                                                                                                  
\begin{table*}                                                                                                                   
\begin{tabular}{ccccccccc}                                                                                                            
\hline\hline                                                                                                                    
Parameters & CMB & CMB+BAO & CMB+Pantheon & CMB+BAO+Pantheon \\ \hline

$\Omega_c h^2$ & $ <0.034\;<0.110 $ & $    0.079_{-    0.014-    0.059}^{+    0.034+    0.043}$ & $    0.079_{-    0.015-    0.054}^{+    0.034+    0.042}$ & $    0.075_{-    0.017-    0.058}^{+    0.037+    0.0445}$ \\

$\Omega_b h^2$ & $    0.02232_{-    0.00016-    0.00031}^{+    0.00015+    0.00030}$ & $    0.02233_{-    0.00015-    0.00027}^{+    0.00014+    0.00029}$  & $    0.02232_{-    0.00015-    0.00030}^{+    0.00017+    0.00029}$  & $    0.02234_{-    0.00015-    0.00029}^{+    0.00015+    0.00029}$ \\

$100\theta_{MC}$ & $    1.0450_{-    0.0036-    0.0042}^{+    0.0020+    0.0049}$ & $    1.0435_{-    0.0023-    0.0030}^{+    0.0008+    0.0044}$  & $    1.0435_{-    0.0023-    0.0029}^{+    0.0010+    0.0040}$  & $    1.0437_{-    0.0025-    0.0031}^{+    0.0010+    0.0043}$ \\

$\tau$ & $    0.0549_{-    0.0075-    0.014}^{+    0.0072+    0.015}$ & $    0.0549_{-    0.0075-    0.015}^{+    0.0076+    0.016}$  & $    0.0547_{-    0.0076-    0.016}^{+    0.0077+    0.016}$  & $    0.0546_{-    0.0073-    0.015}^{+    0.0076+    0.015}$  \\

$n_s$ & $    0.9639_{-    0.0044-    0.0088}^{+    0.0048+    0.0089}$ & $    0.9645_{-    0.0046-    0.0082}^{+    0.0042+    0.0085}$  & $    0.9640_{-    0.0048-    0.0084}^{+    0.0043+    0.0090}$  & $    0.9647_{-    0.0041-    0.0083}^{+    0.0042+    0.0080}$ \\

${\rm{ln}}(10^{10} A_s)$ & $    3.046_{-    0.016-    0.030}^{+    0.015+    0.029}$ & $    3.046_{-    0.016-    0.032}^{+    0.016+    0.033}$  & $    3.046_{-    0.016-    0.033}^{+    0.015+    0.035}$  & $    3.045_{-    0.016-    0.031}^{+    0.016+    0.032}$ \\

$w_q$ & $ < -0.901 \; < -0.783 $ & $ < -0.872 \; < -0.779  $ & $ -0.905_{-    0.088}^{+    0.033}\; < -0.790 $ & $  -0.892_{-    0.097}^{+    0.040}\;  < -0.774 $\\

$\xi_0$ & $ 0.136_{-    0.088}^{+    0.085} \; < 0.254  $ & $ < 0.117 \; < 0.220 $  & $ < 0.125 < 0.212 $ & $  < 0.132 < 0.225 $ \\

$\xi_a$ & $  < 0.022 \; < 0.050  $ & $  < 0.021 \; < 0.044  $  & $  < 0.020 < 0.044 $ & $  < 0.020 < 0.042  $ \\

$\Omega_{m0}$ & $    0.17_{-    0.11-    0.15}^{+    0.07+    0.15}$ & $    0.220_{-    0.036-    0.13}^{+    0.077+    0.10}$  & $    0.217_{-    0.039-    0.12}^{+    0.075+    0.10}$  & $    0.210_{-    0.039-    0.13}^{+    0.079+    0.10}$ \\

$\sigma_8$ & $    1.7_{-    1.0-    1.2}^{+    0.2+    2.1}$ & $    1.18_{-    0.43-    0.6}^{+    0.07+    1.0}$  & $    1.17_{-    0.40-    0.53}^{+    0.09+    0.86}$  & $    1.23_{-    0.47-    0.63}^{+    0.08+    1.07}$   \\

$H_0$ [km/s/Mpc] & $   70.4_{-    3.1-    7.2}^{+    4.2+    6.8}$ & $   68.3_{-    1.2-    2.4}^{+    1.2+    2.5}$ & $   68.5_{-    1.1-    2.1}^{+    1.1+    2.2}$ & $   68.40_{-    0.81-    1.5}^{+    0.79+    1.6}$  \\

$S_8$ & $    1.06_{-    0.27-    0.32}^{+    0.08+    0.50}$ & $    0.94_{-    0.14-    0.19}^{+    0.03+    0.32}$ & $    0.94_{-    0.14-    0.18}^{+    0.04+    0.27}$ & $    0.96_{-    0.15-    0.20}^{+    0.04+    0.33}$ \\

$r_{\rm{drag}}$ [Mpc] & $  147.01_{- 0.30-    0.59}^{+    0.30+    0.59}$ & $  147.07_{-    0.29-    0.56}^{+    0.28+    0.57}$ & $  147.01_{-    0.29-    0.59}^{+    0.29+    0.59}$   & $  147.06_{-    0.27-    0.54}^{+    0.27+    0.52}$ \\
\hline\hline                                        
\end{tabular}                               
\caption{Observational constraints on the interacting  scenario \textbf{``$\xi_0 \xi_a w_q$IDE$Q_A$''} obtained from  several observational datasets, namely, CMB, CMB+BAO, CMB+Pantheon and CMB+BAO+Pantheon are presented.  }                         \label{tab:IDEq-ModelA-xi-dyn}                             
\end{table*}                                                                                    
\end{center}                                                                                    
\endgroup                                                                                       
\begingroup                                                                                                                     
\begin{center}                                                                                                                  
\begin{table*}                                                                                                                   
\begin{tabular}{ccccccc}                                                                                                        
\hline\hline                                                                                                                    
Parameters & CMB & CMB+BAO & CMB+Pantheon & CMB+BAO+Pantheon \\ \hline
$\Omega_c h^2$ & $    0.165_{-    0.012-    0.028}^{+    0.018+    0.027}$ & $    0.145_{-    0.016-    0.023}^{+    0.009+    0.028}$  & $    0.155_{-    0.011-    0.024}^{+    0.014+    0.023}$   & $    0.140_{-    0.015-    0.020}^{+    0.007+    0.026}$ \\

$\Omega_b h^2$ & $    0.02246_{-    0.00015-    0.00030}^{+    0.00015+    0.00031}$ & $    0.02244_{-    0.00016-    0.00030}^{+    0.00015+    0.00030}$  & $    0.02244_{-    0.00015-    0.00030}^{+    0.00015+    0.00030}$  & $    0.02244_{-    0.00016-    0.00028}^{+    0.00014+    0.00030}$ \\

$100\theta_{MC}$ & $    1.03869_{-    0.00085-    0.0014}^{+    0.00066+    0.0015}$ & $    1.03965_{-    0.00058-    0.0015}^{+    0.00086+    0.0014}$ & $    1.03911_{-    0.00078-    0.0013}^{+    0.00059+    0.0014}$  & $    1.03993_{-    0.00048-    0.0014}^{+    0.00081+    0.0012}$ \\

$\tau$ & $    0.0530_{-    0.0077-    0.016}^{+    0.0081+    0.016}$ & $    0.0551_{-    0.0086-    0.015}^{+    0.0078+    0.016}$ & $    0.0543_{-    0.0076-    0.016}^{+    0.0076+    0.016}$ & $    0.0561_{-    0.0076-    0.016}^{+    0.0077+    0.016}$ \\

$n_s$ & $    0.9668_{-    0.0046-    0.0090}^{+    0.0045+    0.0091}$ & $    0.9664_{-    0.0043-    0.0085}^{+    0.0044+    0.0082}$  & $    0.9660_{-    0.0045-    0.0086}^{+    0.0045+    0.0086}$  & $    0.9668_{-    0.0041-    0.0079}^{+    0.0042+    0.0083}$ \\

${\rm{ln}}(10^{10} A_s)$ & $    3.040_{-    0.016-    0.032}^{+    0.016+    0.032}$ & $    3.044_{-    0.016-    0.033}^{+    0.016+    0.033}$  & $    3.044_{-    0.016-    0.034}^{+    0.016+    0.033}$   & $    3.046_{-    0.016-    0.031}^{+    0.016+    0.033}$  \\

$w_p$ & $   -2.10_{-    0.58-    0.84}^{+    0.40+    0.86}$ & $   -1.26_{-    0.10-    0.27}^{+    0.16+    0.24}$   & $   -1.32_{-    0.14-    0.24}^{+    0.14+    0.25}$  & $   -1.143_{-    0.053-    0.15}^{+    0.089+    0.13}$   \\

$\xi_0$ & $ > -0.478 \; > -0.765 $ & $ > -0.289 \; > -0.641  $ & $ > -0.242 > -0.531 $ & $ > -0.252 > -0.604  $  \\

$\xi_a$ & $  < -0.47$\; unconstr.  & $ -0.21_{-    0.08}^{+    0.18} \;  > -0.44 $  & $ -0.50_{-    0.26}^{+    0.26}\;$ unconstr.  & $ > -0.17 > -0.31  $  \\

$\Omega_{m0}$ & $    0.28_{-    0.10-    0.13}^{+    0.04+    0.18}$ & $    0.344_{-    0.031-    0.050}^{+    0.022+    0.056}$  & $    0.405_{-    0.031-    0.073}^{+    0.045+    0.066}$  & $    0.351_{-    0.033-    0.048}^{+    0.018+    0.058}$  \\

$\sigma_8$ & $    0.823_{-    0.086-    0.17}^{+    0.086+    0.16}$ & $    0.759_{-    0.034-    0.085}^{+    0.050+    0.077}$ & $    0.719_{-    0.038-    0.074}^{+    0.038+    0.076}$ & $    0.755_{-    0.026-    0.085}^{+    0.048+    0.069}$  \\

$H_0$ [km/s/Mpc] & $   85_{-    7-   22}^{+   15+   18}$ & $   69.9_{-    2.0-    3.4}^{+    1.7+    3.5}$  & $   66.4_{-    1.4-    2.4}^{+    1.2+    2.6}$ & $   68.21_{-    0.82-    1.6}^{+    0.82+    1.6}$ \\

$S_8$ & $    0.770_{-    0.048-    0.069}^{+    0.029+    0.081}$ & $    0.811_{-    0.015-    0.041}^{+    0.023+    0.036}$  & $    0.833_{-    0.020-    0.039}^{+    0.022+    0.037}$   & $    0.814_{-    0.014-    0.038}^{+    0.020+    0.034}$ \\

$r_{\rm{drag}}$ [Mpc] & $  147.15_{-  0.30-    0.58}^{+    0.30+    0.61}$ & $  147.18_{-  0.28-    0.54}^{+    0.28+    0.57}$   & $  147.10_{-  0.30-    0.58}^{+    0.30+    0.59}$  & $  147.23_{-  0.28-    0.54}^{+    0.27+    0.54}$  \\
\hline\hline                                            
\end{tabular}                                           
\caption{Observational constraints on the interacting  scenario \textbf{``$\xi_0 \xi_a w_p$IDE$Q_B$''} obtained from  several observational datasets, namely, CMB, CMB+BAO, CMB+Pantheon and CMB+BAO+Pantheon are presented.  }                                      \label{tab:IDEp-ModelB-xi-dyn}                                                                                           
\end{table*}                                                                                                                     
\end{center}                                                                                                                    
\endgroup   
\begin{figure*}
\centering
\includegraphics[width=0.8\textwidth]{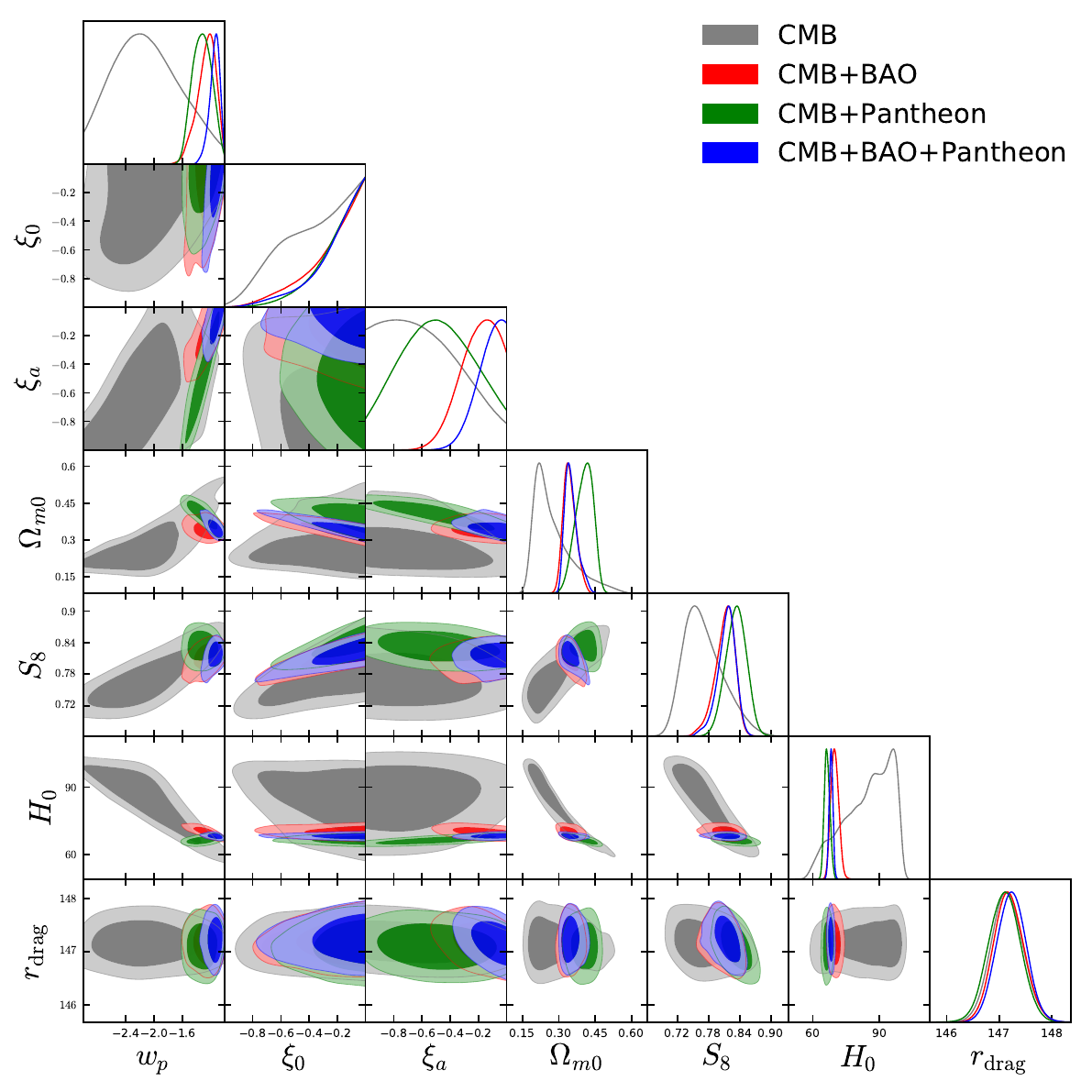}
    \caption{One dimensional marginalized posterior distributions and the two-dimensional joint contours for the most relevant parameters of the interacting  scenario \textbf{``$\xi_0 \xi_a w_p$IDE$Q_B$''} for several datasets, namely, CMB, CMB+BAO, CMB+Pantheon and CMB+BAO+Pantheon. }
    \label{fig:IDEp-ModelB-xi-dyn}
\end{figure*}
\begin{figure*}
\centering
\includegraphics[width=0.8\textwidth]{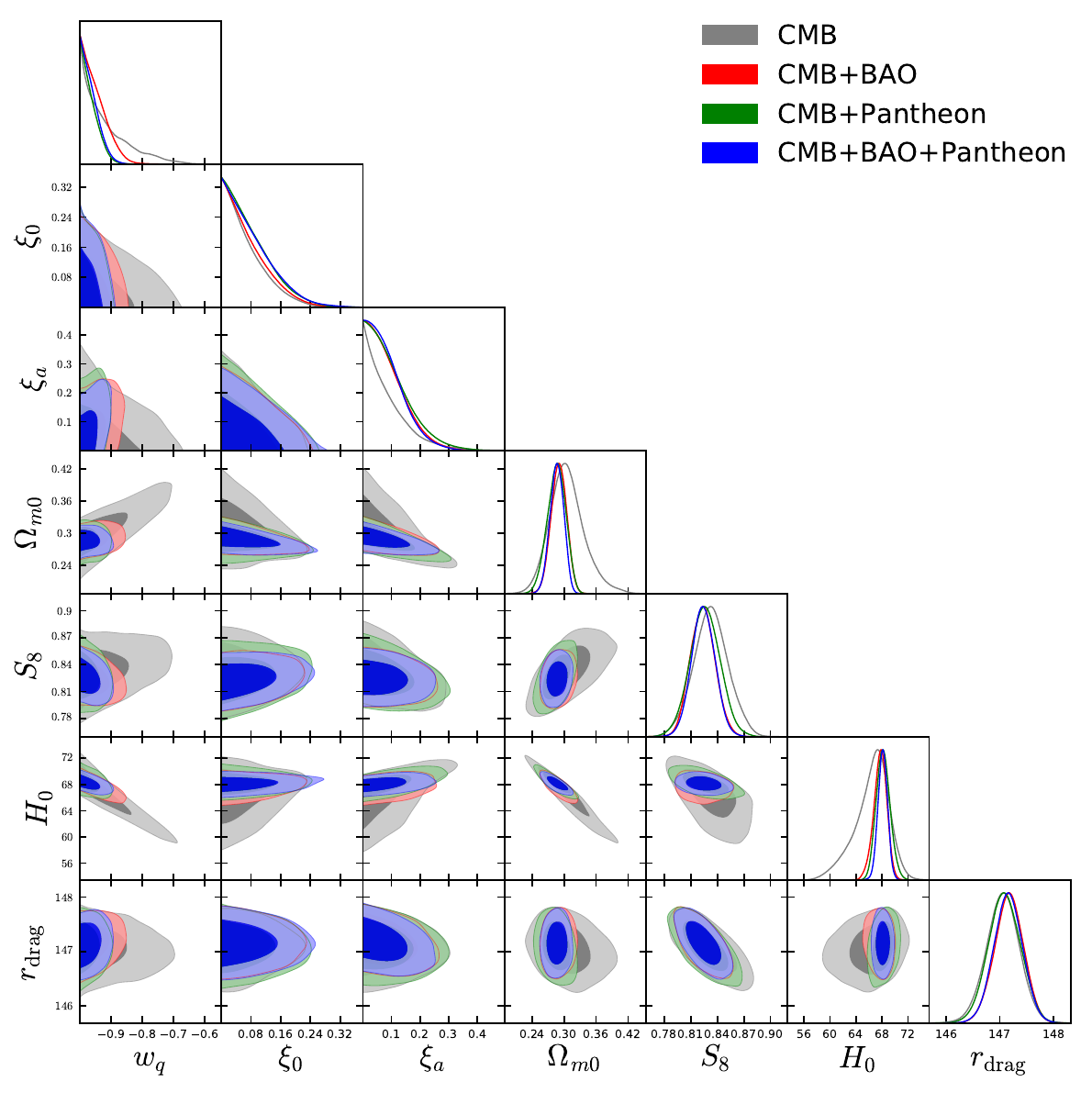}
    \caption{One dimensional marginalized posterior distributions and the two-dimensional joint contours for the most relevant parameters of the interacting  scenario \textbf{``$\xi_0 \xi_a w_q$IDE$Q_B$''} for several datasets, namely, CMB, CMB+BAO, CMB+Pantheon and CMB+BAO+Pantheon.  }
    \label{fig:IDEq-ModelB-xi-dyn}
\end{figure*}
\begingroup                                                                                                                     
\begin{center}                                                                                                                  
\begin{table*}                                                                                                                   
\begin{tabular}{ccccccc}                                                                                                            
\hline\hline                                                            
Parameters & CMB & CMB+BAO & CMB+Pantheon & CMB+BAO+Pantheon \\ \hline
$\Omega_c h^2$ & $    0.1109_{-    0.0044-    0.011}^{+    0.0065+    0.010}$ & $  0.1098_{-    0.0039-    0.0089}^{+    0.0053+    0.0084}$  & $    0.1095_{-    0.0043-    0.0097}^{+    0.0057+   0.0093}$ & $    0.1096_{-    0.0037-    0.0085}^{+    0.0050+    0.0080}$ \\

$\Omega_b h^2$ & $    0.02234_{-    0.00015-    0.00029}^{+    0.00015+    0.00030}$ & $    0.02239_{-    0.00014-    0.00028}^{+    0.00014+    0.00029}$ & $    0.02237_{-    0.00015-    0.00029}^{+    0.00015+    0.00030}$ & $    0.02238_{-    0.00014-    0.00028}^{+    0.00015+    0.00028}$  \\

$100\theta_{MC}$ & $    1.04140_{-    0.00043-    0.00086}^{+    0.00042+    0.00087}$ &  $    1.04152_{-    0.00040-    0.00070}^{+    0.00036+    0.00075}$  & $    1.04152_{-    0.00042-    0.00075}^{+    0.00038+    0.00079}$  & $    1.04151_{-    0.00040-    0.00070}^{+    0.00036+    0.00076}$ \\

$\tau$ & $    0.0537_{-    0.0078-    0.016}^{+    0.0078+    0.016}$ & $    0.0546_{-    0.0083-    0.015}^{+    0.0074+    0.016}$  & $    0.0541_{-    0.0081-    0.015}^{+    0.0076+    0.016}$  & $    0.0545_{-    0.0076-    0.014}^{+    0.0077+    0.016}$ \\

$n_s$ & $    0.9648_{-    0.0044-    0.0091}^{+    0.0044+    0.0087}$ & $    0.9664_{-    0.0040-    0.0075}^{+    0.0039+    0.0080}$  & $    0.9654_{-    0.0043-    0.0090}^{+    0.0044+    0.0087}$ & $    0.9662_{-    0.0040-    0.0081}^{+    0.0040+    0.0080}$ \\

${\rm{ln}}(10^{10} A_s)$ & $    3.043_{-    0.016-    0.032}^{+    0.016+    0.033}$ & $    3.044_{-    0.017-    0.031}^{+    0.015+    0.033}$   & $    3.044_{-    0.016-    0.032}^{+    0.016+    0.032}$  & $    3.044_{-    0.017-    0.030}^{+    0.015+    0.032}$ \\

$w_q$ & $  < -0.900 \; < -0.762  $ & $  < -0.938 \; < -0.879 $ & $  < -0.958 < -0.916 $ & $  < -0.956 < -0.912 $ \\

$\xi_0$ & $  < 0.082 \; < 0.173  $ & $ < 0.088 \; < 0.183 $  & $  < 0.096 < 0.194  $ & $ < 0.098 < 0.198 $ \\

$\xi_a$ & $  < 0.100 \; < 0.234 $ & $ < 0.108 < 0.213  $ & 
$ < 0.116 < 0.238  $  & $ < 0.105 < 0.205 $ \\

$\Omega_{m0}$ & $    0.305_{-    0.036-    0.066}^{+    0.028+    0.070}$ & $    0.290_{-    0.014-    0.029}^{+    0.014+    0.028}$  & $    0.285_{-    0.017-    0.035}^{+    0.019+    0.034}$ & $    0.286_{-    0.013-    0.026}^{+    0.013+    0.025}$ \\

$\sigma_8$ & $    0.828_{-    0.034-    0.073}^{+    0.034+    0.069}$ & $    0.838_{-    0.025-    0.044}^{+    0.021+    0.047}$  & $    0.848_{-    0.026-    0.043}^{+    0.020+    0.047}$   & $    0.844_{-    0.024-    0.042}^{+    0.020+    0.044}$ \\

$H_0$ [km/s/Mpc] & $   66.4_{-    2.0-    5.9}^{+    3.1+    5.0}$ & $   67.7_{-    0.9-    2.1}^{+    1.1+    2.0}$ & $   68.2_{-    1.2-    2.1}^{+    1.0+    2.2}$ & $   68.10_{-    0.72-    1.4}^{+    0.72+    1.4}$ \\

$S_8$ & $    0.833_{-    0.019-    0.039}^{+    0.019+    0.038}$ & $    0.823_{-    0.014-    0.026}^{+    0.014+    0.028}$  & $    0.826_{-    0.016-    0.033}^{+    0.017+    0.033}$ & $    0.824_{-    0.013-    0.026}^{+    0.013+    0.026}$ \\

$r_{\rm{drag}}$ [Mpc] & $  147.05_{-    0.30-    0.62}^{+    0.30+    0.59}$ & $  147.17_{-    0.27-    0.54}^{+    0.27+    0.51}$   & $  147.09_{-    0.30-    0.59}^{+    0.31+    0.58}$  & $  147.15_{-    0.26-    0.53}^{+    0.27+    0.51}$ \\
\hline\hline                                                                                                                    
\end{tabular}                                                                                                                   
\caption{Observational constraints on the interacting  scenario \textbf{``$\xi_0 \xi_a w_q$IDE$Q_B$''} obtained from  several observational datasets, namely, CMB, CMB+BAO, CMB+Pantheon and CMB+BAO+Pantheon are presented.   }                        \label{tab:IDEq-ModelB-xi-dyn}                                                          
\end{table*}                                                                                                                     
\end{center}                                                                                                                    
\endgroup                                                                             
\begingroup                                                                                                                     
\begin{center}                                      
\begin{table*}               
\begin{tabular}{ccccccc}                                                                                                            
\hline\hline                                                                                                                    
Parameters & CMB & CMB+BAO & CMB+Pantheon & CMB+BAO+Pantheon  \\ \hline

$\Omega_c h^2$ & $    0.149_{-    0.007-    0.019}^{+    0.013+    0.017}$ & $    0.1431_{-    0.0044-    0.014}^{+    0.0084+    0.011}$  & $    0.1461_{-    0.0034-    0.012}^{+    0.0070+    0.009}$   & $    0.1429_{-    0.0038-    0.013}^{+    0.0084+    0.010}$ \\

$\Omega_b h^2$ & $    0.02247_{-    0.00016-    0.00030}^{+    0.00015+    0.00030}$ & $    0.02247_{-    0.00015-    0.00029}^{+    0.00015+    0.00029}$  & $    0.02244_{-    0.00016-    0.00030}^{+    0.00015+    0.00030}$  & $    0.02245_{-    0.00015-    0.00030}^{+    0.00015+    0.00031}$  \\

$100\theta_{MC}$ & $    1.03944_{-    0.00065-    0.0010}^{+    0.00048+    0.0011}$ & $    1.03977_{-    0.00049-    0.00078}^{+    0.00039+    0.00090}$  & $    1.03958_{-    0.00044-    0.00073}^{+    0.00035+    0.00086}$  & $    1.03975_{-    0.00047-    0.00079}^{+    0.00041+    0.00088}$ \\

$\tau$ & $    0.0526_{-    0.0074-    0.016}^{+    0.0075+    0.016}$ & $    0.0541_{-    0.0082-    0.015}^{+    0.0074+    0.016}$   & $    0.0532_{-    0.0076-    0.015}^{+    0.0076+    0.016}$  & $    0.0540_{-    0.0079-    0.016}^{+    0.0078+    0.016}$  \\

$n_s$ & $    0.9677_{-    0.0044-    0.0087}^{+    0.0044+    0.0089}$ & $    0.9675_{-    0.0044-    0.0086}^{+    0.0044+    0.0089}$  & $    0.9668_{-    0.0044-    0.0086}^{+    0.0043+    0.0088}$  & $    0.9675_{-    0.0044-    0.0088}^{+    0.0045+    0.0089}$  \\

${\rm{ln}}(10^{10} A_s)$ & $    3.039_{-    0.015-    0.034}^{+    0.016+    0.033}$ & $    3.042_{-    0.016-    0.030}^{+    0.016+    0.032}$  & $    3.041_{-    0.016-    0.032}^{+    0.016+    0.033}$  & $    3.041_{-    0.016-    0.032}^{+    0.015+    0.033}$  \\

$w_0^p$ & $ -2.00_{-    0.62}^{+    0.62}$ \; unconstr. & $ > -1.20 \; > -1.46 $  & $ > -1.24 > -1.45 $  & $ > -1.11 > -1.20 $  \\

$w_e^p$ & $ > - 2.07$ \; unconstr.   &  $ -1.34_{-    0.14-    0.32}^{+    0.21+    0.31}$  & $ -1.27_{-    0.12}^{+    0.22}\;  > -1.56 $  & $   -1.32_{-    0.13-    0.30}^{+    0.19+    0.28}$ \\

$\xi_0$ & $ > -0.033 \;  > -0.072 $ & $ > -0.040 \; > -0.075 $  & $ > -0.033 > -0.067 $ & $ > -0.042 > -0.078 $ \\

$\xi_a$ & $ -0.073_{-    0.031}^{+    0.059}  \; > -0.154 $ &  $ -0.064_{-    0.029}^{+    0.052} \; > -0.131  $  & $   -0.084_{-    0.040-    0.067}^{+    0.040+    0.071}$ & $   -0.065_{-    0.028}^{+    0.052}\;  > -0.136$ \\

$\Omega_{m0}$ & $    0.256_{-    0.080-    0.10}^{+    0.043+    0.12}$ & $    0.345_{-    0.019-    0.042}^{+    0.024+    0.039}$ & $    0.375_{-    0.023-    0.044}^{+    0.023+    0.044}$ & $    0.356_{-    0.014-    0.034}^{+    0.020+    0.030}$  \\

$\sigma_8$ & $    0.863_{-    0.078-    0.12}^{+    0.065+    0.13}$ & $    0.760_{-    0.032-    0.050}^{+    0.025+    0.056}$  & $    0.747_{-    0.024-    0.039}^{+    0.017+    0.043}$  & $    0.749_{-    0.027-    0.047}^{+    0.022+    0.049}$ \\

$H_0$ [km/s/Mpc] & $   84_{-    9.4-   16}^{+    10+   17}$ & $   69.4_{-    1.9-    3.1}^{+    1.2+    3.5}$  & $   67.2_{-    1.4-    2.7}^{+    1.4+    2.7}$  & $   68.26_{-    0.81-    1.6}^{+    0.82+    1.7}$ \\

$S_8$ & $    0.786_{-    0.040-    0.078}^{+    0.040+    0.080}$ & $    0.815_{-    0.016-    0.034}^{+    0.018+    0.031}$  & $    0.834_{-    0.020-    0.044}^{+    0.023+    0.040}$  & $    0.816_{-    0.016-    0.036}^{+    0.020+    0.034}$ \\

$r_{\rm{drag}}$ [Mpc] & $  147.18_{-    0.30-    0.58}^{+    0.30+    0.60}$ & $  147.20_{-    0.30-    0.56}^{+    0.30+    0.58}$   & $  147.13_{-    0.30-    0.60}^{+    0.30+    0.58}$  & $  147.23_{-    0.29-    0.56}^{+    0.29+    0.56}$ \\
\hline\hline                                                                                                                    
\end{tabular}
\caption{Observational constraints of the interacting  scenario \textbf{``$\xi_0 \xi_a w_0^{p} w_e^{p}$IDE$Q_A$''} obtained from  several observational datasets, namely, CMB, CMB+BAO, CMB+Pantheon and CMB+BAO+Pantheon are presented.   }                                  \label{tab:IDEp-ModelA-xi-dyn-w-dyn}       
\end{table*}                                                                
\end{center}                                                          
\endgroup
\begin{figure*}
\centering
\includegraphics[width=0.9\textwidth]{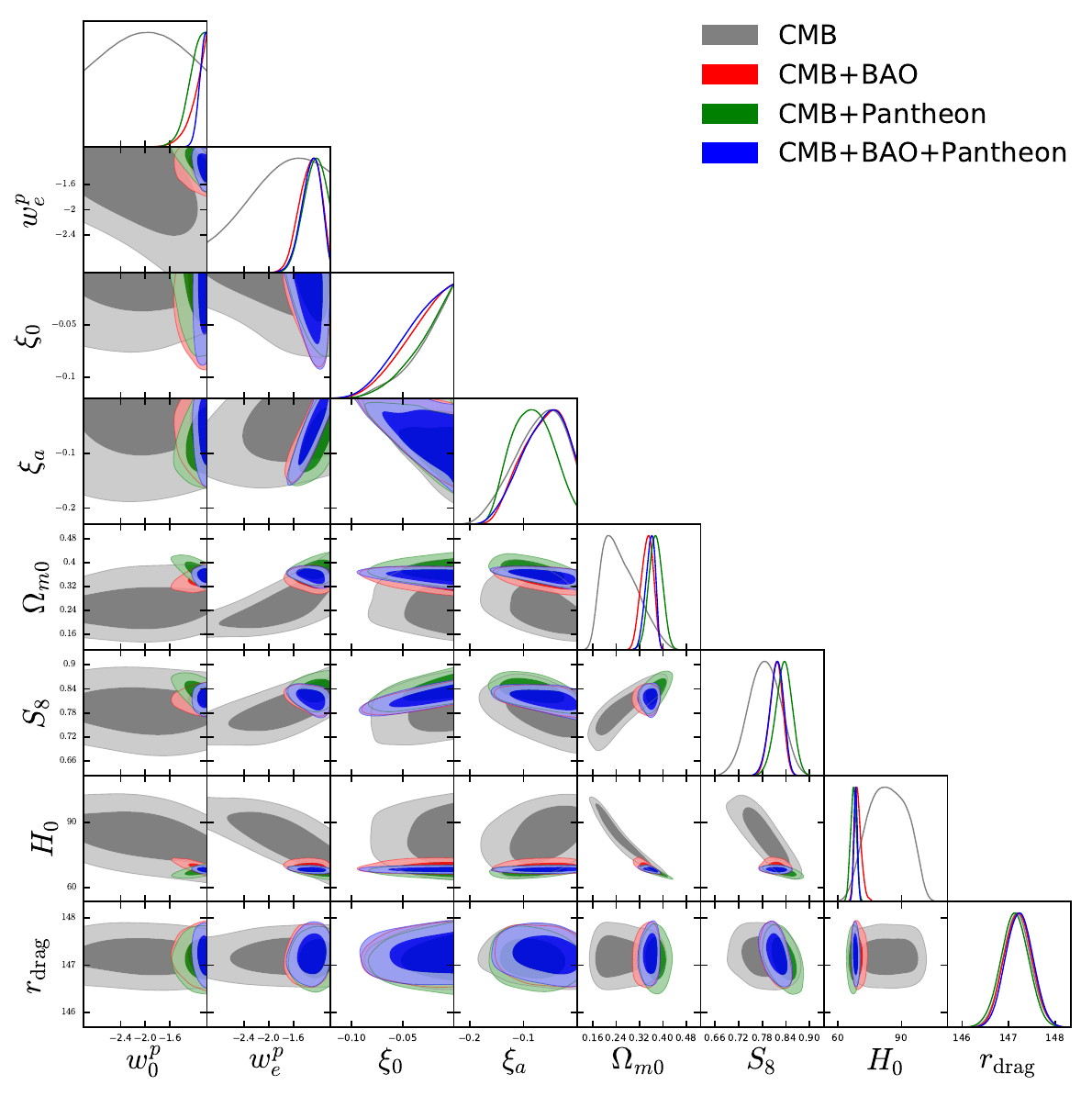}
    \caption{One-dimensional marginalized posterior distributions and two-dimensional joint contours for the most relevant parameters of the interacting  scenario \textbf{``$\xi_0 \xi_a w_0^{p} w_e^{p}$IDE$Q_A$''} for several datasets, namely, CMB, CMB+BAO, CMB+Pantheon and CMB+BAO+Pantheon. }
    \label{fig:IDEp-ModelA-xi-dyn-w-dyn}
\end{figure*}
\begin{figure*}
\centering
\includegraphics[width=0.9\textwidth]{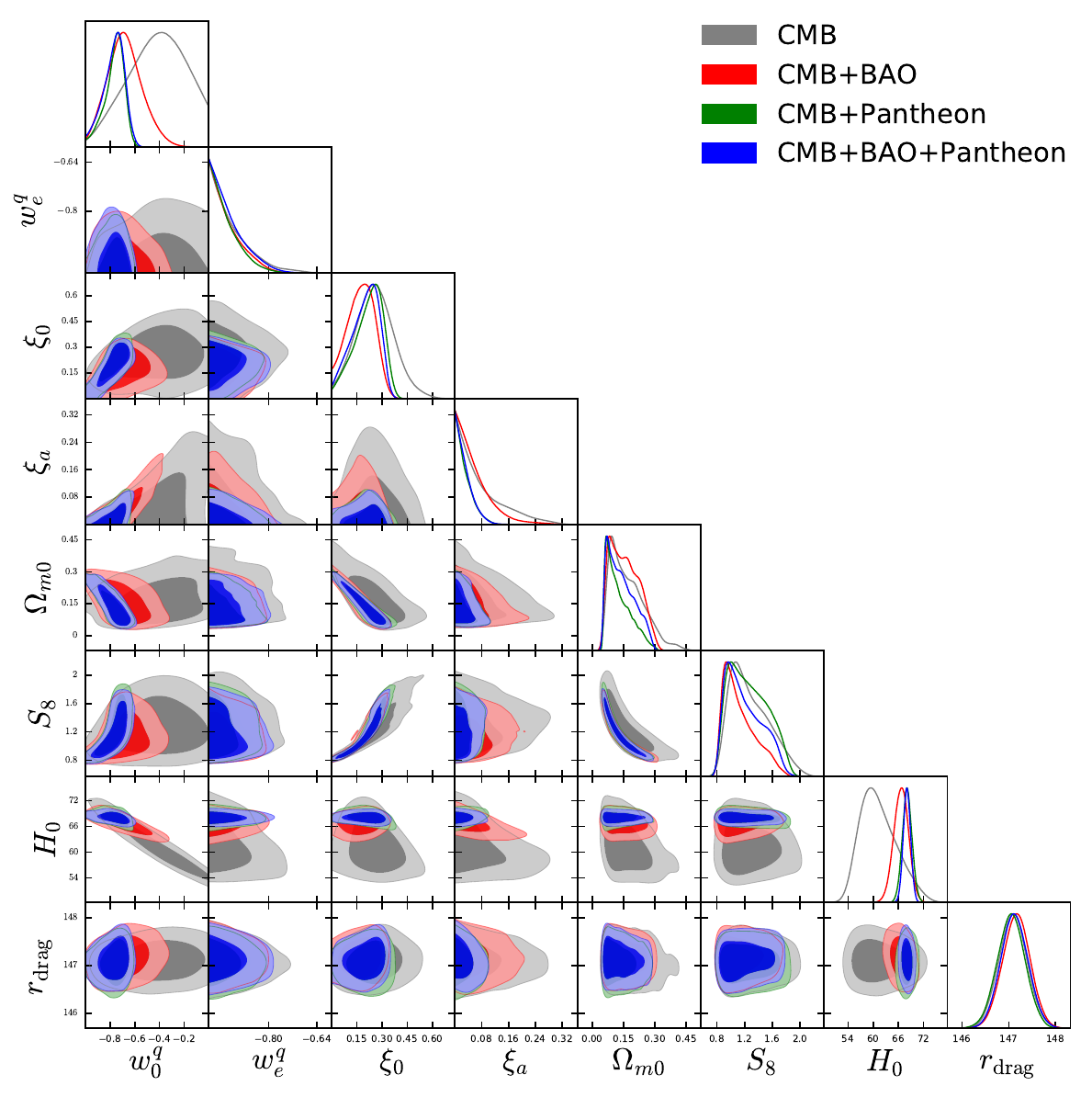}
    \caption{One-dimensional marginalized posterior distributions and two-dimensional joint contours for the most relevant parameters of the interacting  scenario \textbf{``$\xi_0 \xi_a w_0^{q} w_e^{q}$IDE$Q_A$''} for several datasets, namely, CMB, CMB+BAO, CMB+Pantheon and CMB+BAO+Pantheon.  }
    \label{fig:IDEq-ModelA-xi-dyn-w-dyn}
\end{figure*}
\begingroup                                                                                                                     
\begin{center}                                                                                                                  
\begin{table*}                                                                                                                   
\begin{tabular}{cccccccccc}                                                                                                            
\hline\hline                                                                                                                    
Parameters & CMB & CMB+BAO & CMB+Pantheon & CMB+BAO+Pantheon \\ \hline
$\Omega_c h^2$ & $  <0.050\;<0.091 $ & $  <0.060\;<0.095  $ & $ <0.046\;<0.090 $  & $  <0.054\;<0.095$ \\

$\Omega_b h^2$ & $    0.02239_{-    0.00018-    0.00032}^{+    0.00016+    0.00032}$ & $    0.02238_{-    0.00015-    0.00030}^{+    0.00015+    0.00030}$ & $    0.02236_{-    0.00015-    0.00030}^{+    0.00015+    0.00029}$  & $    0.02238_{-    0.00015-    0.00031}^{+    0.00015+    0.00029}$  \\

$100\theta_{MC}$ & $    1.0465_{-    0.0020-    0.0044}^{+    0.0030+    0.0040}$ & $    1.0460_{-    0.0027-    0.0042}^{+    0.0027+    0.0043}$ & $    1.0467_{-    0.0015-    0.0045}^{+    0.0031+    0.0037}$  & $    1.0463_{-    0.0019-    0.0044}^{+    0.0033+    0.0040}$ \\

$\tau$ & $    0.0523_{-    0.0075-    0.016}^{+    0.0076+    0.016}$ & $    0.0549_{-    0.0072-    0.015}^{+    0.0074+    0.015}$ & $    0.0538_{-    0.0077-    0.015}^{+    0.0076+    0.016}$  & $    0.0542_{-    0.0076-    0.014}^{+    0.0075+    0.016}$ \\

$n_s$ & $    0.9662_{-    0.0047-    0.0096}^{+    0.0051+    0.0093}$ & $    0.9660_{-    0.0045-    0.0084}^{+    0.0044+    0.0091}$ & $    0.9651_{-    0.0045-    0.0090}^{+    0.0044+    0.0086}$ & $    0.9660_{-    0.0046-    0.0090}^{+    0.0045+    0.0089}$  \\

${\rm{ln}}(10^{10} A_s)$ & $    3.040_{-    0.015-    0.033}^{+    0.015+    0.031}$ & $    3.045_{-    0.015-    0.030}^{+    0.015+    0.031}$  & $    3.044_{-    0.016-    0.032}^{+    0.016+    0.033}$  & $    3.043_{-    0.016-    0.029}^{+    0.015+    0.032}$ \\ 

$w_0^p$ & $ -0.42_{-    0.20}^{+    0.28}\; > -0.82  $ & $   -0.68_{-    0.16-    0.27}^{+    0.13+    0.28}$  & $   -0.766_{-    0.055-    0.17}^{+    0.097+    0.14}$  & $   -0.770_{-    0.07-    0.19}^{+    0.10+    0.15}$ \\

$w_e^q$ & $ < -0.91\; < -0.80 $ & $  < -0.92 \; < -0.834 $ & $  < -0.93 < -0.85 $  & $  < -0.92 < -0.82  $ \\

$\xi_0$ & $    0.26_{-    0.11-    0.23}^{+    0.11+    0.21}$ & $ 0.17_{-    0.07}^{+    0.10} \; < 0.30  $  & $    0.22_{-    0.06-    0.17}^{+    0.11+    0.14}$ & $    0.20_{-    0.06-    0.16}^{+    0.10+    0.14}$ \\

$\xi_a$ & $  < 0.090 \; < 0.219 $ & $  < 0.068 \; < 0.160 $ & $  < 0.036 < 0.079 $ & $ < 0.036 < 0.080  $ \\

$\Omega_{m0}$ & $    0.17_{-    0.11-    0.13}^{+    0.05+    0.17}$ & $    0.155_{-    0.095-    0.11}^{+    0.049+    0.13}$  & $    0.127_{-    0.078-    0.09}^{+    0.031+    0.12}$ & $    0.139_{-    0.088-    0.10}^{+    0.040+    0.13}$ \\

$\sigma_8$ & $    1.9_{-    1.2-    1.4}^{+    0.5+    1.9}$ & $    1.8_{-    1.0-    1.2}^{+    0.4+    1.8}$  & $    2.2_{-    1.3-    1.5}^{+    0.6+    1.9}$ & $    2.0_{-    1.2-    1.3}^{+    0.5+    1.9}$ \\

$H_0$ [km/s/Mpc] & $   61.3_{-    5.6-    7.8}^{+    3.3+    9.1}$ & $   66.6_{-    1.7-    3.5}^{+    1.7+    3.4}$  & $   68.1_{-    1.1-    2.3}^{+    1.2+    2.3}$ & $   68.03_{-    0.86-    1.7}^{+    0.87+    1.7}$ \\

$S_8$ & $    1.27_{-    0.38-    0.47}^{+    0.19+    0.58}$ & $    1.13_{-    0.31-    0.36}^{+    0.13+    0.49}$  & $    1.26_{-    0.38-    0.45}^{+    0.21+    0.52}$ & $    1.19_{-    0.36-    0.41}^{+    0.17+    0.50}$  \\

$r_{\rm{drag}}$ [Mpc] & $  147.09_{-    0.30-    0.59}^{+    0.30+    0.60}$ & $  147.16_{-    0.30-    0.58}^{+    0.29+    0.58}$  & $  147.05_{-    0.30-    0.59}^{+    0.30+    0.59}$  & $  147.12_{-    0.29-    0.57}^{+    0.29+    0.58}$ \\
\hline\hline                                                                                                                    
\end{tabular}                                                                                                                   
\caption{Observational constraints on the interacting  scenario \textbf{``$\xi_0 \xi_a w_0^{q} w_e^{q}$IDE$Q_A$''} obtained from  several observational datasets, namely, CMB, CMB+BAO, CMB+Pantheon and CMB+BAO+Pantheon are presented.   }                        \label{tab:IDEq-ModelA-xi-dyn-w-dyn}                                            
\end{table*}                                                                                                                     
\end{center}                                                                                                                    
\endgroup 
\begingroup                                                                                                                     
\begin{center}                                                                                                                  
\begin{table*}                                                                                                                   
\begin{tabular}{cccccccc}                                                                                                            
\hline\hline                                                                                                                    
Parameters & CMB & CMB+BAO & CMB+Pantheon & CMB+BAO+Pantheon \\ \hline

$\Omega_c h^2$ & $    0.157_{-    0.015-    0.027}^{+    0.018+    0.026}$ & $    0.151_{-    0.013-    0.023}^{+    0.013+    0.023}$  & $    0.158_{-    0.009-    0.020}^{+    0.011+    0.019}$  & $    0.152_{-    0.013-    0.024}^{+    0.014+    0.023}$ \\

$\Omega_b h^2$ & $    0.02244_{-    0.00017-    0.00032}^{+    0.00016+    0.00033}$ & $    0.02244_{-    0.00016-    0.00029}^{+    0.00016+    0.00030}$ & $    0.02242_{-    0.00016-    0.00030}^{+    0.00015+    0.00032}$  & $    0.02245_{-    0.00016-    0.00030}^{+    0.00016+    0.00030}$   \\

$100\theta_{MC}$ & $    1.03906_{-    0.00094-    0.0013}^{+    0.00070+    0.0015}$ & $    1.03937_{-    0.00070-    0.0013}^{+    0.00067+    0.0013}$  & $    1.03897_{-    0.00061-    0.0011}^{+    0.00052+    0.0011}$ & $    1.03930_{-    0.00075-    0.0012}^{+    0.00064+    0.0013}$  \\

$\tau$ & $    0.0529_{-    0.0072-    0.014}^{+    0.0075+    0.014}$ & $    0.0548_{-    0.0086-    0.015}^{+    0.0071+    0.016}$ & $    0.0538_{-    0.0076-    0.016}^{+    0.0077+    0.016}$  & $    0.0558_{-    0.0076-    0.016}^{+    0.0076+    0.015}$ \\

$n_s$ & $    0.9664_{-    0.0050-    0.0092}^{+    0.0044+    0.0092}$ & $    0.9666_{-    0.0044-    0.0085}^{+    0.0044+    0.0088}$  & $    0.9657_{-    0.0046-    0.0090}^{+    0.0046+    0.0091}$  & $    0.9668_{-    0.0040-    0.0082}^{+    0.0042+    0.0079}$  \\

${\rm{ln}}(10^{10} A_s)$ & $    3.040_{-    0.014-    0.030}^{+    0.016+    0.028}$ & $    3.044_{-    0.017-    0.030}^{+    0.015+    0.032}$  & $    3.042_{-    0.016-    0.034}^{+    0.016+    0.033}$  & $    3.045_{-    0.015-    0.031}^{+    0.015+    0.032}$ \\

$w_0^p$ & $ -2.0_{-    0.54}^{+    0.63}$ \; unconstr. & $ > -1.19 \; > -1.44 $  & $ > -1.31   > -1.60  $  & $ > -1.10 > -2.0  $ \\

$w_e^p$ & $ -1.97_{-    0.54}^{+    0.64}$ \; unconstr.  & $   -1.48_{-    0.21-    0.45}^{+    0.30+    0.43}$ & $   -1.48_{-    0.24-    0.39}^{+    0.22+    0.40}$ & $   -1.47_{-    0.21-    0.39}^{+    0.28+    0.37}$ \\

$\xi_0$ & $  > -0.405 \; > -0.712  $ & $ > -0.264 \; > -0.507 $  & $ > -0.244 > -0.484  $ & $ > -0.272 > -0.521 $ \\

$\xi_a$ & $ -0.44_{-    0.19}^{+    0.38}$ \; unconstr. & $   -0.44_{-    0.18}^{+    0.38}$,\; unconstr. & $  -0.61_{-    0.38}^{+    0.13}\; < 0.19 $ & $ -0.47_{-    0.21}^{+    0.39}$, unconstr.   \\

$\Omega_{m0}$ & $    0.261_{-    0.087-    0.11}^{+    0.040+    0.14}$ & $    0.364_{-    0.035-    0.058}^{+    0.031+    0.062}$  & $    0.405_{-    0.032-    0.064}^{+    0.036+    0.065}$ & $    0.377_{-    0.030-    0.058}^{+    0.034+    0.056}$\\

$\sigma_8$ & $    0.845_{-    0.091-    0.15}^{+    0.078+    0.16}$ & $    0.741_{-    0.041-    0.080}^{+    0.045+    0.076}$  & $    0.715_{-    0.033-    0.066}^{+    0.033+    0.066}$ & $    0.728_{-    0.044-    0.075}^{+    0.038+    0.076}$ \\

$H_0$ [km/s/Mpc] & $   85_{-    7-   19}^{+   13+   16}$ & $   69.3_{-    1.9-    3.0}^{+    1.2+    3.4}$   & $   66.9_{-    1.5-    3.0}^{+    1.5+    3.0}$ & $   68.16_{-    0.81-    1.6}^{+    0.81+    1.5}$ \\

$S_8$ & $    0.774_{-    0.046-    0.068}^{+    0.028+    0.078}$ & $    0.813_{-    0.017-    0.036}^{+    0.019+    0.033}$  & $    0.829_{-    0.020-    0.040}^{+    0.021+    0.039}$ & $    0.814_{-    0.016-    0.036}^{+    0.019+    0.032}$  \\

$r_{\rm{drag}}$ [Mpc] & $  147.15_{-    0.30-    0.60}^{+    0.29+    0.58}$ & $  147.18_{-    0.29-    0.59}^{+    0.30+    0.57}$ & $  147.10_{-    0.30-    0.60}^{+    0.30+    0.60}$  & $  147.20_{-    0.29-    0.55}^{+    0.29+    0.56}$ \\

\hline\hline                                                
\end{tabular}                                                                                                       
\caption{Observational constraints on the interacting  scenario \textbf{``$\xi_0 \xi_a w_0^{p} w_e^{p}$IDE$Q_B$''} obtained from  several observational datasets, namely, CMB, CMB+BAO, CMB+Pantheon and CMB+BAO+Pantheon are presented.  }                         \label{tab:IDEp-ModelB-xi-dyn-w-dyn}                                                       
\end{table*}                                     
\end{center}                                                                                                                
\endgroup                                                                                                                       
\begingroup                                                                                                                     
\begin{center}                                      
\begin{table*}           
\begin{tabular}{ccccccccc}                              
\hline\hline                                                
Parameters & CMB & CMB+BAO & CMB+Pantheon & CMB+BAO+Pantheon  \\ \hline
$\Omega_c h^2$ & $    0.1121_{-    0.0036-    0.011}^{+    0.0065+    0.009}$ & $    0.1086_{-    0.0041-    0.0096}^{+    0.0055+    0.0090}$  & $    0.1085_{-    0.0046-    0.0101}^{+    0.0057+    0.0096}$  & $    0.1085_{-    0.0040-    0.0095}^{+    0.0054+    0.0087}$ \\

$\Omega_b h^2$ & $    0.02233_{-    0.00015-    0.00028}^{+    0.00015+    0.00030}$ & $    0.02239_{-    0.00015-    0.00030}^{+    0.00015+    0.00029}$ & $    0.02236_{-    0.00016-    0.00029}^{+    0.00015+    0.00030}$  & $    0.02238_{-    0.00014-    0.00028}^{+    0.00014+    0.00028}$ \\

$100\theta_{MC}$ & $    1.04134_{-    0.00048-    0.00086}^{+    0.00039+    0.00088}$ & $    1.04160_{-    0.00042-    0.00071}^{+    0.00036+    0.00078}$ & $    1.04158_{-    0.00042-    0.00080}^{+    0.00041+    0.00084}$ & $    1.04159_{-    0.00040-    0.00071}^{+    0.00035+    0.00077}$ \\

$\tau$ & $    0.0533_{-    0.0074-    0.015}^{+    0.0075+    0.015}$ & $    0.0552_{-    0.0082-    0.015}^{+    0.0073+    0.016}$ & $    0.0542_{-    0.0081-    0.015}^{+    0.0072+    0.016}$ & $    0.0547_{-    0.0078-    0.015}^{+    0.0078+    0.016}$  \\

$n_s$ & $    0.9648_{-    0.0047-    0.0083}^{+    0.0042+    0.0089}$ & $    0.9670_{-    0.0042-    0.0083}^{+    0.0042+    0.0082}$   & $    0.9657_{-    0.0044-    0.0088}^{+    0.0044+    0.0086}$  & $    0.9664_{-    0.0040-    0.0082}^{+    0.0039+    0.0080}$  \\

${\rm{ln}}(10^{10} A_s)$ & $    3.043_{-    0.015-    0.032}^{+    0.016+    0.031}$ & $    3.045_{-    0.015-    0.031}^{+    0.015+    0.032}$  & $    3.044_{-    0.016-    0.031}^{+    0.016+    0.032}$  & $    3.044_{-    0.017-    0.031}^{+    0.016+    0.033}$  \\

$w_0^p$ & $  < -0.706 \; < -0.333 $ & $ < -0.846 \; < -0.704  $  & $  < -0.944 < -0.884 $ & $ < -0.938 < -0.877 $ \\

$w_e^q$ & $  < -0.879\; < -0.731 $ & $ < -0.933 \; < -0.856  $  & $  < -0.929 < -0.847 $  & $ < -0.933 < -0.846 $ \\

$\xi_0$ & $  < 0.070 < 0.158 $ & $  < 0.082  < 0.168 $  & $   < 0.094 < 0.188 $ & $ < 0.089 < 0.183 $ \\

$\xi_a$ & $  < 0.096 < 0.228 $ & $   < 0.140 < 0.266 $  & $    < 0.135 < 0.259  $ & $  < 0.132 < 0.243  $ \\

$\Omega_{m0}$ & $    0.338_{-    0.054-    0.088}^{+    0.037+    0.102}$ & $    0.296_{-    0.018-    0.034}^{+    0.018+    0.036}$  & $    0.286_{-    0.017-    0.035}^{+    0.020+    0.034}$ & $    0.286_{-    0.013-    0.028}^{+    0.015+    0.027}$  \\

$\sigma_8$ & $    0.795_{-    0.041-    0.094}^{+    0.049+    0.084}$ & $    0.828_{-    0.025-    0.050}^{+    0.024+    0.053}$  & $    0.846_{-    0.026-    0.044}^{+    0.021+    0.046}$  & $    0.843_{-    0.025-    0.041}^{+    0.019+    0.043}$ \\

$H_0$ [km/s/Mpc] & $   63.6_{-    2.9-    7.5}^{+    4.2+    6.4}$ & $   66.8_{-    1.2-    2.9}^{+    1.6+    2.6}$  & $   67.9_{-    1.2-    1.9}^{+    1.0+    2.2}$  & $   67.87_{-    0.74-    1.5}^{+    0.72+    1.5}$ \\

$S_8$ & $    0.839_{-    0.020-    0.040}^{+    0.020+    0.039}$ & $    0.822_{-    0.013-    0.027}^{+    0.013+    0.027}$  & $    0.825_{-    0.017-    0.032}^{+    0.017+    0.032}$   & $    0.822_{-    0.014-    0.025}^{+    0.013+    0.026}$  \\

$r_{\rm{drag}}$ [Mpc] & $  147.04_{-    0.31-    0.59}^{+    0.31+    0.62}$ & $  147.21_{-    0.28-    0.55}^{+    0.28+    0.56}$  & $  147.11_{-    0.29-    0.58}^{+    0.29+    0.58}$ & $  147.16_{-    0.27-    0.54}^{+    0.28+    0.51}$  \\
\hline\hline                                                                                                                    
\end{tabular}                                                                                                                   
\caption{Observational constraints on the interacting  scenario \textbf{``$\xi_0 \xi_a w_0^{q} w_e^{q}$IDE$Q_B$''} obtained from  several observational datasets, namely, CMB, CMB+BAO, CMB+Pantheon and CMB+BAO+Pantheon are presented.  }                 \label{tab:IDEq-ModelB-xi-dyn-w-dyn}                                 
\end{table*}                                                                                                                     
\end{center}                                                                                                                    
\endgroup

\subsubsection{Constant dark energy equation of state}

The simplest interacting cosmological scenario in this framework is the one where the dark energy equation of state mimics the vacuum energy, corresponding to $w= -1$.  In Tabs.~\ref{tab:IVS-ModelA-xi-dyn} and~\ref{tab:IVS-ModelB-xi-dyn} we present the constraints for the two interacting scenarios driven by the functions $Q_A$ and $Q_B$, respectively.

Table~\ref{tab:IVS-ModelA-xi-dyn} summarizes the constraints of the interacting scenario $\xi_0\xi_a${\bf IVS}$Q_A$ using the different observational datasets, namely,  CMB, CMB+BAO, CMB+Pantheon and CMB+BAO+Pantheon. A common feature which remains valid for all the observational datasets is that we do not find any preference for ($\xi_0, \xi_a$) $\neq  \;(0, 0)$, i.e. we do not find any evidence for a non-zero coupling in the dark sector. Nevertheless there are large differences in the cosmological parameter constraints arising from CMB alone and from CMB plus other external datasets, such as BAO or Pantheon. For CMB alone, we see a very low value of the matter density parameter which is due to the energy flow from the dark matter sector to the dark energy one, which is always associated to a larger value of the Hubble constant. However, when BAO observations are added to CMB, the mean values of all the parameters shift to closer values to the $\Lambda$CDM framework. The matter density parameter increases significantly (compared to the CMB alone case) because in this case the one probability distribution of $\xi_0$ is shifted towards negative values of this parameter, changing the energy flow at present times (i.e. from the dark energy to the dark  matter one). That means the transfer of energy is reversed after the inclusion of BAO data to CMB, fact that is clear from the mean value of $\xi_0$: $\xi_0 = -0.05_{- 0.14}^{+    0.10}$ at 68\% CL for CMB+BAO, while $\xi_0 = 0.09_{- 0.15}^{+ 0.15}$ at 68\% CL for CMB alone.  When Pantheon SNIa survey measurements are added to CMB observations, no significant changes appear except a mild reduction of the matter density parameter, and a slight increase in $\sigma_8$ parameter compared to the CMB+BAO case. The final analysis with CMB+BAO+Pantheon leads  to very similar constraints to the CMB+BAO dataset. 

Table~\ref{tab:IVS-ModelB-xi-dyn} summarizes the constraints of the interacting scenario $\xi_0\xi_a${\bf IVS$Q_B$} using the observational datasets, namely, CMB, CMB+BAO, CMB+Pantheon and CMB+BAO+Pantheon. For CMB alone, we do not find any evidence of $(\xi_0, \xi_a) \neq (0, 0)$ and therefore this scenario is very close to the non-interacting one. When the BAO observations are added to CMB, the mean values of some of the parameters slightly change, and the most interesting change appears in $\xi_a$ which prefers a non-zero value at 68\% CL ($\xi_a = 0.37_{-    0.33}^{+    0.41}$ at 68\% CL, CMB+BAO) and hence a mild evidence of a dynamical $\xi (a)$ is suggested for CMB+BAO. The inclusion of Pantheon to CMB slightly changes the mean values of the parameters, however, we do not find any indication for a non-zero dynamical coupling in the dark sector. 
Finally, for the combined dataset CMB+BAO+Pantheon we find, again, a mild evidence of a non-zero $\xi_a$ ($= 0.38_{-  0.34}^{+    0.37}$) at 68\% CL, that means, a dynamical coupling in the dark sector is mildly preferred also for this dataset. This feature is mainly driven by the BAO data.

The consideration of $w$ other than $-1$ but constant generalizes the previous interacting scenarios (for which $w =  - 1$). Here we explore the constraints on the interacting scenarios with a dynamical coupling parameter when the dark energy equation of state is constant but it is other than $-1$. As already mentioned,  the nature of $w$ may affect the stability of the interacting scenario. Therefore, following the doom factor analysis of Ref.~\cite{Gavela:2009cy}, one can select accordingly the parameter space where the interacting scenario will be free of instabilities.  

In Tabs.~\ref{tab:IDEp-ModelA-xi-dyn} and~\ref{tab:IDEq-ModelA-xi-dyn} we have summarized the constraints on the interacting scenarios driven by the interaction function $Q_A$ of Eq.~(\ref{modelA}) assuming two different regions of the parameter space. 

Table~\ref{tab:IDEp-ModelA-xi-dyn} presents the observational constraints of the model $\xi_0 \xi_a w_{p}${\bf IDE}$Q_A$ characterized by the parameter space  $\xi_0 <0$, $\xi_a <0$, $w_p < -1$ for CMB, CMB+BAO, CMB+Pantheon and CMB+BAO+Pantheon.  For CMB alone we do not find any preference for a non-zero and dynamical coupling parameters, however, we observe a preference for the phantom dark energy at 68\% CL ($w_p =  -1.83_{-    0.38}^{+    0.54}$ at 68\% CL) and a higher value of the Hubble constant, due to the strong correlation between these two parameters. Additionally, we find a lower (higher) value of  $\Omega_{m0}$ ($\sigma_8$) compared to its estimation within the canonical $\Lambda$CDM picture, due to the fact that the energy flows from the dark matter sector to the dark energy one. When BAO observations are added to CMB data we have no evidence for a non-zero dynamical coupling in the dark sector but in this case the preference for  $w_p< -1$ is much more significant, as $w_p$ remains in the phantom region at more than 95\% CL ($w_p = -1.21_{- 0.22}^{+ 0.20}$ at 95\% CL).  The inclusion of Pantheon data to CMB  observations leads to a non-zero value of $\xi_a$ ($ =  -0.071_{-    0.040}^{+    0.049}$ at 68\% CL, CMB+Pantheon) although the present day value of $\xi (a)$, i.e. $\xi_0$, does not show any preference for a non-vanishing value. The dark energy equation of state remains in the phantom regime at more than 95\% CL ($w_p = -1.20_{-    0.16}^{+0.16}$ at 95\% CL). We also find in this case that $\Omega_{m0}$ and $\sigma_8$ take larger and smaller values compared to their values in the canonical $\Lambda$CDM picture, respectively. Similar results are obtained for the combined dataset CMB+BAO+Pantheon except in the constraints on $H_0$ and $S_8$: there is no indication for a non-zero dynamical coupling in the dark sector, and the dark energy equation of state remains again in the phantom regime at more than 95\% CL ($w_p =  -1.13_{- 0.11}^{+0.11}$ at 95\% CL). As observed in the earlier cases with CMB plus either BAO or Pantheon data, in this case, $\Omega_{m0}$ ($\sigma_8$) adopts a higher (lower) value compared to the $\Lambda$CDM picture (however, it should be noted that $\Omega_{m0}$ takes relatively a lower value compared to the CMB+Pantheon case), due to the energy flow. Lastly, we notice that $H_0$ is increased slightly, alleviating the Hubble tension within $3\sigma$, and additionally we obtain a smaller value of $S_8$ compared to the Planck's estimation (within the canonical $\Lambda$CDM picture~\cite{Planck:2018vyg}), indicating a better consistency with the weak lensing measurements~\cite{Heymans:2020gsg,KiDS:2020ghu,DES:2021vln,DES:2022ygi}.

Table~\ref{tab:IDEq-ModelA-xi-dyn} presents the observational constraints of the model $\xi_0\xi_a w_{q}${\bf IDE}$Q_A$, characterized by the parameter space $\xi_0 >0$, $\xi_a >0$, $w_q >-1$ using the observational datasets CMB, CMB+BAO, CMB+Pantheon and CMB+BAO+Pantheon. Our observations for CMB alone are as follows. We find no evidence of dynamical coupling in the dark sector, but an indication for a non-zero coupling today at $1\sigma$ ($\xi_0=0.136^{+0.085}_{-0.088}$) at 68\% CL for CMB alone. The matter density parameter takes a very low value due to the exchange of energy from the dark matter to dark energy and also due to the fact that the dark energy equation of state is required to lie within the quintessence region. Despite the very strong anti-correlation between $w_q$ and $H_0$, a higher value of $H_0$ compared to the $\Lambda$CDM picture is obtained to increase the $\Omega_m h^2$ parameter, governing the angular location of the CMB acoustic peaks.  When BAO observations are added to CMB data, we find no preference for either a non-zero or a dynamical coupling in the dark sector. Even if the value of the matter density parameter increases, $\Omega_{m0} =  0.220_{- 0.036}^{+    0.077}$ at 68\% CL, the value is low if compared to its estimated figure within the $\Lambda$CDM picture.  The inclusion of the Pantheon catalogue to CMB (i.e. CMB+Pantheon) does not lead to any significant changes in the constraints (the same is true for CMB+BAO+Pantheon dataset).

Tables~\ref{tab:IDEp-ModelB-xi-dyn} and~\ref{tab:IDEq-ModelB-xi-dyn} together with Figs.~\ref{fig:IDEp-ModelB-xi-dyn} and~\ref{fig:IDEq-ModelB-xi-dyn} summarize the observational constraints on the interacting scenarios driven by the interaction function $Q_B$, Eq.~(\ref{modelB}), for two different regions of the parameter space. 

Firstly, Tab.~\ref{tab:IDEp-ModelB-xi-dyn} presents the constraints on the model $\xi_0\xi_a w_{p}${\bf IDE}$Q_B$, characterized by the parameter space $\xi_0 <0$, $\xi_a <0$, $w_p < -1$ for CMB, CMB+BAO, CMB+Pantheon and CMB+BAO+Pantheon. For CMB alone, we find no indication for a non-zero $\xi_0$ and a non-zero $\xi_a$, but the dark energy equation of state deeply lies in the  phantom region,  $w_p = -2.10^{+0.86}_{-0.84}$ at more than 95\% CL. Due to the phantom nature of  $w_p$, this scenario leads to a high value of the Hubble constant, see Fig.~\ref{fig:IDEp-ModelB-xi-dyn}. The addition of BAO observations to CMB data shifts the parameter values. First of all, we get an evidence for a non-zero $\xi_a$ at more than 68\% CL ($\xi_a = -0.21_{-    0.08}^{+    0.18}$ at 68\% CL) while within 95\% CL, we do not have any evidence for $\xi_a \neq 0$. Even though the dark energy equation of state remains in the phantom regime at more than 95\% CL ($w_p = -1.26_{- 0.27}^{+    0.24}$ at 95\% CL, CMB+BAO), the indication for a phantom nature is slightly decreased (compared to what it is observed in the CMB data alone case).  Additionally, we find a very high (low) value of $\Omega_{m0}$ ($\sigma_8$), and the requirement of a lower value of $H_0$, see the one-dimensional probability distribution for $\Omega_{m0}$ in Fig.~\ref{fig:IDEp-ModelB-xi-dyn}. The inclusion of Pantheon measurements to CMB also offers almost similar results to the CMB+BAO case: we find $\xi_a \neq 0$ at 68\% CL but it is found to be unconstrained within our prior at 95\%; $w_p$  remains in the phantom region at more than 95\% CL but also the preference for a phantom nature has decreased compared to the CMB alone case.  The final combination CMB+BAO+Pantheon does not exhibit any indication for $\xi_0 \neq 0 $ and $\xi_a \neq 0$, but the dark energy equation of state behaves such as  $w_p < -1$ (at more than 95\% CL). The other parameter values are very close to the CMB+BAO constraints.

Secondly, Tab.~\ref{tab:IDEq-ModelB-xi-dyn} shows the constraints on the model $\xi_0\xi_a w_{q}${\bf IDE}$Q_B$ characterized by the parameter space $\xi_0 >0$, $\xi_a >0$, $w_q > -1$ for CMB, CMB+BAO, CMB+Pantheon and CMB+BAO+Pantheon. For all the four datasets, we find no indication for a non-zero and dynamical coupling in the dark sector. And the remaining parameters take values very similar to those in the $\Lambda$CDM cosmology: in practise, this scenario  is perfectly consistent with the non-interacting standard cosmological scenario, see also the one-dimensional probability distributions in Fig.~\ref{fig:IDEq-ModelB-xi-dyn}.

\subsubsection{Dynamical dark energy equation of state}        

In the following, we shall present the constraints on the most general scenario so far discussed in this work. Despite the dimension of the parameter space, we restrict ourselves to the regions allowed by the doom factor analyses provided in Ref.~\cite{Gavela:2009cy}. We have therefore explored the interacting scenarios driven by Eqs.~(\ref{modelA}) and~(\ref{modelB}) dividing the entire parameter space into two disjoint regions. 

Namely, Tab.~\ref{tab:IDEp-ModelA-xi-dyn-w-dyn}, together with Fig.~\ref{fig:IDEp-ModelA-xi-dyn-w-dyn}, summarize the observational constraints on the interacting model $\xi_0\xi_a w_0^{p} w_e^{p}${\bf IDE}$Q_A$, characterized by the parameter space $w_0^p < -1,\; w_e^p < -1,\; \xi_0 < 0, \; \xi_a < 0$ for the CMB, CMB+BAO, CMB+Pantheon and CMB+BAO+Pantheon datasets. For CMB alone, we find a preference for a non-zero $\xi_a$ at 68\% CL ($\xi_a = -0.073_{-    0.031}^{+    0.059}$) which is diluted at 95\% CL. The dark energy remains in the phantom region at 68\% CL ($w_0^p = -2.00_{-    0.62}^{+    0.62}$), and it is unconstrained at 95\% CL. The Hubble constant assumes a higher value which is, as usual, driven by the phantom dark energy. As a consequence, a relatively small value of the matter density parameter compared to what we see in the $\Lambda$CDM picture is required (see Fig.~\ref{fig:IDEp-ModelA-xi-dyn-w-dyn}). When the BAOs are added to CMB, we find no evidence for $w_0^p < -1$ but an indication for an early dark energy component different from a cosmological constant at early times is suggested at more than 95\% CL ($w_e^p  =  -1.34_{-0.32}^{+ 0.31}$ at 95\% CL, CMB+BAO). Interestingly, we notice that $\xi_a$ is non-zero at 68\% CL ($\xi_a  = -0.064_{-    0.029}^{+    0.052}$ at 68\% CL) but this indication is diluted at a higher significance. Notice that the value of $\Omega_{m0} $ ($\sigma_8$) is still mildly higher (lower) than in the $\Lambda$CDM model. When the Pantheon catalogue is added to CMB, no evidence for $w_0^p < -1$ is found, as already observed in CMB+BAO case, and the evidence for an early dark energy fluid and not for the simplest vacuum energy scenario is reduced down to 68\% CL ($w_e^p = -1.27_{-    0.12}^{+    0.22}$ at 68\% CL for CMB+Pantheon). Interestingly, we find a strong evidence for a dynamical $\xi_a$ at more than 95\% CL ($=   -0.084_{-    0.067}^{+ 0.071}$ at 95\% CL, CMB+Pantheon) and a very large (small) value of $\Omega_{m0}$ ($\sigma_8$) (while the remaining parameters take almost similar results than the $\Lambda$CDM values). For the final combination, i.e. CMB+BAO+Pantheon, our results are similar to the CMB+BAO analysis, that means, no preference for $w_0^p < -1$ and $\xi_a<0$ at 68\% CL are obtained, but an indication for an early dark energy component different from the cosmological constant picture is suggested with a significance above 95\% CL, see the one-dimensional distribution for $w_e^p$ in Fig.~\ref{fig:IDEp-ModelA-xi-dyn-w-dyn}.

Table~\ref{tab:IDEq-ModelA-xi-dyn-w-dyn} and  Fig.~\ref{fig:IDEq-ModelA-xi-dyn-w-dyn} show the constraints on the interacting scenario $\xi_0\xi_a w_0^{q} w_e^{q}${\bf IDE}$Q_A$,  characterized by the parameter space $w_0^q > -1,\; w_e^q > -1,\; \xi_0 > 0, \; \xi_a > 0$ for CMB, CMB+BAO, CMB+Pantheon and CMB+BAO+Pantheon data analyses.  For CMB alone, we find a strong evidence for a non-zero interaction ($\xi_0 = 0.26_{- 0.23}^{+ 0.21}$ at 95\% CL) while $\xi_a$ is consistent with a null value, i.e. there is no evidence for a dynamical coupling for this dataset. The matter density parameter takes a very low value due to the values of both $w_0^q$ an $\xi_0$, see the parameter degeneracies illustrated in the two-dimensional contours in Fig.~\ref{fig:IDEq-ModelA-xi-dyn-w-dyn}.  As a result, this leads to a very high value of the $\sigma_8$ parameter. When BAO are added to CMB, the mean values of the parameters are shifted: namely, the evidence for a non-zero $\xi_0$  is reduced down to 68\% CL ($\xi_0 = 0.17_{-    0.07}^{+    0.10}$ at 68\% CL). For CMB+Pantheon instead we find a strong evidence for a non-zero $\xi_0$ at more than 95\% CL while no indication for a non-zero $\xi_a$ is observed. Finally, for the entire combined dataset,  i.e. for CMB+BAO+Pantheon, an evidence for a non-zero $\xi_0$ is found at more than 95\% CL, and the same significance is present for a quintessence dark energy today, while the behaviours of the other parameters are similar to the previous cases, including we a very small value of the matter density parameter and a very large value of $\sigma_8$.

Tables~\ref{tab:IDEp-ModelB-xi-dyn-w-dyn} and \ref{tab:IDEq-ModelB-xi-dyn-w-dyn} summarize the observational constraints on the interacting scenarios driven by the interaction function $Q_B$ of Eq.~(\ref{modelB}) for two disjoint parameter spaces,  ensuring the stability of the scenarios at the level of perturbations. 

Table~\ref{tab:IDEp-ModelB-xi-dyn-w-dyn} corresponds to the constraints on the interacting scenario $\xi_0\xi_a w_0^{p} w_e^{p}${\bf IDE}$Q_B$ characterized by the parameter space $w_0^p < -1,\; w_e^p < -1,\; \xi_0 < 0, \; \xi_a < 0$ for the observational datasets CMB, CMB+BAO, CMB+Pantheon and CMB+BAO+Pantheon.  For CMB alone we find an evidence for a dynamical coupling at 68\% CL: $\xi_a = -0.44_{-    0.19}^{+    0.38}$ at 68\% CL (even though at 95\% CL, this parameter is found to be unconstrained) together with an evidence for a  phantom dark energy at present ($w_0^p < -1$ at 68\% CL) and also for an early dark energy component ($w_e^p < -1$). Due to the phantom nature of $w_0$, we find a higher value of the Hubble constant. When BAO are added to CMB, the evidence for a dynamical coupling is still found at more than 68\% CL (although it remains unconstrained at 95\% CL) together with an early dark energy fluid instead of the minimal cosmological constant, present at more than 95\% CL ($w_e^p =   -1.48_{-0.45}^{+ 0.43}$ at 95\% CL, CMB+BAO). Additionally, a higher (lower) value of $\Omega_{m0}$ ($\sigma_8$) is found, due to both the energy flow and the parameter space of $w(a)$.  The inclusion of the Pantheon catalogue to CMB shifts the mean values of the parameters a bit. Notice that, similarly to the CMB+BAO case, we find an evidence for an early dark energy component at more than 95\% CL together with an evidence for a dynamical coupling in the dark sector at 68\% CL, which is diluted at 95\% CL (unlike to the previous cases, where $\xi_a$ has been found to be unconstrained). The final dataset combination, i.e. CMB+BAO+Pantheon, leads to very  similar results to the CMB+BAO case,  except for (very mild) changes in the mean parameter values.       

To conclude, Tab.~\ref{tab:IDEq-ModelB-xi-dyn-w-dyn} corresponds to the constraints on the interacting scenario $\xi_0\xi_a w_0^{q} w_e^{q}${\bf IDE}$Q_B$, characterized by  the parameter space $w_0^q > -1,\; w_e^q > -1,\; \xi_0 > 0, \; \xi_a > 0$, for the observational datasets CMB, CMB+BAO, CMB+Pantheon and CMB+BAO+Pantheon. For all four cases, the results agree completely with the $\Lambda$CDM scenario: we find no evidence for $\xi_0 \neq 0$, $\xi_a\neq 0$, while statistically we cannot exclude the possibility for a non-zero coupling,  and for a dark energy instead of the cosmological constant. Further, for CMB data alone we obtain a relatively higher value of $\Omega_{m0}$ (caused due to the exchange of energy from the dark energy sector to the dark matter one), and consequently, a lower value of the Hubble constant is found. For the remaining three analyses, i.e. CMB+BAO, CMB+Pantheon and CMB+BAO+Pantheon, we notice that both free and derived parameters, adopt very similar values to the $\Lambda$CDM case.

\section{Bayesian evidence analysis}
\label{sec-BE}

In this section we present a further statistical test of interacting dark energy cosmologies by means of  Bayesian evidence analysis~\cite{Trotta:2008qt,Trotta:2017wnx}. The main goal is to examine whether the interacting scenarios with a constant coupling parameter in the coupling functions are preferred over the interacting scenarios with a variable coupling parameter in the interaction function, or vice-versa.  Since in the present work we have considered two interaction functions (namely, $Q_A$ of Eq.~(\ref{modelA}) and $Q_B$ of Eq.~(\ref{modelB})),   for the calculations of the Bayes factors of the interacting scenarios driven by the interaction function $Q_A$, we have chosen $\xi_0$\textbf{IVS}$Q_A$ as the reference model and for the interacting scenarios driven by the interaction function $Q_B$ we have chosen $\xi_0$\textbf{IVS}$Q_B$ as the reference model. For numerical purposes we exploit the cosmological package MCEvidence~\cite{Heavens:2017hkr,Heavens:2017afc}, a publicly available code, and compute the evidences in terms
of the logarithm of the Bayes factor $\ln B_{ij}$ of the interacting scenarios for all the observational datasets. 
The numerical values of $\ln B_{ij}$ quantify the observational support of the underlying model with respect to the reference model. Following the revised Jeffreys scale ~\cite{Kass:1995loi,Trotta:2008qt},   for $0 \leq | \ln B_{ij}|  < 1$, the model has an inconclusive evidence; (ii) for $1 \leq | \ln B_{ij}|  < 2.5$, the model has  a weak evidence; (iii) for $2.5 \leq | \ln B_{ij}|  < 5$, the model has  a moderate evidence, and, finally, (iv) for $| \ln B_{ij} | \geq 5$, the model has a strong evidence.  In Tabs.~\ref{tab1:BE-cons-vs-dyn} and~\ref{tab2:BE-cons-vs-dyn}, we respectively show the $\ln B_{ij}$ values for all the interacting scenarios considering $\xi_0$\textbf{IVS}$Q_A$ and $\xi_0$\textbf{IVS}$Q_B$ as the reference models.  Note that a negative value of $\ln B_{ij}$ obtained for a particular dataset indicates that the data prefer the reference model $M_j$ (either $\xi_0$\textbf{IVS}$Q_A$ or $\xi_0$\textbf{IVS}$Q_B$ according to the choice) with respect to the interacting scenario $M_i$ for this dataset.

Looking at the numbers of $\ln B_{ij}$ as displayed in  Tabs.~\ref{tab1:BE-cons-vs-dyn} and \ref{tab2:BE-cons-vs-dyn}, one can clearly see that the nature of the dark energy equation-of-state parameter plays a crucial role in this context. We observe that interacting scenarios in which either $w_q >-1$ or $w_0^q > -1, w_e^q > -1$ are not preferred over the reference models (except for the interacting scenario $\xi_0 w_0^q w_e^q$\textbf{IDE}$Q_A$ where for the CMB alone case we notice $\ln B_{ij} = 0.5$ while for the remaining datasets we find $\ln B_{ij} < 0$). For the interacting scenarios in which $w_p <-1$ or $w_0^q < -1, w_e^q < -1$, we have mixed evidences, that means from the Bayesian point of view, both the scenarios are equally preferred. That further implies that the interacting models where the interaction function allows a variable coupling parameter can compete equally with the interaction functions having a constant coupling parameter.

\begin{table*}
	\centering
	\begin{tabular}{c c c c c}
		\hline
		\hline
		Model ~&~  CMB ~&~ CMB+BAO ~&~ CMB+Pantheon ~&~ CMB+BAO+Pantheon\\
		\hline
		\hline

		\bigskip 
		
		$\xi_0w_p$\textbf{IDE}$Q_A$ & $0.9$   & $-0.7$  & $-2.2$ & $-2.2$\\
		
		\bigskip 
		
		$\xi_0w_q$\textbf{IDE}$Q_A$ &  $-1.0$ & $-1.7$ & $-2.9$ & $-2.6$ \\
		
		\bigskip 
		
		$\xi_0 w_0^p w_e^p$\textbf{IDE}$Q_A$ &   $0.8$ & $0.9$ &  $0.9$ &  $-0.1$\\
		
		\bigskip

		$\xi_0 w_0^q w_e^q$\textbf{IDE}$Q_A$ & $0.5$ & $-0.3$ &  $-0.9$ &  $-1.0$\\

		\bigskip 
		
    $\xi_0 \xi_a$\textbf{IVS}$Q_A$ &  $0.2$ & $-1.0$   & $0.6$ &  $-2.0$\\

     \bigskip 
    
    $\xi_0 \xi_a w_p$\textbf{IDE}$Q_A$ &  $1.4$ & $-2.0$ & $-0.8$ & $-2.1$ \\

    \bigskip 
        
    $\xi_0 \xi_a w_q$\textbf{IDE}$Q_A$ & $-3.2$ & $-4.3$ & $-4.3$ & $-4.4$ \\    
    
     \bigskip 
    
    $\xi_0 \xi_a w_0^p w_e^p$\textbf{IDE}$Q_A$ & $1.0$ & $1.4$ & $-1.4$ &   $-2.5$ \\
    
    \bigskip 
    
    $\xi_0 \xi_a w_0^q w_e^q$\textbf{IDE}$Q_A$ & $-2.6$ & $-2.7$ &  $-3.3$ &  $-3.9$\\

     \hline 
     \hline 
     
	\end{tabular}
 \caption{The values of $\ln B_{ij}$  computed for the interacting scenarios driven by the interaction function $Q_A$ of eqn.~(\ref{modelA}) with respect to the reference model $\xi_0$\textbf{IVS}$Q_A$, considering all the observational datasets. We note that here $i$ refers to the interacting scenario $M_i$ and $j$ refers to the reference model $M_j$ (i.e. $\xi_0$\textbf{IVS}$Q_A$). According to the sign convention, negative value of $\ln B_{ij}$ obtained for a particular observational data indicates that the reference model $\xi_0$\textbf{IVS}$Q_A$ is favored over the interacting scenario for this dataset.   }
 \label{tab1:BE-cons-vs-dyn}
\end{table*}

\begin{table*}
	\centering
	\begin{tabular}{c c c c c}
		\hline
		\hline
		Model ~&~  CMB ~&~ CMB+BAO ~&~ CMB+Pantheon ~&~ CMB+BAO+Pantheon\\
		
		\hline
		\hline	
		
		\bigskip 
		
		$\xi_0w_p$\textbf{IDE}$Q_B$ & $0.6$   & $0$  & $-2.1$ & $-0.6$\\
		
		\bigskip 
		
		$\xi_0w_q$\textbf{IDE}$Q_B$ &  $-3.0$ & $-2.7$ & $-5.5$ & $-4.4$ \\
		
		\bigskip 
		
		$\xi_0 w_0^p w_e^p$\textbf{IDE}$Q_B$ &   $0.7$ & $0.6$ &  $-1.2$ &  $-0.7$\\
		
		\bigskip

		$\xi_0 w_0^q w_e^q$\textbf{IDE}$Q_B$ & $-1.9$ & $-2.1$ &  $-5.8$ &  $-5.3$\\

		 \bigskip 
        
    $\xi_0 \xi_a$\textbf{IVS}$Q_B$ &  $-0.1$  &   $0.6$ & $-0.8$  & $0.5$ \\
    
    \bigskip

    $\xi_0 \xi_a w_p$\textbf{IDE}$Q_B$ &  $0.2$  & $0.5$ &  $-0.1$ & $-4.1$ \\
    
    \bigskip 
    
    $\xi_0 \xi_a w_q$\textbf{IDE}$Q_B$ &  $-4.2$ & $-3.4$ & $-5.8$ & $-5.0$ \\

    \bigskip 
    
    $\xi_0 \xi_a w_0^p w_e^p$\textbf{IDE}$Q_B$ &  $0.6$ & $-0.8$ & $-0.5$ & $-2.0$\\

     \bigskip 
    
    $\xi_0 \xi_a w_0^q w_e^q$\textbf{IDE}$Q_B$ &  $-3.9$ &  $-4.3$  & $-7.6$ & $-6.2$ \\

     \hline  \hline 
     
	\end{tabular}
 \caption{The values of $\ln B_{ij}$  computed for the interacting scenarios driven by the interaction function $Q_B$ of eqn.~(\ref{modelB}) with respect to the reference model $\xi_0$\textbf{IVS}$Q_B$, considering all the observational datasets. We note that here $i$ refers to the interacting scenario $M_i$ and $j$ refers to the reference model $M_j$ (i.e. $\xi_0$\textbf{IVS}$Q_B$). According to the sign convention, negative value of $\ln B_{ij}$ obtained for a particular observational data indicates that the reference model $\xi_0$\textbf{IVS}$Q_B$ is favored over the interacting scenario for this dataset.  }
 \label{tab2:BE-cons-vs-dyn}
\end{table*}

\section{Discussion and Summary}
\label{sec-conclu}

Interacting dark energy models are very appealing cosmologies which play a crucial role in explaining several important observational issues in modern cosmology. Ranging from the `why now?' problem in cosmology,  where interacting theories have shown their ability to offer a possible explanation~\cite{Amendola:1999er,Huey:2004qv,Cai:2004dk,Pavon:2005yx,Berger:2006db,delCampo:2006vv,delCampo:2008sr,delCampo:2008jx} to current cosmological tensions~\cite{Kumar:2017dnp,DiValentino:2017iww,Yang:2018euj,Yang:2018uae,Pan:2019gop,Pan:2020bur,DiValentino:2021izs}, these particular theories have a very rich phenomenology. The underlying mechanism of interaction between dark matter and dark energy is driven by an energy and/or momentum transfer between these dark fluids and this  is usually quantified by an interaction function, $Q$. The function $Q$ modifies the expansion history of the universe both at the background and the perturbation levels, and, as a consequence, cosmological parameters are directly affected. Nevertheless, and for the shake of simplicity, almost \emph{all} works in the literature have assumed  a constant coupling parameter. Considering this issue,  the possibility of dynamical nature in the coupling parameter has been raised and investigated by a few authors in a few articles~\cite{Li:2011ga,Guo:2017deu,Wang:2018azy,Yang:2019uzo,Yang:2020tax}. Note that interacting scenarios with dynamical coupling parameter offer a much more general picture of the universe in which a constant coupling parameter would be a special case of the former. On the other hand, there is no reason to exclude the possibility of the dynamical coupling parameter in the interaction functions. In this article we have investigated this issue with a special focus on the dark energy sector, making the analyses as general as possible. We have therefore considered three different types of dark energy, quantified through their equation-of-state, namely, vacuum energy (characterized by $w = -1$), a dark energy fluid with constant  $w \neq -1 $, and a dynamical $w (a)$ with $w (a) = w_0 a + w_e (1-a)$, in which $w_0$ refers to its present value and $w_e$ refers to its value in the early time. Additionally, following Refs.~\cite{Yang:2019uzo,Yang:2020tax}, we have assumed that  the coupling parameter takes the following expression, $\xi (a) = \xi_0 + \xi_a (1-a)$, which as one can recognizes quickly, is the first two terms of the Taylor expansion of $\xi (a)$ around the present value of the scale factor, i.e. unity. We note that the above choice of the coupling parameter is the most natural one in order to detect the deviation, if any, of the coupling parameter from $\xi (a) = \xi_0  = $ constant. Finally, we further note that we have considered two interaction functions, $Q_A$ and $Q_B$, given by  Eqs.~(\ref{modelA}) and (\ref{modelB}) respectively.

We summarize below the most relevant results of our analyses, focusing on the key issue of the article, namely, the constant coupling parameter versus the dynamical coupling parameter.

\begin{itemize}
 \item {\textbf{Constant dark energy - dark matter coupling}:} 
 
 The simplest interacting scenario in this class of models is the one where the dark energy equation of state $w$ is equal to $-1$. We find only mild evidence for a coupling parameter (quantified through $\xi_a \neq 0$) at 68\% CL for model A, that is completely diluted when considering BAO observations. For the case in which  $w \neq -1$, we have divided the parameter space into two regions, to satisfy the stability requirements. For model A and $w<-1$, i.e. for the interacting scenario \textbf{$\xi_0 w_p${\bf IDE}$Q_A$}, $\xi_0$ is required to be such that $\xi_0<-1$. In this case, there is a strong  preference for $w<-1$ together with a very large value of $H_0$ without BAO data. Once BAO measurements are considered, the data suggest a phantom-interacting scenario at $1\sigma$, that persists and increases with the addition of the Pantheon data. For model A and dark energy within the  quintessence region, \textbf{$\xi_0 w_q${\bf IDE}$Q_A$}, there is a preference for a (constant) dark coupling that  is not diluted after including BAO and Pantheon measurements, but rather exacerbated at $2\sigma$. Model B gives  no indication for a coupling for all the combination of datasets considered, contrarily to model A, in the phantom region, while in the quintessence region it is very  consistent with a $\Lambda$CDM non-interacting scenario. Finally, we have also explored the case of a time-varying dark energy equation of state governed by two parameters, $w_0$, which states its current value, and $w_e$, which refers to its component at early times. For model A and $\xi_0<0$ (and therefore both $w_0$ and $w_e$ in the phantom region), for the most complete combination of data sets, i.e. CMB+BAO+Pantheon, the coupling parameter is found to be non-zero at $95\%$~CL, and a slight ($68\%$~CL) evidence for an early dark energy component (i.e. a dark energy fluid different from a cosmological constant) is also found. In this case the Hubble tension is alleviated at $2.5\sigma$ and the $S_8$ tension is relaxed. For the very same model but for $\xi_0>0$ (and therefore both $w_0$ and $w_e$ in the quintessence region) and also for the most complete combination of data sets, i.e. CMB+BAO+Pantheon, there is  evidence for a quintessence interacting scenario today at more than 95\% CL, but the value of $\Omega_{m0}$ is much lower than its $\Lambda$CDM-like value. This case exacerbates the so-called $S_8$ tension, as the value of the clustering parameter $\sigma_8$ is much higher than in the canonical $\Lambda$CDM scheme. For model B and $\xi_0<0$ (and therefore both $w_0$ and $w_e$ in the phantom region), we only find slight evidence for a non-zero coupling and for an early dark energy component different from a cosmological constant  at $1\sigma$ for the full combination CMB+BAO+Pantheon, while if we restrict the model to the quintessence region and $\xi>0$, we just note a $1\sigma$ indication for $\xi\neq0$ for the most complete data set.

 \item {\textbf{Dynamical dark energy - dark matter coupling}:}

In the following, we shall summarize our main results  for a dynamical coupling of the form  $\xi_0 + \xi_a \; (1-a)$, see Eq.~(\ref{dynamical-xi}). For vacuum dark energy and model A, the parameter values are shifted accordingly to the sign of $\xi_0$: if $\xi_0<0$ ($\xi_0>0$), the energy flows from the dark energy (matter) sector to the dark matter (energy) one, resulting in higher (lower) values of $\Omega_{m0}$ and lower (higher) values of $\sigma_8$. However, all the datasets are in agreement with no interaction in the dark sector. For model B, BAO data seems to prefer $\xi_a\neq 0$ at 68\% CL, and therefore a mild  evidence for a dynamical coupling arises when considering the former observations. If instead $w \neq -1$, we have as usual the phantom (associated to $\xi_0 <0$ and $\xi_a <0$) and quintessence (associated to $\xi_0 >0$ and $\xi_a >0$) regions. For model A in the phantom region $w>-1$ at more than $95\%$~CL and there is mild evidence for $\xi_a\neq0$ when Pantheon observations are added to CMB data. For this very same model but within the quintessence region, the dark matter energy density is very low due to the energy flow and no evidence for a dynamical dark sector is found. For Model B, we find again strong evidence for $w<-1$ within the phantom region (\textbf{$\xi_0\xi_aw_{0}^{p} w_e^{p}$IDE$Q_{B}$}) and a mild evidence for $\xi_a\neq 0$ in some parameter combinations. The quintessence region (\textbf{$\xi_0\xi_aw_{0}^{q} w_e^{p}$IDE$Q_{B}$}) is extremely in agreement to a non-interacting model.

Finally, we consider the most generalized interacting scenarios in which both the coupling and the dark energy equation of state are dynamical. Following the stability criteria~\cite{Gavela:2009cy} to have divided the parameter space. For model A in its phantom region ($w_0^p<-1$ and $w_e^p<-1$, i.e. \textbf{$\xi_0 \xi_a w_0^p w_e^p${\bf IDE}$Q_A$}), the coupling parameters are required to be negative, and we find evidence at $95\%$~CL ($68\%$~CL) significance for an early dark energy component (a dynamical coupling) for CMB+BAO+Pantheon. In the case of the model \textbf{$\xi_0 \xi_a w_0^q w_e^q${\bf IDE}$Q_A$}, the energy flow is reversed and therefore $\Omega_{m0}$ ($\sigma_8$) takes much lower (higher) values than within the minimal $\Lambda$CDM scenario. For CMB+BAO+Pantheon a non-zero value of $\xi_0$ and $w_0^q$ are found at more than $95\%$~CL, or an indication for a quintessence interacting model. Concerning model B within its phantom region (\textbf{$\xi_0 \xi_a w_0^p w_e^p${\bf IDE}$Q_B$}) the results are very similar to model A in this very same region, finding evidence at $95\%$~CL ($68\%$~CL) significance for an early dark energy component (a dynamical coupling) for CMB+BAO+Pantheon. Model 
\textbf{$\xi_0 \xi_a w_0^q w_e^q${\bf IDE}$Q_B$} instead provides results in agreement with the canonical non-interacting case except for slightly larger (smaller) values of $\Omega_{m0}$ ($\sigma_8$) for the CMB alone data, due to the energy exchange among the dark sectors.

\end{itemize}

Our summary above clearly states that interacting cosmologies have a very rich phenomenology. In particular, really interesting are those cases, where a dark energy component different from the cosmological constant at early times, and a coupling different from zero today are explored, that for the full combination CMB+BAO+Pantheon can alleviate both the $H_0$ and $S_8$ tensions, as for example for the \textbf{``$\xi_0w_0^{p}w_e^{p}$IDE$Q_A$''} model. Scenarios in which the coupling parameter is time dependent are highly promising, even from the Bayesian point of view, and are expected to be strongly constrained with upcoming cosmological observations.

\section{Acknowledgments}
WY was supported by the National Natural Science Foundation of China under Grants No. 12175096 and No. 11705079, and Liaoning Revitalization Talents Program under Grant no. XLYC1907098.
SP acknowledges the financial support from the Department of Science and Technology (DST), Govt. of India under the Scheme   ``Fund for Improvement of S\&T Infrastructure (FIST)'' (File No. SR/FST/MS-I/2019/41).  OM is supported by the Spanish grants PID2020-113644GB-I00, PROMETEO/2019/083 and by the European ITN project HIDDeN (H2020-MSCA-ITN-2019//860881-HIDDeN). EDV is supported by a Royal Society Dorothy Hodgkin Research Fellowship.

\bibliography{biblio}
\end{document}